\documentclass[aps,nofootinbib,showpacs,preprintnumbers,amsmath,amssymb]{revtex4}
\def\be{\begin{equation}}
\def\ee{\end{equation}}
\def\bea{\begin{eqnarray}}
\def\eea{\end{eqnarray}}
\def\bes{\begin{subequations}}
\def\ees{\end{subequations}}
\newcommand{\A}{{\mathcal{A}}}
\newcommand{\tA}{{\widetilde {\mathcal{A}}}}
\newcommand{\ta}{{\widetilde a}}
\newcommand{\tal}{{\widetilde \alpha}}
\newcommand{\tu}{{\widetilde u}}

\newcommand{\tk}{{\widetilde k}}
\newcommand{\td}{{\widetilde d}}
\newcommand{\tb}{{\widetilde b}}
\newcommand{\bb}{{\overline b}}
\newcommand{\tK}{{\widetilde K}}

\newcommand{\tD}{{\widetilde {\cal D}}}
\newcommand{\tg}{{\widetilde {\gamma}}}
\newcommand{\bg}{{\overline {\gamma}}}
  
\newcommand{\MSbar}{\overline{\rm MS}}

\usepackage{graphics}
\usepackage{graphicx}
\usepackage{dcolumn}
\usepackage{bm}
\usepackage{epsfig}
\usepackage{graphicx,color}
\usepackage{multirow}

\begin{document}

\vspace{1cm}

\preprint{USM-TH-358}

\begin{flushright} {\bf} \end{flushright}
 
\title{Renormalon-motivated evaluation of QCD observables}

\author{Gorazd Cveti\v{c}}
 \email{gorazd.cvetic@usm.cl}
\affiliation{Department of Physics,
Universidad T\'ecnica Federico Santa Mar\'{\i}a, Casilla 110-V, 
Valpara\'{\i}so, Chile}

\date{\today}

\begin{abstract}
  A method of evaluation of spacelike QCD observables ${\cal D}(Q^2)$ is developed, motivated by the renormalon structure of these quantities. A related auxiliary quantity ${\widetilde {\cal D}}(Q^2)$ is introduced, which is renomalization scale independent only at the one-loop level, and a large-$\beta_0$-type renormalon motivated ansatz is made for the Borel transform of this quantity. This leads to a ``dressed'' Borel transform of the considered observable ${\cal D}(Q^2)$. Subsequently, a Neubert-type distribution is obtained for the observable. The described method is then applied to the massless Adler function and the related decay ratio of the $\tau$ lepton semihadronic decays. Comparisons are then made with an evaluation method at higher truncated orders, developed in our earlier works, which is a renormalization scale invariant extension of the diagonal Pad\'e approximants.
\end{abstract}

\maketitle

\section{Introduction}
\label{sec:intr}

The perturbative QCD (pQCD) running coupling $a(Q^2) \equiv \alpha_s(Q^2)/\pi$ (where $Q^2 \equiv -q^2$), in the usual theoretical known renormalization schemes, has the peculiar property of having a significantly different regime of holomorphic (analytic) behavior in the $Q^2$-complex plane than the spacelike QCD observables ${\cal D}(Q^2)$ such as current correlators, nucleon structure functions and their sum rules. Namely, on the one hand, the general principles of Quantum Field Theories (QFTs) imply \cite{BS,Oehme} that these observables ${\cal D}(Q^2)$ are holomorphic functions in the $Q^2$-complex plane with the exception of a part of the negative semiaxis, $Q^2 \in  \mathbb{C} \backslash (-\infty, -M_{\rm thr}^2]$, where $M_{\rm thr} \sim 0.1$ GeV is a threshold scale comparable with the lightest meson mass.\footnote{
The indicated regime is usually called the (generalized) spacelike regime; and the semiaxis $Q^2 \in  (-\infty, -M_{\rm thr}^2]$ is called the timelike regime. Here, $Q^2 \equiv - q^2 = - (q^0)^2 + {\vec q} \cdot {\vec q}$, where $q$ is the relevant 4-momentum in the considered process.} 
On the other hand, the pQCD coupling $a(Q^2)$ has in general singularities along the negative semiaxis and, in addition, singularities outside the negative semiaxis, usually on the positive semiaxis $0 \leq Q^2 \leq \Lambda^2_{\rm Lan.}$. These are called Landau singuarities or Landau ghosts, and the point $Q^2_{\rm br.}=\Lambda^2_{\rm Lan.} \sim 10^{-2}$-$1 \ {\rm GeV}^2$ is the Landau branching point. This difference between the singularities of ${\cal D}(Q^2)$ and $a(Q^2)$ represents a problem in QCD from the theoretical, and from the practical point of view. Theoretically, in pQCD the leading-twist part (and the Wilson coefficients of the higher-twist part) of ${\cal D}(Q^2)$ is a function of $a( \kappa Q^2)$, where $0 < \kappa \sim 1$ is a chosen fixed renormalization scale parameter. Hence the evaluated total observable ${\cal D}(a(\kappa Q^2); Q^2)_{\rm ev.}$ does not fulfill the holomorphic properties required by the QFT principles. Practically, for low spacelike scales $Q^2$, $|Q^2| \lesssim 1 \ {\rm GeV}^2$, the argument $\mu^2=\kappa Q^2$ in the coupling $a(\kappa Q^2)$ is either close to or within the regime of the Landau singularities, making the evaluation of $a(\kappa Q^2)$ and thus of  ${\cal D}(a(\kappa Q^2); Q^2)_{\rm ev.}$ either unreliable or impossible.

This problem was addressed systematically first in Refs.~\cite{ShS,MS,Sh1Sh2,KS}, with Analytic Perturbation Theory (APT). It consists of replacing the pQCD coupling $a(Q^2)$ by a related holomorphic running coupling $\A(Q^2)^{\rm (APT)}$ which is holomorphic in  $Q^2 \in  \mathbb{C} \backslash (-\infty, 0]$ and has the same discontinuity (spectral) function $\rho_{\A}(\sigma) \equiv {\rm Im} \A(\sigma e^{-i \pi})$ across the cut along the negative semiaxis as the corresponding (underlying) pQCD coupling $a$: $\rho_{\A}(\sigma)=\rho_{a}(\sigma)$. The APT analogs $\A_n^{\rm (APT)}(Q^2)$ of the powers $a(Q^2)^n$, and their explicit expressions at one-loop, were obtained and used also for $n$ noninteger \cite{BMS} (Fractional Analytic Perturbation Theory - FAPT). For reviews of (F)APT we refer to Refs.~\cite{Shirkov,Prosperi,Bakulev,Stefanis}, and for further applications to Refs.~\cite{APTappl1,APTappl2,APTappl3}.

  Other holomorphic variants of QCD have been proposed and used since then, cf.~Refs.~\cite{Nest1,Webber,Boucaud,Alekseev,Nest2,GCrev,CV12,1danQCD,2dAQCD,2dCPC,anOPE,anOPE2,Brod,Brodrev,ArbZaits,Shirkovmass,KKS,Luna,NestBook,3l3dAQCD,4l3dAQCD}.\footnote{Mathematical packages for evaluation of specific couplings $\A$ and their power analogs are in Refs.~\cite{BK,2dCPC,2dCPCb} and on the web page \cite{MathPrgs}. Theoretically, the construction for the analogs $\A_n(Q^2)$ of the powers $a(Q^2)^n$ in general $\A$QCD variants was performed in Ref.~\cite{CV12} for integer $n$, and in \cite{GCAK} for noninteger $n$.}
  The significant difference between most of them and the (F)APT is usually the behavior of the spectral function of the coupling $\rho_{\A}(\sigma)$ in the low-momentum regime $0 \leq \sigma \lesssim 1 \ {\rm GeV}^2$. The coupling of Refs.~\cite{Nest1} is infinite at $Q^2=0$, but in most of the other works the coupling is finite nonzero there, $0 < \A(0) < \infty$. Nonetheless, in some of the works the coupling has zero value at $Q^2=0$, $\A(0)=0$, cf.~Refs.~\cite{ArbZaits,Boucaud,mes2,3l3dAQCD,4l3dAQCD}. On the other hand, large volume lattice calculations of the dressing functions of the Landau gauge gluon and ghost propagators \cite{LattcouplNf0,LattcouplNf0b,LattcouplNf24,LattcouplDiscr,Latt3gluon} give at low $Q^2$ the results which can be interpreted as implying $A(0)=0$, when the coupling is defined in a natural way as the product of the obtained gluon and the square of the ghost dressing function, $\A(Q^2) = {\rm const} \times Z_{\rm gl}(Q^2) Z_{\rm gh}(Q^2)^2$. Similar behavior of the mentioned dressing functions is also obtained in the Gribov-related and Dyson-Schwinger Equations (DSE) approaches  \cite{Gribovdecoup,Latt3gluon,DSEdecoup}. A different definition \cite{DSEdecoupFreez,PTBMF} of the low-$Q^2$ running coupling $\A(Q^2)$, which involves an additional factor $Q^2/[Q^2 + M(Q^2)^2]$, where the parameter function $M(Q^2)$ is usually called dynamical gluon mass, leads with such dressing functions to a coupling with $\A(0)>0$. 
  
  In most of the mentioned QCD variants ($\A$QCD), the coupling $\A(Q^2)$ and its power analogs $\A_n(Q^2)$ are, or can be, constructed by dispersive methods, automatically ensuring the wanted holomorphic properties of the coupling. In such dispersive approaches, nonperturbative contributions are naturally generated in the couplings $\A(Q^2)$ and $\A_n(Q^2)$. Similar kind of dispersive approaches can be applied also directly to the spacelike QCD observables ${\cal D}(Q^2)$ \cite{MSS1,MSS2,MagrGl,mes2,DeRafael,MagrTau,Nest3a,Nest3b,NestBook}, i.e., without referring to the coupling.

In this work, a method motivated by the renormalon structure of spacelike QCD observables ${\cal D}(Q^2)$ is developed. In order to obtain a practical and unambiguous evaluation of the observables ${\cal D}(Q^2)$ with such a method, the coupling $\A(Q^2)$ should not have Landau singularities. As the starting point, an auxiliary quantity ${\widetilde {\cal D}}(Q^2)$ is introduced in Sec.~\ref{sec:BTs}, which is renormalization scale invariant only at the one-loop level. Physically motivated one-loop-type (large-$\beta_0$-type) renormalon ans\"atze are made for the Borel transform ${\rm B}[{\widetilde {\cal D}}](u)$ of the latter quantity,  leading to ``dressed'' renormalon expression for the Borel transform ${\rm B}[{\cal D}](u)$ of the original observable. In Sec.~\ref{sec:Adl}, the parameters in the expression for  ${\rm B}[{\widetilde {\cal D}}](u)$ are fixed by requiring the reproduction of the known perturbation series coefficients of the observable ${\cal D}(Q^2)$, specifically in the case of the Adler function. Subsequent application of the Neubert approach \cite{Neubert} to the Borel transform of this observable gives us the characteristic distribution function $G_D(t)$ of this observable. With $G_D(t)$ we can evaluate the value of ${\cal D}(Q^2)$, the evaluation being without infrared (IR) ambiguity in $\A$QCD variants with IR-safe couplings $\A(Q^2)$. In the usual perturbative QCD (pQCD) the evaluation has ambiguity due the concurrence of the Landau singularities of the pQCD coupling $a(Q^2)= \alpha_s(Q^2)/\pi$ and the IR renormalon. This approach is then applied to the evaluation of the massless Adler function ${\cal D}(Q^2)$ in Sec.~\ref{subs:ft} and the related (timelike) $\tau$ lepton semihadronic decay ratio $r_{\tau}$ in Sec.~\ref{subs:rtau}, in two versions of QCD with IR-safe coupling ($2\delta$ \cite{2dAQCD,2dCPC} and $3\delta$ $\A$QCD \cite{4l3dAQCD}), and in pQCD in the corresponding schemes. In Sec.~\ref{sec:dBG} the obtained results are compared with those obtained with a generalization of the diagonal Pad\'e method at high truncated orders, a method developed in our earlier works. The conclusions are presented in Sec.~\ref{sec:concl}. Some additional details are presented in Appendices: in Appendix \ref{app:ktkrr} specific recursion relations for the coefficients $k_m(n)$ and $\tk_m(n)$ introduced in Sec.~\ref{subs:lpt}; in Appendix \ref{app:ndAQCD} the two used $\A$QCD variants; in Appendix \ref{app:calC} the method of obtaining specific constants ${\cal C}_{i,j}^{\rm (X)}$ appearing in Sec.~\ref{subs:fulloneloop} is explained; in Appendix \ref{app:FjFjSL} explicit expressions are given for specific integrals of the Adler characteristic functions needed in Sec.~\ref{subs:rtau}.

\section{Bare and dressed Borel transforms}
\label{sec:BTs}

\subsection{Logarithmic perturbation expansion}
\label{subs:lpt}

A spacelike QCD observable ${\cal D}(Q^2)$ is considered whose perturbation expansion in powers of $a(Q^2) \equiv \alpha_s(Q^2)/\pi$, in a given renormalization scheme, has the form
\be
{\cal D}_{\rm pt}(Q^2) = d_0 a(Q^2) + d_1 a(Q^2)^2 + \ldots + d_n a(Q^2)^{n+1} + \ldots
\label{Dpt}
\ee
This power series can be reorganized into a series in logarithmic derivatives
\be
\ta_{n+1}(Q^2) \equiv \frac{(-1)^n}{\beta_0^n n!} \left( \frac{d}{d \ln Q^2} \right)^n a(Q^2) \qquad (n=0,1,2,\ldots),
\label{tan}
\ee
where $\beta_0$ is the first coefficient in the beta function of the perturbative renormalization group equation (RGE)
\bes
\label{RGE}
\bea
\frac{d a(Q^2)}{d \ln Q^2} \equiv \beta(a(Q^2)) & = & - \beta_0 a(Q^2)^2 - \beta_1 a(Q^2)^3 - \beta_2 a(Q^2)^4 - \ldots
\label{RGEa}
\\
& = &  - \beta_0 a(Q^2)^2 \left[ 1 + c_1 a(Q^2) +  c_2 a(Q^2)^2 + \ldots \right].
\label{RGEb}
\eea
\ees
Mass-independent schemes will be considered here, where the first two coefficients are universal (scheme independent): $\beta_0 = (1/4)(11- 2 N_f/3)$ and $\beta_1=(1/16)(102 - 38 N_f/3)$, where $N_f$ is the number of effective quark flavors. This work will concentrate on the low-$|Q^2|$ regime ($|Q^2| \lesssim 1 \ {\rm GeV}^2$) where $N_f=3$. The higher beta coefficients $c_j \equiv \beta_j/\beta_0$ ($j=2,3,\ldots$) characterize the renormalization scheme. The usual $\MSbar$ scaling convention for momenta $Q^2$ will be used throughout ($\Leftrightarrow \Lambda^2_{\rm QCD} ={\overline \Lambda}^2$).

Repeated use of the RGE (\ref{RGE}) makes it possible to express the logarithmic derivatives Eq.~(\ref{tan}) as a power series
\be
\ta_n(Q^2) = a(Q^2)^n + \sum_{m=1}^{\infty} k_m(n) \; a(Q^2)^{n+m}.
\label{tanan1}
\ee
For example, we have $\ta_1=a$ and
\bes
\label{tanan2}
\bea
\ta_2 &=& a^2 + c_1 a^3 + c_2 a^4 + c_3 a^5 + \ldots,
\label{ta2a}
\\
\ta_3 & = & a^3 + \frac{5}{2} c_1 a^4 + \left( 3 c_2 + \frac{3}{2} c_1^2 \right) a^5 + \ldots,
\label{ta3a}
\\
\ta_4 & = & a^4 + \frac{13}{3} c_1 a^5 + \ldots,
\label{ta4a}
\\
\ta_5 & = & a^5 + \ldots, \qquad {\rm etc.},
\label{ta5a}
\eea
\ees
where the dependence on $Q^2$ was omitted for simplicity of notation [$a^n \equiv a(Q^2)^n$, $\ta_n \equiv \ta_n(Q^2)$].
Step-by-step inversion of these relations makes it possible to express separate powers in terms of the logarithmic derivatives
\be
a(Q^2)^n = \ta_n(Q^2) + \sum_{m=1}^{\infty} \tk_m(n) \; \ta_{n+m}(Q^2) \ .
\label{antan1}
\ee
For example, we have
\bes
\label{antan}
\bea
a^5 &=&  \ta_5 + \ldots,
\label{a5ta}
\\
a^4 & = & \ta_4 - \frac{13}{3} c_1 \ta_5 + \ldots,
\label{a4ta}
\\
a^3 & = & \ta_3 - \frac{5}{2} c_1 \ta_4 + \left( -3 c_2 + \frac{28}{3} c_1^2 \right) \ta_5 + \ldots,
\label{a3ta}
\\
a^2 &=& \ta_2 - c_1 \ta_3 + \left(-c_2 +\frac{5}{2} c_1^2 \right) \ta_4 + \left(- c_3 a^5 + \frac{22}{3} c_1 c_2 - \frac{28}{3} c_1^3 \right) \ta_5 + \ldots,
\label{a2ta}
\eea
\ees
etc. When these expressions for each of the powers $a^n$ are substituted in the power expansion Eq.~(\ref{Dpt}) of ${\cal D}(Q^2)$, the rearranged expansion in the logarithmic detivatives (logarithmic perturbation expansion - 'lpt') is obtained
\be
{\cal D}_{\rm lpt}(Q^2) = \td_0 a(Q^2) + \td_1 \ta_2(Q^2) + \td_2 \ta_3(Q^2) + \ldots + \td_n \ta_{n+1}(Q^2) + \ldots,
\label{Dlpt}
\ee
where the new coefficients $\td_n$ are specific combinations of $d_n$, $d_{n-1}$,$\ldots$, $d_0$
\be
\td_n = \sum_{s=0}^{n-1} \tk_s(n+1-s) \; d_{n-s}  \quad (n=1,2,\ldots; \tk_0(m)=0),
\label{tdndn}
\ee
and where the coefficients $\tk_s(n+1-s)$ are those appearing in the relations (\ref{antan1}). The more explicit form of Eqs.~(\ref{tdndn}) is
\bes
\label{tdns}
\bea
\td_0 & = & d_0, \quad \td_1=d_1, \quad \td_2 = d_2 - c_1 d_1,
\label{td012}
\\
\td_3 & = & d_3 - \frac{5}{2} c_1 d_2 + \left( - c_2 + \frac{5}{2} c_1^2 \right) d_1,
\label{td3}
\\
\td_4 & = & d_4 - \frac{13}{3} c_1 d_3 + \left(-3 c_2 + \frac{28}{3} c_1^2 \right) d_2 + \left(-c_3 + \frac{22}{3} c_1 c_2 - \frac{28}{3} c_1^3 \right) d_1,
\label{td4}
\\
\td_5 & = & d_5 -\frac{77}{12} c_1 d_4  + \left(-6 c_2 + \frac{791}{36} c_1^2 \right) d_3 + \left(-\frac{7}{2} c_3 + \frac{123}{4} c_1 c_2 - \frac{1631}{36} c_1^3 \right) d_2
\nonumber\\
&& + \left(-c_4 + \frac{119}{12} c_1 c_3 + 6 c_2^2 - \frac{949}{18} c_1^2 c_2 + \frac{1631}{36} c_1^4 \right) d_1,
\label{td5}
\eea
\ees
etc. The inverse relations are
\be
d_n = \sum_{s=0}^{n-1} k_s(n+1-s) \; \td_{n-s} \quad (n=1,2,\ldots; k_0(m)=0),
\label{dntdn}
\ee
where the coefficients $k_s(n+1-s)$ are those appearing in the relations (\ref{tanan1}). We refer to Appendix \ref{app:ktkrr} for recursion relations which allow us to obtain the coefficients $k_m(n)$ and $\tk_m(n)$ to any order in any given renormalization scheme.

Having obtained the modified coefficients $\td_n$ [via Eqs.~(\ref{tdndn})-(\ref{tdns})], an auxiliary quantity $\tD(Q^2)$ is constructed whose power expansion is obtained by the formal replacements $\ta_n(Q^2) \mapsto a(Q^2)^n$ in the logarithmic perturbation expansion (\ref{Dlpt})
\be
\tD(Q^2) = \td_0 a(Q^2) + \td_1 a(Q^2)^2 + \td_2  a(Q^2)^3 + \ldots + \td_n a(Q^2)^{n+1} + \ldots.
\label{tD1}
\ee
While the observable ${\cal D}(Q^2)$ is independent of the renormalization scale $\mu^2$ (i.e., $\kappa$-independent where $\kappa \equiv \mu^2/Q^2$)
\bes
\label{Dkapptlpt}
\bea
{\cal D}_{\rm pt}(Q^2) &=& d_0 a(\mu^2) + d_1(\kappa)  a(\mu^2)^2 + d_2(\kappa) a(\mu^2)^3 + \ldots + d_n(\kappa) a(\mu^2)^{n+1} + \ldots,
\label{Dkappt}
\\
{\cal D}_{\rm lpt}(Q^2) &=& \td_0 a(\mu^2) + \td_1(\kappa)  \ta_2(\mu^2) + \td_2(\kappa) \ta_3(\mu^2) + \ldots + \td_n(\kappa) \ta_{n+1}(\mu^2) + \ldots,
\label{Dkaplpt}
\eea
\ees
the quantity $\tD$ is not an observable because it is $\kappa$-independent only at the one-loop level.\footnote{In this work the $\kappa$-dependent quantities at $\kappa=1$ (i.e., for $\mu^2=Q^2$) are simply denoted without reference to $\kappa$; e.g., $d_n(\kappa=1) \equiv d_n$, $\tD(Q^2; \kappa=1) \equiv \tD(Q^2)$.}$^{,}$\footnote{\label{tdndnkappa} It is straightforward to see that the relations (\ref{tdndn}) and (\ref{dntdn}) are valid at any $\kappa$ ($>0$), not just $\kappa=1$, i.e., the coefficients $\tk_s(m)$ and $k_s(m)$ are independent of $\kappa$: $\td_n(\kappa)=\sum_{s=0}^{n-1} \tk_s(n+1-s) \; d_{n-s}(\kappa)$ and $d_n(\kappa)=\sum_{s=0}^{n-1} k_s(n+1-s) \; \td_{n-s}(\kappa)$.}
Namely, the dependence of the coefficients $d_n(\kappa)$ and thus of $\td_n(\kappa)$ on the renormalization scale parameter $\kappa$ is determined uniquely by the $\mu^2$-independence of ${\cal D}_{\rm pt}(Q^2)$, and then the quantity
\be
\tD(Q^2;\kappa) = \td_0 a(\mu^2) + \td_1(\kappa) a(\mu^2)^2 + \td_2(\kappa) a(\mu^2)^3 + \ldots + \td_n(\kappa) a(\mu^2)^{n+1} + \ldots
\label{tDkap}
\ee
is $\kappa$-dependent at the level beyond one-loop. More specifically, it is possible to check that $\mu^2$-independence of the series (\ref{Dkaplpt}) implies the following $\kappa$-dependence of the coefficients $\td_n(\kappa)$:
\bes
\label{tdnkap}
\bea
\frac{d}{d \ln \kappa}  \td_0(\kappa) &=& 0,
\label{td0kap}
\\
\frac{d}{d \ln \kappa} \td_n(\kappa) &=& n \beta_0 \td_{n-1}(\kappa)
\quad (n \geq 1).
\label{tdnot0kap}
\eea
\ees
As a consequence, $\td_0(\kappa)$ is $\kappa$-independent, and
\be
\td_n(\kappa) = \td_n + \sum_{k=1}^n \binom{n}{k} (\beta_0 \ln \kappa)^k \td_{n-k} \quad (n \geq 1),
\label{tdnkapsum}
\ee
where $\td_n \equiv \td_n(1)$. This implies that, if $a(\mu^2)$ were obtained from $a(Q^2)$ by one-loop RGE running, the expansion (\ref{tDkap}) would be $\mu^2$-independent (note: $\kappa \equiv \mu^2/Q^2$).

Based on the relations (\ref{tdnkap}), it is the straightforward to show that the Borel transform of the quantity $\tD(Q^2;\kappa)$ of Eq.~(\ref{tDkap}) 
\bea
{\rm B} [\tD](u;\kappa) & \equiv & \td_0(\kappa) + \frac{\td_1(\kappa)}{1! \beta_0} u + \frac{\td_2(\kappa)}{2! \beta_0^2} u^2 + \ldots +  \frac{\td_n(\kappa)}{n! \beta_0^n} u^n + \ldots
\label{BtD1}
\eea
has a simple (``one-loop-type'') $\kappa$-dependence
\be
{\rm B} [\tD](u;\kappa) = \kappa^u {\rm B} [\tD](u) .
\label{BtDkap}
\ee
We point out that this is not an approximate, but exact $\kappa$-dependence. On the other hand, the Borel transform of the perturbation series Eq.~(\ref{Dkappt}) of the observable ${\cal D}(Q^2)$ has the $\kappa$-dependence that is only approximately of the type (\ref{BtDkap}), namely at the one-loop level approximation.

These considerations suggest that the Borel transform ${\rm B}[\tD](u)$ of the auxiliary quantity $\tD(Q^2)$ has a one-loop (large-$\beta_0$) type renormalon structures
\be
{\rm B}[\tD](u) =  \frac{1}{(p-u)^k}, \; \frac{1}{(p+u)^k},
\label{BtDpoles}
\ee
where $p$ and $k$ are positive integers. The first structure represents a large-$\beta_0$-type $k$-fold IR renormalon at $u=p$, and the second a large-$\beta_0$-type $k$-fold UV renormalon at $u=-p$. For example, the IR and UV $p$ double pole renormalons (DP, $k=2$) generate coefficients $\td_n \propto (n+1)! (\beta_0/p)^n$ and $(n+1)! (-\beta_0/p)^n$, respectively; the single pole (SP, $k=1$) variant generates $\td_n \propto (n)! (\beta_0/p)^n$ and $(n)! (-\beta_0/p)^n$, respectively. In addition to these structures, the IR and UV $p$ subleading renormalon (SL, '$k=0$') can (and will) be included
\be
{\rm B}[\tD](u) =  -\ln(1 - u/p), \; \ln(1+u/p),
\label{BtDSLs}
\ee
which, although not meromorphic functions, generate the analogous ``subleading'' type of coefficients $\td_n = (n-1)! (\beta_0/p)^n$ and $(n-1)! (-\beta_0/p)^n$, respectively.

In practice, for QCD spacelike observables ${\cal D}(Q^2)$ we know the exact values of only the first three or four coefficients $d_n$ and $\td_n$ (i.e., $n=0,1,2,3$). One of the aims of the present work is to obtain a physically motivated estimate of all the other higher-order coefficients $d_n$ and $\td_n$ ($n \geq 4$). The idea is to make a physically motivated ansatz for the Borel transform ${\rm B}[\tD](u)$ as a sum of the (one-loop type) renormalon terms of the form Eqs.~(\ref{BtDpoles})-(\ref{BtDSLs}), and adjust the weights of these renormalon terms so that the (known) values of the first three or four coefficients $\td_n$ are reproduced.\footnote{It is mentioned here in advance that, knowing the Borel transform of the auxiliary quantity $\tD(Q^2)$, allows one to construct the Neubert-type of characteristic function $G_D(t)$ for the observable ${\cal D}(t)$, and thus to evaluate the (leading-twist) part of ${\cal D}(Q^2)$ as an integral over $t$ involving $G_D(t)$ and the running coupling $a(t Q^2)$.} However, before embarking on this, it has to be checked first whether the forms Eqs.~(\ref{BtDpoles})-(\ref{BtDSLs}) generate physically acceptable coefficients $\td_n$. Namely, the generated coefficients $\td_n$ lead, via the relations (\ref{dntdn}), to the coefficients $d_n$ for the power expansion (\ref{Dpt}) of the observable ${\cal D}(Q^2)$,\footnote{Note that the relations (\ref{dntdn}), with the same coefficients $k_s(m)$, are valid at any renormalization scale $\mu^2 = \kappa Q^2$, i.e. $d_n(\kappa)=\sum_{s=0}^{n-1} k_s(n+1-s) \; \td_{n-s}(\kappa)$, cf.~also footnote \ref{tdndnkappa}.}
and the question is whether the (higher-order) coefficients $d_n$ obtained in this way fulfill the physically motivated expectations. The latter expectations are contained in the expected renormalon structure of the Borel transform of the observable ${\cal D}(Q^2)$
\be
{\rm B}[{\cal D}](u; \kappa) \equiv  d_0(\kappa) + \frac{d_1(\kappa)}{1! \beta_0} u + \frac{d_2(\kappa)}{2! \beta_0^2} u^2 + \ldots +  \frac{d_n(\kappa)}{n! \beta_0^n} u^n + \ldots .
\label{BD1}
\ee

\subsection{Full renormalon structure of ${\cal D}(Q^2)$}
\label{subs:renstr}

In many cases of  spacelike observables $D(Q^2)$ such as Adler function, the expected structure of ${\rm B}[D](u;\kappa)$ is known and comes from the IR and UV renormalons. The IR renormalon leads to an ambiguity of the Borel integral, the origin of this ambiguity is the low-momentum regime of the Feynman integrals, and consequently \cite{Mueller,Renormalons,BenBra} the $Q^2$-dependence of these terms must agree with the power-suppressed (higher-twist) terms appearing in the Operator Product Expansion (OPE). A $D$-dimensional OPE term of the spacelike observable ${\cal D}(Q^2)$ has the form
\be
{\hat C}_{O_D}(a_Q) \frac{\langle {\hat O}_D \rangle}{(Q^2)^{D/2}} = {\cal C}_{O_D}(a_Q) \frac{\langle O_D(Q^2) \rangle}{(Q^2)^{(D/2)}},
\label{OPED}
\ee
where $a_Q \equiv a(Q^2)$, $ {\cal C}_{O_D}(a_Q)$ is the Wilson coefficient of the operator [usually of the form $1 + {\cal O}(a_Q)$], $\langle {\hat O}_D \rangle$ is a scale-invariant operator, and $\langle O_D(Q^2) \rangle$ is the $Q^2$-dependent operator with anomalous dimension 
\be
\gamma_{O_D} \equiv - \frac{d \ln \langle O_D(Q^2) \rangle}{d \ln Q^2} =
\gamma_{O_D}^{(1)} a_Q + \gamma_{O_D}^{(2)} a_Q^2 + \ldots
\label{andimdef}
\ee
which then gives the relation between $\langle O_D(Q^2) \rangle$ and $\langle {\hat O}_D \rangle$
\bea
\langle O_D(Q^2) \rangle & = & \langle {\hat O}_D \rangle \exp \left[ - \int_{\rm const}^{a_Q} \frac{\gamma_{O_D}(a_{\mu})}{\beta(a_{\mu})} d a_{\mu} \right]
\nonumber\\
& = & {\rm const} \times (a_Q)^{\gamma_{O_D}^{(1)}/\beta_0} \left[ 1 +
  \frac{1}{\beta_0} ( \gamma_{O_D}^{(2)} - \gamma_{O_D}^{(1)} c_1 ) a_Q + {\cal O}(a_Q^2) \right] .
\label{andimint}
\eea
Furthermore, RGE gives for the inverse powers of $Q^2$ the expression
\be
\frac{1}{(Q^2)^{D/2}} = {\rm const} \times \exp \left(- \frac{D}{2 \beta_0 a_Q} \right) (a_Q)^{-(D/2) (c_1/\beta_0)} \left[ 1 + b_1^{(D)} a_Q + b_2^{(D)} a_Q^2 + {\cal O}(a^3) \right],
\label{invpow}
\ee
where the overall constant is $Q^2$-independent and renormalization scheme independent, and the $b_j^{(D)}$ coefficients are
\bes
\label{bjDs}
\bea
b_1^{(D)} &=& \frac{D}{2 \beta_0} ( c_1^2 - c_2),
\label{b1D}
\\
b_2^{(D)} &=& \frac{1}{2} (b_1^{(D)})^2 - \frac{D}{4 \beta_0} ( c_1^3 - 2 c_1 c_2 + c_3).
\label{b2D}
\eea
\ees
The $Q^2$-dependence of the OPE operator term is then
\bea
{\cal D}^{(D)}_{\rm OPE}(Q^2) & = & {\rm const}  \times (a_Q)^{\gamma^{(1)}_{O_D}/\beta_0} \left[ 1 + {\hat c}_1^{(D)} a_Q + {\hat c}_2^{(D)} a_Q^2 + {\cal O}(a^3) \right] \langle {\hat O}_D \rangle
\nonumber\\ && \times
\exp \left(- \frac{D}{2 \beta_0 a_Q} \right) (a_Q)^{-(D/2) (c_1/\beta_0)} \left[ 1 + b_1^{(D)} a_Q + b_2^{(D)} a_Q^2 + {\cal O}(a^3) \right],
\label{Q2depOPE}
\eea
where the first series, with coefficients ${\hat c}_j^{(D)}$, represents the product of the Wilson coefficient ${\cal C}_{O_D}(a_Q)={\rm const} \times (1 + {\cal C}_1 a_Q + \ldots)$ and of the subleading effects of the exponent of the integral with the anomalous dimension (\ref{andimint})
\bea
\left[1 + {\cal C}_1 a_Q + \ldots \right] \left[ 1 +
  \left( \frac{\gamma_{O_D}^{(2)}}{\beta_0} - \gamma_{O_D}^{(1)} \frac{\beta_1}{\beta_0^2} \right) a_Q + \ldots \right] & = & \left[ 1 + \sum_{j=1}^{\infty} {\hat c}_j^{(D)} a_Q^j \right].
\label{Wcandim}
\eea
On the other hand, the $Q^2$-dependence coming from the IR renormalon ambiguity of the IR $u=p$ renormalon can be obtained by evaluating the imaginary part ${\rm Im} {\cal D}(Q^2)^{(\pm)}_{{\rm IR},p,{\rm BI}}$ of the generalized Principal Value of the Borel integral
\be
{\cal D}(Q^2)^{(\pm)}_{{\rm IR},p,{\rm BI}} = \frac{1}{\beta_0} \int_{\pm i \epsilon}^{+\infty \pm i \epsilon} du \exp \left( - \frac{u}{\beta_0 a_Q} \right) B[{\cal D}_{{\rm IR},p}](u),
\label{PVBI}
\ee
where the Borel transform of the IR $p$ renormalon takes the ansatz
\be
B[{\cal D}_{{\rm IR},p}](u) = \frac{\pi d_p^{\rm IR}}{(p - u)^{\tg_p^{(1)}}} \left[ 1 + \tb_1^{(2p)} (p-u) + \tb_2^{(2p)} (p-u)^2 + \ldots \right].
\label{BDIRp}
\ee
Direct evaluation gives for this ambiguity (cf., e.g., \cite{PRD67GC})
\bea
\frac{1}{\pi} {\rm Im} {\cal D}(Q^2)^{(\pm)}_{{\rm IR},p,{\rm BI}} &=&
\mp d_p^{\rm IR} \frac{\Gamma(-\tg_p^{(1)}+1) \sin (\pi (\tg_p^{(1)}-1))}{\beta_0^{\tg_p^{(1)}}}
\exp \left( - \frac{p}{\beta_0 a_Q} \right) a_Q^{-\tg_p^{(1)}+1}
\nonumber\\ &&\times
\left[ 1 + \left( \tb_1^{(2p)} (\tg_p^{(1)}-1) \beta_0 \right) a_Q + \left( \tb_2^{(2p)} (\tg_p^{(1)}-1) (\tg_p^{(1)}-2) \beta_0^2 \right) a_Q^2 + {\cal O}(a^3) \right].
\label{renamb}
\eea
As mentioned, the $Q^2$-dependence of the OPE term Eq.~(\ref{Q2depOPE}) is the same as that of the renormalon ambiguity Eq.~(\ref{renamb}), leading to the following expressions of the Borel renormalon parameters in terms of the corresponding OPE operator parameters:
\bes
\label{OPErenamb}
\bea
p & = & \frac{D}{2}, \quad \tg_p^{(1)}=1 - \frac{\gamma_{O_D}^{(1)}}{\beta_0} + p \frac{c_1}{\beta_0},
\label{OPErenamb1}
\\
\tb_1^{(2 p)} &=& \frac{1}{\beta_0 (\tg_p^{(1)}-1)} (b_1^{(2 p)} + {\hat c}_1^{(2p)}),
\quad
\tb_2^{(2 p)} = \frac{1}{\beta_0^2 (\tg_p^{(1)}-1) (\tg_p^{(1)}-2)} (b_2^{(2p)} + {\hat c}_1^{(2p)} b_1^{(2 p)} + {\hat c}_2^{(2p)}).
\label{OPErenamb2}
\eea
\ees
The coefficient $p$ is a positive integer for spacelike observables.
The corresponding expressions for the coefficients $(d_n)_{{\rm IR},p}$ at $a_Q^{n+1}$ in the perturbation expansion ('pt')
\be
{\cal D}(Q^2)_{{\rm IR},p,{\rm pt}} = \sum_{n=0}^{\infty} (d_n)_{{\rm IR},p} a(Q^2)^{n+1},
\label{DIRppt}
\ee
can then be obtained directly from the Borel transform (\ref{BDIRp}), using the expansion (\ref{BD1}) for $\kappa=1$
\bea
(d_n)_{{\rm IR},p} &=& \frac{\pi d_p^{\rm IR}}{p^{\tg_p^{(1)}} \Gamma[\tg_p^{(1)}]} \Gamma(\tg_p^{(1)}+n)
\left( \frac{\beta_0}{p} \right)^n {\Big [} 1 + (b_1^{(2p)} + {\hat c}_1^{(2p)}) \frac{p}{\beta_0} \frac{1}{(\tg_p^{(1)}-1 + n)}
\nonumber\\ &&
+ (b_2^{(2p)} + {\hat c}_1^{(2p)} b_1^{(2 p)} + {\hat c}_2^{(2p)}) \left( \frac{p}{\beta_0} \right)^2 \frac{1}{(\tg_p^{(1)}-1 + n)(\tg_p^{(1)}-2 + n)} + {\cal O}\left( \frac{1}{n^3} \right) {\Big ]}.
\label{dnIRp}
\eea

When considering the UV $u=-p$ renormalon contribution ${\cal D}(Q^2)_{{\rm UV},p}$, whose Borel transform has the ansatz
\be
B[{\cal D}_{{\rm UV},p}](u) = \frac{\pi d_p^{\rm UV}}{(p + u)^{\bg_p^{(1)}}} \left[ 1 + \bb_1^{(2p)} (p+u) + \bb_2^{(2p)} (p+u)^2 + \ldots \right],
\label{BDUVp}
\ee
the previous analysis can be repeated analogously \cite{BeJa08} (cf.~also \cite{BeJa13}), considering $a_Q$ as negative, $a_Q=-|a_Q|$, and in the Borel integral the integration over $u$ now goes from $\pm i \epsilon$ to $\pm i \epsilon - \infty$. The corresponding (formal) OPE operators have dimension $D = - 2 p$ ($<0$), and the relations analogous to those of Eqs.~(\ref{OPErenamb}) are
\bes
\label{OPErenambUV}
\bea
p & = & -\frac{D}{2}, \quad \bg_p^{(1)}=1 - \frac{\gamma_{O_D}^{(1)}}{\beta_0} - p \frac{c_1}{\beta_0},
\label{OPErenambUV1}
\\
\bb_1^{(2 p)} &=& \frac{1}{(-\beta_0) (\bg_p^{(1)}-1)} (b_1^{(-2 p)} + {\hat c}_1^{(-2p)}),
\quad
\bb_2^{(2 p)} = \frac{1}{\beta_0^2 (\bg_p^{(1)}-1) (\bg_p^{(1)}-2)} (b_2^{(-2p)} + {\hat c}_1^{(-2p)} b_1^{(-2 p)} + {\hat c}_2^{(-2p)}),
\label{OPErenambUV2}
\eea
\ees
and the corresponding coefficients  $(d_n)_{{\rm UV},p}$ at $a_Q^{n+1}$ in the perturbation expansion ('pt') of ${\cal D}_{{\rm UV},p}$ are
\bea
(d_n)_{{\rm UV},p} &=& \frac{\pi d_p^{\rm UV}}{p^{\bg_p^{(1)}} \Gamma[\bg_p^{(1)}]} \Gamma(\bg_p^{(1)}+n)
\left( \frac{\beta_0}{-p} \right)^n {\Big [} 1 + (b_1^{(-2p)} + {\hat c}_1^{(-2p)}) \left( \frac{-p}{\beta_0} \right) \frac{1}{(\bg_p^{(1)}-1 + n)}
\nonumber\\ &&
+ (b_2^{(-2p)} + {\hat c}_1^{(-2p)} b_1^{(-2 p)} + {\hat c}_2^{(-2p)}) \left( \frac{-p}{\beta_0} \right)^2 \frac{1}{(\bg_p^{(1)}-1 + n)(\bg_p^{(1)}-2 + n)} + {\cal O}\left( \frac{1}{n^3} \right) {\Big ]}.
\label{dnUVp}
\eea

\subsection{Generation of the full renormalon structure from the one-loop-type renormalon structure}
\label{subs:fulloneloop}

This Section addresses the question of what kind of the renormalon structure for the observable ${\cal D}(Q^2)$ is obtained when the one-loop-type of renormalon structures Eqs.~(\ref{BtDpoles})-(\ref{BtDSLs}) are adopted for the auxiliary quantity ${\cal \tD}(Q^2)$ Eq.~(\ref{tD1}). We will see that the full renormalon structures of the type Eqs.~(\ref{BDIRp}) with (\ref{OPErenamb}) are obtained in the case of IR renormalons, and those of Eqs.~(\ref{BDUVp}) with (\ref{OPErenambUV}) in the case of UV renormalons. The numerical investigation will be performed in more detail in two specific renormalization schemes: the Lambert scheme with $c_2=-4.9$, and (four-loop) Lambert MiniMOM (LMM) scheme. These are the two schemes in which IR-safe (and holomorphic) QCD couplings $\A(Q^2)$ were constructed \cite{2dAQCD, 4l3dAQCD} which can be used naturally later on in the numerical evaluations of the observables ${\cal D}(Q^2)$, giving unambiguous numerical results. This is so because these couplings, although coinciding practically with the underlying perturbative coupling $a(Q^2)$ at high $|Q^2|$, do not have Landau singularities at lower $|Q^2| \lesssim 1 \ {\rm GeV}^2$. For more details on these couplings, the reader is referred to Appendix \ref{app:ndAQCD}.  

In practice, the Borel transforms of the form Eqs.~(\ref{BtDpoles})-(\ref{BtDSLs}) generate $\td_n$ coefficients from which, in practice, the coefficients $d_n$ up to $n=n_{\rm max}=70$ were generated via the relations (\ref{dntdn}), using Mathematica software \cite{Math} (cf.~also Appendix \ref{app:ktkrr}). These $d_n$ coefficients, at large $n$, are then compared with the expressions of the form of Eqs.~(\ref{dnIRp}) and (\ref{dnUVp}) originating from the form of the full renormalon Borel transforms ${\rm B}[{\cal D}](u)$ as suggested by the theory, Eqs.~(\ref{BDIRp})-(\ref{OPErenamb}) and (\ref{BDUVp})-(\ref{OPErenambUV}). It will be seen that the numerical results confirm that these structures are really reproduced.

Specifically, for the $p=2$ and $p=3$ IR renormalons, four cases were considered (in the LMM, Lambert, and $\MSbar$ schemes) for the one-loop-type ans\"atze  for the Borel transforms ${\rm B}[\tD](u)$ of the auxiliary quantity $\tD(Q^2)$: double pole ('DP'), single pole ('SP'), subleading ('SL'), subsubleading ('SSL').
\bes
\label{BtDp}
\bea
{\rm B}[\tD](u)_{{\rm IR},p,{\rm DP}} & = & \frac{\pi \td_{p,2}^{\rm IR}}{(p-u)^2},
\label{BtDpDP}
\\
{\rm B}[\tD](u)_{{\rm IR},p,{\rm SP}} & = & \frac{\pi \td_{p,1}^{\rm IR}}{(p-u)},
\label{BtDpSP}
\\
{\rm B}[\tD](u)_{{\rm IR},p,{\rm SL}} & = & \pi \td_{p,0}^{\rm IR} (-1) \ln \left(1 - \frac{u}{p} \right),
\label{BtDpSL}
\\
{\rm B}[\tD](u)_{{\rm IR},p,{\rm SSL}} & = & \pi \td_{p,-1}^{\rm IR} (p-u) \ln \left(1 - \frac{u}{p} \right),
\label{BtDpSSL}
\eea
\ees
These Borel transforms generate the following perturbation expansion coefficients $\td_n$ of the corresponding auxiliary quantities $\tD(Q^2)$:
\bes
\label{BtDtdns}
\bea
(\td_n)_{{\rm IR},p,{\rm DP}}{\big |}_{\td_{p,2}^{\rm IR}=p^2/\pi} & = &
(n+1)! \left( \frac{\beta_0}{p} \right)^n,
\label{tdnIRpDP}
\\
(\td_n)_{{\rm IR},p,{\rm SP}}{\big |}_{\td_{p,1}^{\rm IR}=p/\pi} & = &
n! \left( \frac{\beta_0}{p} \right)^n,
\label{tdnIRpSP}
\\
(\td_n)_{{\rm IR},p,{\rm SL}}{\big |}_{\td_{p,0}^{\rm IR}=1/\pi} & = &
(n-1)! \left( \frac{\beta_0}{p} \right)^n \quad (n \geq 1),
\label{tdnIRpSL}
\\
(\td_n)_{{\rm IR},p,{\rm SSL}}{\big |}_{\td_{p,-1}^{\rm IR}=1/(p \pi)} & = &
(n-2)! \left( \frac{\beta_0}{p} \right)^n \quad (n \geq 2),
\label{tdnIRpSSL}
\eea
\ees
The corresponding coefficients $d_n$ of the observable ${\cal D}(Q^2)$ that are obtained numerically from these $\td_n$'s by the aforementioned relations (\ref{dntdn}), turn out to agree at high $n$ with the following expressions to a high precision ($n \leq n_{\rm max}$ with $n_{\rm max}=70$ was used):
\bes
\label{BDdns}
\bea
( d_n )_{{\rm IR},p,{\rm DP}} & = & \frac{\pi d_{p,2}^{\rm IR}}{p^{\tg_p+1} \Gamma(\tg_p+1)} \Gamma(\tg_p+1+n) \left( \frac{\beta_0}{p} \right)^n {\Bigg \{} 1 + \left( b_1^{(2p)} + {\cal C}_{1,2}^{(2p)} \right) \left( \frac{p}{\beta_0} \right) \frac{1}{(\tg_p+n)}
\nonumber\\ &&
+ \left( b_2^{(2p)} + b_1^{(2p)} {\cal C}_{1,2}^{(2p)} + {\cal C}_{2,2}^{(2p)} \right)  \left( \frac{p}{\beta_0} \right)^2 \frac{1}{(\tg_p+n)(\tg_p-1 + n)} + {\cal O} \left( \frac{1}{n^3} \right) {\Bigg \}},
\label{dnIRpDP}
\\
( d_n )_{{\rm IR},p,{\rm SP}} & = & \frac{\pi d_{p,1}^{\rm IR}}{p^{\tg_p} \Gamma(\tg_p)} \Gamma(\tg_p+n) \left( \frac{\beta_0}{p} \right)^n {\Bigg \{} 1 + \left( b_1^{(2p)} + {\cal C}_{1,1}^{(2p)} \right) \left( \frac{p}{\beta_0} \right) \frac{1}{(\tg_p-1+n)}
\nonumber\\ &&
+ \left( b_2^{(2p)} + b_1^{(2p)} {\cal C}_{1,1}^{(2p)} + {\cal C}_{2,1}^{(2p)} \right)  \left( \frac{p}{\beta_0} \right)^2 \frac{1}{(\tg_p-1+n)(\tg_p-2 + n)} + {\cal O} \left( \frac{1}{n^3} \right) {\Bigg \}},
\label{dnIRpSP}
\\
( d_n )_{{\rm IR},p,{\rm SL}} & = & \frac{\pi d_{p,0}^{\rm IR}}{p^{\tg_p-1} \Gamma(\tg_p-1)} \Gamma(\tg_p-1+n) \left( \frac{\beta_0}{p} \right)^n {\Bigg \{}
1 + \left( b_1^{(2p)} + {\cal C}_{1,0}^{(2p)} \right) \left( \frac{p}{\beta_0} \right) \frac{1}{(\tg_p-2+n)}  + {\cal O} \left( \frac{1}{n^2} \right) {\Bigg \}},
\label{dnIRpSL}
\\
( d_n )_{{\rm IR},p,{\rm SSL}} & = & \frac{\pi d_{p,-1}^{\rm IR}}{p^{\tg_p-2} \Gamma(\tg_p-2)} \Gamma(\tg_p-2+n) \left( \frac{\beta_0}{p} \right)^n {\Bigg \{}
1 + \left( b_1^{(2p)} + {\cal C}_{1,-1}^{(2p)} \right) \left( \frac{p}{\beta_0} \right) \frac{1}{(\tg_p-3+n)}  + {\cal O} \left( \frac{1}{n^2} \right) {\Bigg \}},
\label{dnIRpSSL}
\eea
\ees
where the coefficients $b_j^{(D)}$ are given in Eqs.~(\ref{bjDs}), and for the index $\tg_p$ the following notation is used [cf.~also Eq.~(\ref{OPErenamb1})]:
\be
\tg_p \equiv 1 + p \frac{c_1}{\beta_0}.
\label{tgp}
\ee
The coefficients ${\cal C}^{(2p)}_{j,k}$ ($j=1,2$) and the ratios $d_{p,k}^{\rm IR}/\td_{p,k}^{\rm IR}$ are given in the upper part of Table \ref{tabcalCrat} (IR, $p=2,3$) for the mentioned (four-loop) LMM scheme and the Lambert scheme (with $c_2=-4.9$). In Appendix \ref{app:calC} it is explained how the values and the uncertainties of the coefficients ${\cal C}^{(2p)}_{j,k}$ were extracted.
\begin{table}
  \caption{The numerically extracted values of the coefficients  ${\cal C}^{(D)}_{1,k}$,  ${\cal C}^{(D)}_{2,k}$, and the ratios  $d_{p,k}^{\rm X}/\td_{p,k}^{\rm X}$, for X=IR with $p=2,3$ ($D=2 p$), and X=UV and $p=1$ ($D=-2  p$) in two renormalization schemes: (four-loop) Lambert MiniMOM (LMM) scheme, and Lambert scheme (Lamb., with $c_2=-4.9$).}
\label{tabcalCrat}
\begin{ruledtabular}
\begin{tabular}{l|rrr|rrr}
  type & ${\rm LMM:} \;\; {\cal C}^{(D)}_{1,k}$ &  ${\cal C}^{(D)}_{2,k}$ & $d_{p,k}^{\rm X}/\td_{p,k}^{\rm X}$ &  ${\rm Lamb.:} \;\; {\cal C}^{(D)}_{1,k}$ & ${\cal C}^{(D)}_{2,k}$ & $d_{p,k}^{\rm X}/\td_{p,k}^{\rm X}$
\\
\hline
X=IR, $p=2, {\rm DP}(k=2)$ & $(-9.3 \pm 2.0)$ & $(+14. \pm 20.)$ & $(15.9 \pm 0.5)$ & $(-4.8 \pm 0.8)$ & $(+47. \pm 6.)$ & $(0.738 \pm 0.008)$ 
\\
X=IR, $p=2, {\rm SP}(k=1)$ & $(-0.30 \pm 0.27)$ & $(+14.5 \pm 2.0)$ & $(6.28 \pm 0.02)$ & $(-0.38 \pm 0.33)$ & $(+22.5 \pm 5.5)$ & $(0.290 \pm 0.001)$ 
\\
X=IR, $p=2, {\rm SL}(k=0)$ & $(+9.3 \pm 0.2)$ & - & $(+4.04 \pm 0.01)$ & $(+3.9 \pm 0.6)$ & - & $(0.186 \pm 0.002)$  
\\
X=IR, $p=2, {\rm SSL}(k=-1)$ & $(+20. \pm 1.)$ & - & $(+7.03 \pm 0.10)$ & $(+8.2 \pm 0.8)$ & - & $(+0.325 \pm 0.006)$  
\\
\hline
X=IR, $p=3, {\rm DP}(k=2)$ & $(-7.3 \pm 1.6)$ & $(-30. \pm 12.)$ & $(+260. \pm 9.)$ & $(-4.6 \pm 1.6)$ & $(+43. \pm 7.)$ & $(+1.12 \pm 0.04)$ 
\\
X=IR, $p=3, {\rm SP}(k=1)$ & $(-1.1 \pm 1.1)$ & $(+37. \pm 10.)$ & $(+81.3 \pm 1.7)$ & $(-0.8 \pm 0.7)$ & $(+41. \pm 14.)$ & $(+0.338 \pm 0.005)$ 
\\
X=IR, $p=3, {\rm SL}(k=0)$ & $(+8.8 \pm 0.1)$ & - & $(+35.1 \pm 0.1)$ & $(+3.6 \pm 1.9)$ & - & $(+0.146 \pm 0.005)$
\\
\hline
X=UV, $p=1, {\rm DP}(k=2)$ & $(-9.0 \pm 1.3)$ & $(-52. \pm 12.)$ & $(+1.361 \pm 0.011)$ & $(+7.5 \pm 1.2)$ & $(-64. \pm 25.)$ & $(-0.489 \pm 0.004)$ 
\\
X=UV, $p=1, {\rm SP}(k=1)$ & $(0.1 \pm 0.1)$ & $(8.2 \pm 0.5)$ & $(+6.444 \pm 0.004)$ & $(0.0 \pm 0.2)$ & $(0. \pm 8.)$ & $(-2.310 \pm 0.003)$ 
\\
X=UV, $p=1, {\rm SL}(k=0)$ & $(+8.6 \pm 1.9)$ & - & $(-8.11 \pm 0.12)$ & $(-6.5 \pm 0.6)$ & - & $(+2.91 \pm 0.01)$ 
\end{tabular}
\end{ruledtabular}
\end{table}
  
The Borel transforms corresponding to the expressions (\ref{BDdns}) are
\bes
\label{BDp}
\bea
    {\rm B}[D](u)_{{\rm IR},p,{\rm DP}} &=&  \frac{\pi d_{p,2}^{\rm IR}}{(p-u)^{\tg_p+1}} {\Bigg \{} 1 + \frac{\left( b_1^{(2p)} + {\cal C}_{1,2}^{(2p)} \right)}{\beta_0 \tg_p} (p - u) + \frac{\left( b_2^{(2p)} + b_1^{(2p)} {\cal C}_{1,2}^{(2p)} + {\cal C}_{2,2}^{(2p)} \right)}{\beta_0^2 \tg_p (\tg_p-1)} (p-u)^2 + \ldots {\Bigg \}},
\label{BDpDP}    
\\
{\rm B}[D](u)_{{\rm IR},p,{\rm SP}} & = & \frac{\pi d_{p,1}^{\rm IR}}{(p-u)^{\tg_p}} {\Bigg \{} 1 + \frac{\left( b_1^{(2p)} + {\cal C}_{1,1}^{(2p)} \right)}{\beta_0 (\tg_p-1)} (p - u) + \frac{\left( b_2^{(2p)} + b_1^{(2p)} {\cal C}_{1,1}^{(2p)} + {\cal C}_{2,1}^{(2p)} \right)}{\beta_0^2 (\tg_p-1) (\tg_p-2)} (p-u)^2 + \ldots  {\Bigg \}},
\label{BDpSP}
\\
{\rm B}[D](u)_{{\rm IR},p,{\rm SL}} & = & \frac{\pi d_{p,0}^{\rm IR}}{(p-u)^{\tg_p-1}} {\Bigg \{} 1 + \frac{\left( b_1^{(2p)} + {\cal C}_{1,0}^{(2p)} \right)}{\beta_0 (\tg_p-2)} (p - u) + \ldots {\Bigg \}},
\label{BDpSL}
\\
{\rm B}[D](u)_{{\rm IR},p,{\rm SSL}} & = & \frac{\pi d_{p,-1}^{\rm IR}}{(p-u)^{\tg_p-2}} {\Bigg \{} 1 + \frac{\left( b_1^{(2p)} + {\cal C}_{1,-1}^{(2p)} \right)}{\beta_0 (\tg_p-3)} (p - u) + \ldots {\Bigg \}}.
\label{BDpSSL}
\eea
\ees
These Borel transforms have, at least at the leading order, the structure of the theoretically expected Borel transforms of the IR $p$ renormalons, Eqs.~(\ref{BDIRp})-(\ref{OPErenamb}), if the leading anomalous dimension coefficient  $\gamma_{O_{2p}}^{(1)}/\beta_0$ in Eq.~(\ref{OPErenamb1}) is an integer. The latter condition appears to be satisfied for spacelike observables. 

Analogously, also the UV $p=1$ renormalons were considered numerically, in the mentioned three renormalization schemes, for the double-pole (DP), single-pole (SP), and the subleading (SL) cases:
\bea
{\rm B}[\tD](u)_{{\rm UV},p=1} & = & \frac{\pi \td_{p,2}^{\rm UV}}{(p+u)^2};
    \;  \frac{\pi \td_{p,1}^{\rm UV}}{(p+u)}; \;
    \pi \td_{p,0}^{\rm UV} (-1) \ln \left( 1 + \frac{u}{p} \right) {\Bigg |}_{p=1}.
\label{BtDp1UV}
\eea    
They generate the coefficients
\bea
(\td_n)_{{\rm UV},p=1} & = & (n+1)! (-\beta_0)^n; \; n! (-\beta_0)^n; \; (n-1)! (-\beta_0)^n,
\label{tdnUVp1}
\eea
respectively,\footnote{
When changing the renormalization scale parameter $\kappa \equiv \mu^2/Q^2$ to $\kappa \not=1$, the generated coefficients (\ref{BtDtdns}) can be shown, by the use of the relation (\ref{tdnkapsum}), to change (in the IR $p$ case): $\td_n(\kappa)/\td_n = \kappa^p  \left( 1 - p (\ln \kappa)/(n+1) \right)$, $\kappa^p$, $\kappa^p \left( 1 + p (\ln \kappa)/n + {\cal O}(1/n^2) \right)$, for the cases DP, SP, SL, respectively. And in the UV $p$ case they change: $\td_n(\kappa)/\td_n = \kappa^{-p} \left( 1 + p (\ln \kappa)/(n+1) \right)$, $\kappa^{-p}$,  $\kappa^{-p} \left( 1 - p (\ln \kappa)/n + {\cal O}(1/n^2) \right)$, for the cases DP, SP, SL. The relative corrections to these relations, due to the finiteness of $n$, are ${\cal O}\left( (p e (\ln \kappa)/n)^{n+3/2} \right)$ for DP and SP, and  ${\cal O}\left( (p e (\ln \kappa)/n)^{n+1/2} \right)$ for SL.} 
for $\td_{1,j}^{\rm UV}=1/\pi$. From these coefficients $\td_n$, the coefficients $d_n$ generated by the relations (\ref{dntdn}) agree numerically with the following expressions to a high precision:
\bes
\label{BDdnsUV}
\bea
( d_n )_{{\rm UV},p=1,{\rm DP}} & = & \frac{\pi d_{1,2}^{\rm UV}}{\Gamma(\bg_1+1)} \Gamma(\bg_1+1+n) (-\beta_0)^n {\Bigg \{} 1 + \left( b_1^{(-2)} + {\cal C}_{1,2}^{(-2)} \right) \frac{1}{(-\beta_0)} \frac{1}{(\bg_1+n)}
\nonumber\\ &&
+ \left( b_2^{(-2)} + b_1^{(-2)} {\cal C}_{1,2}^{(-2)} + {\cal C}_{2,2}^{(-2)} \right)  \frac{1}{(-\beta_0)^2}  \frac{1}{(\bg_1+n)(\bg_1-1 + n)} + {\cal O} \left( \frac{1}{n^3} \right) {\Bigg \}},
\label{dnUVp1DP}
\\
( d_n )_{{\rm UV},p=1,{\rm SP}} & = & \frac{\pi d_{1,1}^{\rm UV}}{\Gamma(\bg_1)} \Gamma(\bg_1+n) (-\beta_0)^n {\Bigg \{} 1 + \left( b_1^{(-2)} + {\cal C}_{1,1}^{(-2)} \right) \frac{1}{(-\beta_0)} \frac{1}{(\bg_1-1+n)}
\nonumber\\ &&
+ \left( b_2^{(-2)} + b_1^{(-2)} {\cal C}_{1,1}^{(-2)} + {\cal C}_{2,1}^{(-2)} \right)  \frac{1}{(-\beta_0)^2}  \frac{1}{(\bg_1-1+n)(\bg_1-2 + n)} + {\cal O} \left( \frac{1}{n^3} \right) {\Bigg \}},
\label{dnUVp1SP}
\\
( d_n )_{{\rm UV},p=1,{\rm SL}} & = & \frac{\pi d_{1,0}^{\rm UV}}{\Gamma(\bg_1-1)} \Gamma(\bg_1-1+n) (-\beta_0)^n {\Bigg \{} 1 + \left( b_1^{(-2)} + {\cal C}_{1,0}^{(-2)} \right) \frac{1}{(-\beta_0)} \frac{1}{(\bg_1-2+n)} +  {\cal O} \left( \frac{1}{n^2} \right) {\Bigg \}},
\label{dnUVp1SL}
\eea
\ees
and for the index $\bg_p$ (with $p=1$) the following notation was used  [cf.~Eq.~(\ref{OPErenambUV1})]:
\be
\bg_p \equiv 1 - p \frac{c_1}{\beta_0}.
\label{bgp}
\ee
The generated Borel transforms are 
\bes
\label{BDp1UV}
\bea
{\rm B}[D](u)_{{\rm UV},1,{\rm DP}} & = & \frac{\pi d_{1,2}^{\rm UV}}{(1+u)^{\bg_1+1}} {\Bigg \{} 1 + \frac{\left( b_1^{(-2)} + {\cal C}_{1,2}^{(-2)} \right)}{(-\beta_0) \bg_1} (1+ u) + \frac{\left( b_2^{(-2)} + b_1^{(-2)} {\cal C}_{1,2}^{(-2)} + {\cal C}_{2,2}^{(-2)} \right)}{(-\beta_0)^2 \bg_1 (\bg_1-1)} (1+u)^2 + \ldots {\Bigg \}},
\label{BDp1UVDP}
\\
{\rm B}[D](u)_{{\rm UV},1,{\rm SP}} & = & \frac{\pi d_{1,1}^{\rm UV}}{(1+u)^{\bg_1}} {\Bigg \{} 1 + \frac{\left( b_1^{(-2)} + {\cal C}_{1,1}^{(-2)} \right)}{(-\beta_0) (\bg_1-1)} (1+u) + \frac{\left( b_2^{(-2)} + b_1^{(-2)} {\cal C}_{1,1}^{(-2)} + {\cal C}_{2,1}^{(-2)} \right)}{(-\beta_0)^2 (\bg_1-1) (\bg_1-2)} (1+u)^2 + \ldots {\Bigg \}},
\label{BDp1UVSP}
\\
{\rm B}[D](u)_{{\rm UV},1,{\rm SL}} & = & \frac{\pi d_{1,0}^{\rm UV}}{(1+u)^{\bg_1-1}} {\Bigg \{} 1 + \frac{\left( b_1^{(-2)} + {\cal C}_{1,0}^{(-2)} \right)}{(-\beta_0) (\bg_1-2)} (1+u) + \ldots {\Bigg \}},
\label{BDp1UVSL}
\eea
\ees
The numerically determined coefficients ${\cal C}_{1,j}^{(-2)}$ and ratios $d_{1,j}^{\rm UV}/\td_{1,j}^{\rm UV}$ are given in the lower part of Table \ref{tabcalCrat} (UV, $p=1$), in the two mentioned renormalization schemes.

One can also ask how the values of the coefficients and ratios in  Table \ref{tabcalCrat} would be affected if the renormalization scheme were truncated, e.g., $c_j = 0$ for $j \geq 5$. The results for the LMM scheme thus truncated are presented in Table \ref{tabcalCratT}, where also the results are included for the $\MSbar$ scheme (with $N_f=3$) which is known to $c_4$.
\begin{table}
  \caption{The same as in Table \ref{tabcalCrat}, but now for the LMM scheme truncated at $c_4$ (TLMM: $c_j=0$ for $j \geq 5$), and for the $\MSbar$ scheme (${\bar c}_j=0$ for $j \geq 5$). In both schemes $N_f=3$ was taken.}
\label{tabcalCratT}
\begin{ruledtabular}
\begin{tabular}{l|rrr|rrr}
  type & ${\rm TLMM:} \;\; {\cal C}^{(D)}_{1,k}$ &  ${\cal C}^{(D)}_{2,k}$ & $d_{p,k}^{\rm X}/\td_{p,k}^{\rm X}$ &  $\MSbar : \;\; {\cal C}^{(D)}_{1,k}$ & ${\cal C}^{(D)}_{2,k}$ & $d_{p,k}^{\rm X}/\td_{p,k}^{\rm X}$
\\
\hline
X=IR, $p=2, {\rm DP}(k=2)$ & $(-9.1 \pm 1.9)$ & $(+11. \pm 14.)$ & $(+14.5 \pm 0.4)$ & $(-7.8 \pm 1.4)$ & $(+35. \pm 10.)$ & $(+4.59 \pm 0.09)$ 
\\
X=IR, $p=2, {\rm SP}(k=1)$ & $(-0.29 \pm 0.26)$ & $(+14.0 \pm 2.2)$ & $(+5.70 \pm 0.02)$ & $(-0.03 \pm 0.02)$ & $(+1.7 \pm 0.3)$ & $(+1.7995 \pm 0.0001)$ 
\\
X=IR, $p=2, {\rm SL}(k=0)$ & $(+8.85 \pm 0.15)$ & - & $(+3.67 \pm 0.01)$ & $(+7.7 \pm 0.4)$ & - & $(+1.155 \pm 0.005)$  
\\
\hline
X=IR, $p=3, {\rm DP}(k=2)$ & $(-7.0 \pm 1.6)$ & $(-26. \pm 12.)$ & $(+199. \pm 7.)$ & $(-7.2 \pm 1.2)$ & $(+20. \pm 9.)$ & $(+29.7 \pm 0.8)$ 
\\
X=IR, $p=3, {\rm SP}(k=1)$ & $(-1.0 \pm 1.0)$ & $(+37. \pm 10.)$ & $(+61.9 \pm 1.2)$ & $(-0.07 \pm 0.06)$ & $(+3.0 \pm 0.8)$ & $(+9.03 \pm 0.01)$ 
\\
X=IR, $p=3, {\rm SL}(k=0)$ & $(8.03 \pm 0.05)$ & - & $(+26.84 \pm 0.03)$ & $(+9.1 \pm 1.9)$ & - & $(+3.82 \pm 0.13)$
\\
\hline
X=UV, $p=1, {\rm DP}(k=2)$ & $(-9.1 \pm 1.3)$ & $(-52. \pm 11.)$ & $(+1.372 \pm 0.011)$ & $(-10.1 \pm 2.1)$ & $(-83. \pm 8.)$ & $(+1.056 \pm 0.014)$ 
\\
X=UV, $p=1, {\rm SP}(k=1)$ & $(0.1 \pm 0.1)$ & $(8.3 \pm 0.5)$ & $(+6.500 \pm 0.004)$ & $(+0.0 \pm 0.0)$ & $(+0.5 \pm 0.1)$ & $(+5.0098 \pm 0.0001)$ 
\\
X=UV, $p=1, {\rm SL}(k=0)$ & $(+8.8 \pm 1.9)$ & - & $(-8.18 \pm 0.12)$ & $(+8.8 \pm 1.7)$ & - & $(-6.30 \pm 0.08)$ 
\end{tabular}
\end{ruledtabular}
\end{table}
Comparing the results for LMM in Table \ref{tabcalCrat} and for its truncated version in Table  \ref{tabcalCratT}, one can see that the truncation does not appreciably affect them. The UV $p=1$ results are almost unaffected by this truncation.

The results in Tables \ref{tabcalCrat} and \ref{tabcalCratT} indicate that, in the cases of SP (simple pole of ${\rm B}[\tD](u)$), the correction coefficients ${\cal C}_{j,1}^{(D)}$ are close to zero or even compatible with zero. Furthermore, the ratios $d_{p,k}^{\rm X}/\td_{p,k}^{\rm X}$ in each of the considered cases (X,$p$) are approximately (but not exactly) proportional to $\Gamma(\tg_p+k-1)$ ($k=2,1,0,-1$). For example, in the case IR $p=2$ we have for LMM scheme $(d_{2,k}^{\rm IR}/\td_{2,k}^{\rm IR})/\Gamma(\tg_2+k-1)=(4.39 \pm 0.13)$, $(4.46 \pm 0.02)$, $(4.53 \pm 0.01)$,  $(4.58 \pm 0.06)$, for $k=2,1,0,-1$, respectively; in the Lambert ($c_2=-4.9$) scheme, these values are $(0.203 \pm 0.002)$, $(0.206 \pm 0.001)$, $(0.209 \pm 0.002)$, $(0.212 \pm 0.004)$, respectively.

Table \ref{tabIRp2SP} shows, in three different renormalization schemes (LMM, $c_2=-4.9$ Lambert, and $\MSbar$), the convergence of the generated $d_n$ coefficients in the case of IR $p=2$ SP, with increasing $n$. Specifically, $\td_n$ coefficients are those given in Eq.~(\ref{tdnIRpSP}) with $p=2$, the coefficients $d_n$ are then generated via the relations (\ref{dntdn}) up to\footnote{{\it Mathematica} software \cite{Math} was used to generate the coefficients $\tk_s$ appearing in Eq.~(\ref{tdndn}), based on specific recursion relations which follow from the relations (\ref{tanan1}) and (\ref{antan1}), cf.~Appendix \ref{app:ktkrr}.} 
$n=70$, and they are divided by the truncated versions of the $n$-dependent part $J(n)$ of Eq.~(\ref{dnIRpSP}) with $p=2$
\bes
\label{Jns}
\bea
J(n)^{\rm (LO)} & = &  \Gamma(\tg_2+n) \left( \frac{\beta_0}{2} \right)^n
\label{JnLO}
\\
J(n)^{\rm (NLO)} & = & \Gamma(\tg_2+n) \left( \frac{\beta_0}{2} \right)^n {\Bigg \{} 1 + \left( b_1^{(4)} + {\cal C}_{1,1}^{(4)} \right) \left( \frac{2}{\beta_0} \right) \frac{1}{(\tg_2-1+n)} {\Bigg \}},
\label{JnNLO}
\\
J(n)^{\rm (NNLO)} & = & \Gamma(\tg_2+n) \left( \frac{\beta_0}{2} \right)^n {\Bigg \{} 1 + \left( b_1^{(4)} + {\cal C}_{1,1}^{(4)} \right) \left( \frac{2}{\beta_0} \right) \frac{1}{(\tg_2-1+n)}
\nonumber\\ &&
+ \left( b_2^{(4)} + b_1^{(4)} {\cal C}_{1,1}^{(4)} + {\cal C}_{2,1}^{(4)} \right)  \left( \frac{2}{\beta_0} \right)^2 \frac{1}{(\tg_2-1+n)(\tg_2-2 + n)} {\Bigg \}},
\label{JnNNLO}
\eea
\ees
The values of the coefficients ${\cal C}_{j,1}^{(4)}$ ($j=1,2$) are the central values given in Tables \ref{tabcalCrat} and \ref{tabcalCratT}.
\begin{table}
\caption{The coefficients $\td_n$ of the IR $p=2$ SP case with $\td^{\rm IR}_{2,1}=2/\pi$ [Eq.~(\ref{tdnIRpSP}) with $p=2$], and various ratios involving the corresponding coefficients $d_n$ generated via the relations (\ref{dntdn}): $d_n/\td_n$, and $d_n/J(n)^{\rm (X)}$ (X=LO, NLO, NNLO); up to $n=70$. These results are for the LMM, $c_2=-4.9$ Lambert (Lamb.), and $\MSbar$ renormalization schemes (all with $N_f=3$).}
\label{tabIRp2SP}
\begin{ruledtabular}
\begin{tabular}{r|r|rrrrr}
  scheme & $n$ & $\td_n$ & $d_n/\td_n$ & $d_n/J(n)^{\rm (LO)}$ & $d_n/J(n)^{\rm (NLO)}$  & $d_n/J(n)^{\rm (NNLO)}$
\\
\hline
LMM & 1 &  1.125              &  1.    & 0.27512 & -0.28003 & 0.22252 \\
& 10 &  $1.1784 \times 10^{7}$ & 41.034 & 0.89116 & 1.5963  & 1.4098\\
& 40 & $9.0729 \times 10^{49}$ & 470.03  & 1.3144 &  1.4987 & 1.4896 \\
& 50 & $1.0982 \times 10^{67}$ & 679.36 & 1.3485 & 1.4969 & 1.4912\\ 
& 60 & $9.7572 \times 10^{84}$ & 915.61  & 1.3715 & 1.4958  & 1.4918\\
& 70 & $4.5612 \times 10^{103}$ & 1176.7 & 1.3881 & 1.4950 & 1.4921\\ 
\hline
Lamb. & 1 &  1.125           & 1.      & 0.27512 &0.082436 &0.030582 \\
& 10 & $1.1784 \times 10^{7}$ & 5.0885 & 0.11051 &0.072666 &0.064660 \\
& 40 & $9.0729 \times 10^{49}$ & 28.448 & 0.079550 &0.069473 & 0.068653\\
& 50 & $1.0982 \times 10^{67}$ & 38.996 & 0.077404 &0.069301 &0.068757 \\
& 60 & $9.7572 \times 10^{84}$ & 50.715 & 0.075968 &0.069191 &0.068804 \\
& 70 & $4.5612 \times 10^{103}$ & 63.525 & 0.074940 &0.069116 &0.068826\\
\hline
$\MSbar$. & 1       &  1.125 & 1.     & 0.27512& 0.46762& 2.0610 \\
& 10 & $1.1784 \times 10^{7}$ & 17.514 & 0.38035& 0.41876& 0.42586\\
& 40 & $9.0729 \times 10^{49}$ & 148.67& 0.41572& 0.42662& 0.42710 \\
& 50 & $1.0982 \times 10^{67}$ & 210.62& 0.41805& 0.42684& 0.42715\\
& 60 & $9.7572 \times 10^{84}$ & 280.12& 0.41960 & 0.42696& 0.42718\\
& 70 & $4.5612 \times 10^{103}$ & 356.62& 0.42069& 0.42703& 0.42719\\
\end{tabular}
\end{ruledtabular}
\end{table}
The results presented in Table \ref{tabIRp2SP} show strong convergence of the mentioned ratios, especially $d_n/J(n)^{\rm (NNLO)}$, toward $n$-independent values when $n$ increases.

\section{Adler function}
\label{sec:Adl}

\subsection{Construction of the generating Borel transforms}
\label{subs:BtD}

The massless Adler function, which is a logarithmic derivative of the light-quark current correlator, is a  specific example of a spacelike QCD observable for which we have a large amount of theoretical information available. Namely, its perturbation expansion (\ref{Dpt}) is known up to $\sim a^4$, i.e., the coefficients $d_n$ for $n \leq 3$ are exactly known \cite{d1,d2,d3}. Further, the large-$\beta_0$ (LB) expansion of its Borel transform is also known \cite{LBAdl1,LBAdl2,Renormalons},
\bes
\label{BtDLB}
\bea
{\rm B}[\tD](u; \kappa)^{\rm (LB)} & = & \frac{32}{3} \frac{ \kappa^u \exp\left[+(5/3)u \right]}{(2-u)}
\sum_{k=2}^{\infty} \frac{ (-1)^k k}{(k-1+u)^2 (k+1-u)^2}
\label{BtdLBa}
\\
& = & 1 + \frac{\td_1^{\rm (LB)}(\kappa)}{1! \beta_0} u + \ldots + \frac{\td_n^{\rm (LB)}(\kappa)}{n! \beta_0^n} u^n + \ldots,
\label{BtdLBb}
\eea
\ees
showing\footnote{
  The $\MSbar$ scale convention is used throughout this work. We recall that $\td_n(\kappa)$ can be expanded in powers of $\beta_0$, namely $\td_n(\kappa) = c_{n,n}(\kappa) \beta_0^n + c_{n,n-1}(\kappa) \beta_0^{n-1} + \ldots + c_{n,-1}(\kappa) \beta_0^{-1}$, and $\td_n^{\rm (LB)}(\kappa) = c_{n,n}(\kappa) \beta_0^n$, where $c_{n,n}(\kappa)$ is renormalization scale dependent and renormalization scheme independent \cite{CV12,Techn}, i.e., independent of the scheme parameters $c_j=\beta_j/\beta_0$ ($j \geq 2$).}
that in the large-$\beta_0$ (resummed one-loop) approximation the IR renormalon poles are double for $p=3,4,\ldots$ [$\Rightarrow -\gamma^{(1)}_{O_D}/\beta_0=1$ in Eq.~(\ref{OPErenamb1})], single for $p=2$ ($\gamma^{(1)}_{O_4}=0$), and the UV renormalon poles ($p=1,2,\ldots$) are all double [$-\gamma^{(1)}_{O_D}/\beta_0=1$ in Eq.~(\ref{OPErenambUV1})]. In addition, for the IR $p=2$ renormalon pole the subleading coefficient ${\hat c}_1^{(4)}$ [cf.~Eqs.~(\ref{Q2depOPE})-(\ref{Wcandim})] is also known \cite{CPS,Renormalons}
\be
{\hat c}_1^{(4)} = \frac{7}{6} - c_1  \left( = -\frac{11}{18}  \; {\rm when \ } N_f=3 \right).
\label{hatc1}
\ee
Using all this information, a physically motivated ansatz for the Borel transform ${\rm B}[\tD](u)$ of the auxiliary quantity $\tD(Q^2)$ Eq.~(\ref{tD1}) of the Adler function will be written, where the IR $p=2$ leading and subleading renormalons are included, as well as the IR $p=3$ and UV $p=1$ leading renormalons
\bea
{\rm B}[\tD](u)^{\rm (4 P)} & = & \exp \left( \tK u \right) \pi {\Big \{}
\td_{2,1}^{\rm IR} \left[ \frac{1}{(2-u)} + \tal (-1) \ln \left( 1 - \frac{u}{2} \right) \right] + \frac{ \td_{3,2}^{\rm IR} }{(3 - u)^2} + \frac{ \td_{1,2}^{\rm UV} }{(1 + u)^2} {\Big \}}.
\label{BtD4P}
\eea
Here, the superscript '(4 P)' indicates that the ansatz contains four adjustable parameters: the ``scaling'' parameter $\tK$ and the renormalon residue parameters $\td_{2,1}^{\rm IR}$, $\td_{3,2}^{\rm IR}$ and $\td_{1,2}^{\rm UV}$. It turns out that the IR $p=2$ subleading parameter $\tal$ is fixed by the knowledge of the subleading coefficient ${\hat c}_1^{(4)}$ Eq.~(\ref{hatc1}). Each one-loop-type IR renormalon term in the ansatz (\ref{BtD4P}) generates the Borel transforms of the type (\ref{BDp}) and the corresponding contributions (\ref{BDdns}) to the perturbation coefficients $d_n$. When assuming $\tK=0$ and requiring that the two IR $p=2$ renormalons in Eq.~(\ref{BtD4P}) together generate the $(d_n)_{\rm IR,p=2}$ coefficient of the form Eq.~(\ref{dnIRp}) with the subleading part there having ${\hat c}_1^{(4)}$ as given in Eq.~(\ref{hatc1}), the following condition is obtained:
\be
\tal  =  \alpha \left( \frac{d_{2,1}^{\rm IR}}{\td_{2,1}^{\rm IR}} \right) \left( \frac{d_{2,0}^{\rm IR}}{\td_{2,0}^{\rm IR}} \right)^{-1},
\label{tal1}
\ee
where
\be
\alpha= \frac{ ({\hat c}_1^{(4)} - {\cal C}_{1,1}^{(4)}) }{\beta_0 (\tg_2 -1)}.
\label{al}
\ee
The ratios and the parameter ${\cal C}_{1,1}^{(4)}$  appearing in these expressions are given in Table \ref{tabcalCrat}. On the basis of the results of the previous Section, the expression (\ref{BtD4P}) generates the following Borel transform of the full Adler function ${\cal D}$, at the renormalization scale $\mu^2=\kappa Q^2 = \exp(-\tK) \; Q^2$:
\bea
\lefteqn{
 {\rm B}[{\cal D}](u;\kappa=e^{-\tK})^{\rm (4 P)} =  \pi {\Bigg \{} \frac{d_{2,1}^{\rm IR}}{(2 - u)^{\tg_2}} \left[ 1 + \frac{(b_1^{(4)}+{\cal C}_{1,1}^{(4)})}{\beta_0 (\tg_2-1)} (2-u) + \frac{(b_2^{(4)}+{\cal C}_{1,1}^{(4)} b_1^{(4)} +{\cal C}_{2,1}^{(4)})}{ \beta_0^2 (\tg_2-1)(\tg_2-2)} (2-u)^2 + \ldots \right]
}
\nonumber\\ &&
+ \frac{d_{2,1}^{\rm IR} \alpha}{(2 - u)^{\tg_2-1}}  \left[ 1 + \frac{(b_1^{(4)}+{\cal C}_{1,0}^{(4)})}{\beta_0 (\tg_2-2)} (2-u) + \ldots \right]
\nonumber\\ &&
+ \frac{d_{3,2}^{\rm IR}}{(3 - u)^{\tg_3+1}} \left[ 1 + \frac{(b_1^{(6)}+{\cal C}_{1,2}^{(6)})}{\beta_0 \tg_3} (3-u) + \frac{(b_2^{(6)}+{\cal C}_{1,2}^{(6)} b_1^{(6)} +{\cal C}_{2,2}^{(6)})}{ \beta_0^2 \tg_3 (\tg_3-1)} (3 - u)^2 + \ldots \right]
\nonumber\\ &&
+ \frac{d_{1,2}^{\rm UV}}{(1+ u)^{\bg_1+1}} \left[ 1 + \frac{(b_1^{(-2)}+{\cal C}_{1,2}^{(-2)})}{(-\beta_0) \bg_1} (1+u) + \frac{(b_2^{(-2)}+{\cal C}_{1,2}^{(-2)} b_1^{(-2)} +{\cal C}_{2,2}^{(-2)})}{ (-\beta_0)^2 \bg_1 (\bg_1-1)} (1+ u)^2 + \ldots \right] {\Bigg \}},
\label{BD4P}
\eea
and the notations Eqs.~(\ref{tgp}) and (\ref{bgp}) for $\tg_p$ and $\bg_p$ were used. The above Borel transform generates the Adler function coefficients $d_n(\kappa)$ at the value of the renormalization scale parameter $\kappa$ ($\equiv \mu^2/Q^2$) $=\exp(-\tK)$. As mentioned, if the value of the renormalization scale parameter were $\kappa=1$ ($\tK=0$), then the requirement of the reproduction of the correct contribution $(d_n)_{\rm IR,p=2}$ of Eq.~(\ref{dnIRp}), at subleading order, with the known value of ${\hat c}_1^{(4)}$ Eq.~(\ref{hatc1}), would imply that the $\alpha$ coefficient at $d_{2,1}^{\rm IR}$ in Eq.~(\ref{BD4P}) must have the value as given in Eq.~(\ref{al}). It turns out that the effect of $\kappa \not= 1$ ($\tK \not=0$) does not change this relation and the relation (\ref{tal1}). This is so because
\be
(d_n)_{\rm IR, p=2} = e^{2 \tK} \left[ 1 + {\cal O}(1/n^2) \right] \; (d_n(\kappa))_{\rm IR, p=2}|_{\kappa=\exp(-\tK)} \ ,
\label{dnIRp2kap}
\ee
i.e., the change of renormalization scale changes the coefficient only by a constant factor, with no subleading corrections (but with subsubleading corrections). This relation can be understood if the IR $p=2$ part of the Borel ${\cal B}[\tD](u)$ of Eq.~(\ref{BtD4P}) is reexpressed by expanding the exponential $\exp(\tK u)$ around $u=2$, which gives
\bea
{\rm B}[\tD](u)_{IR,p=2} & = & \exp \left( \tK u \right) \pi 
\td_{2,1}^{\rm IR} \left[ \frac{1}{(2-u)} + \tal (-1) \ln \left( 1 - \frac{u}{2} \right) \right]
\nonumber\\
& = & \pi 
\td_{2,1}^{\rm IR} \exp(2 \tK) {\Big [} \frac{1}{(2-u)} + \tal (-1) \ln \left( 1 - \frac{u}{2} \right)
  - \tK + \tal 2 \tK \left(1 - \frac{u}{2} \right) \ln \left( 1 - \frac{u}{2} \right) + {\cal O}(2-u) {\Big ]}.
\label{BtDIRp2}
\eea
The term $(1 - u/2) \ln(1 -u/2)$ in the brackets is subsubleading (SSL); comparison of the results Eqs.~(\ref{dnIRpSSL}) with (\ref{dnIRpSP}) then gives that this term gives relative corrections $\sim 1/n^2$, i.e., the relation (\ref{dnIRp2kap}). Hence, in retrospect, we see that the relations (\ref{tal1})-(\ref{al}), which are subleading in their nature, are not affected by $\tK \not= 0$ which is a subsubleading effect [apart from the overall factor $\exp(2 \tK)$].

In the case of the mentioned renormalization schemes LMM and Lambert $c_2=-4.9$ schemes, applicable in the $3\delta$ $\A$QCD and $2\delta$ $\A$QCD, respectively, and the truncated (at $c_4$) TLMM and $\MSbar$ schemes, the relations (\ref{tal1})-(\ref{al}) and the numerical results of Tables \ref{tabcalCrat} and  \ref{tabcalCratT} give us the values
\bes
\label{tals}
\bea
\tal_{\rm LMM} & = & -0.14 \pm 0.12,
\label{talLMM}
\\
\tal_{\rm Lamb.} & = &-0.10 \pm 0.14
\label{talLamb}
\\
\tal_{\rm TLMM} & = & -0.14 \pm 0.11,
\label{talTLMM}
\\
\tal_{\rm \MSbar} & = & -0.255 \pm 0.010.
\label{talMSbar}
\eea
\ees
By far the most dominant source of uncertainty of $\tal$ is the uncertainty $\delta {\cal C}_{1,1}^{(4)}$ (cf.~Tables \ref{tabcalCrat} and \ref{tabcalCratT}).

Since the Borel transform ${\rm B}[\tD](u)$ Eq.~(\ref{BtD4P}) has four free parameters, and its power expansion generates the coefficients $\td_j$, the four free parameters can be determined by the knowledge of the first four coefficients $\td_j$ ($j=0,1,2,3$) in the considered scheme, whose values are given in Table \ref{tabdjtdj} (second line).
\begin{table}
  \caption{The known perturbation coefficients $d_j$ and $\td_j$ in the $\MSbar$, LMM and Lambert $c_2=-4.9$ (Lamb.) renormalization schemes (all with $N_f=3$). The canonical convention $d_0=\td_0=1$ was used.}
\label{tabdjtdj}
\begin{ruledtabular}
\begin{tabular}{r|rrr|rrr}
  scheme & $d_1$ & $d_2$ & $d_3$ & $\td_1$ & $\td_2$ & $\td_3$
  \\
\hline
$\MSbar$ & 1.63982 & 6.37101 & 49.0757 & 1.63982 & 3.45578 & 26.3849
\\
LMM  & 1.63982 & 1.54508 & 8.01658 &  1.63982 & -1.37016 & -1.13924
\\
Lamb. &  1.63982 & 15.7421 & 83.5517 & 1.63982 & 12.8268 & 34.5787
\end{tabular}
\end{ruledtabular}
\end{table}
The ansatz Eq.~(\ref{BtD4P}) was applied in the LMM renormalization scheme, this led to three different solutions. These solutions then give us predictions for the next perturbation coefficient $\td_4$, which can be transformed into the coefficient $d_4$ in the $\MSbar$ scheme, with the values: $d_4(\MSbar)= 338.2; 130.4; 3028$. However, the effective charge (ECH) method \cite{ECH} leads to the estimate $d_4(\MSbar)_{\rm ECH}=275$ \cite{KatStar,BCK}; the conservative estimate of Ref.~\cite{BeJa08} is $0 < d_4(\MSbar) < 642$ [their preferred value is: $d_4(\MSbar)=283$]; recent estimates \cite{Boitoetal} based on Pad\'e approximants give  $d_4(\MSbar)=277 \pm 51$. For these reasons, the solution which gives $d_4(\MSbar)= 338.2$ was chosen here. The results are given in Table \ref{tab4P5P} (first line).
\begin{table}
  \caption{The obtained values of $\tK$ and of the renormalon residues $\td_{i,j}^{\rm X}$ (X=IR,UV) for the four-parameter ansatz (\ref{BtD4P}) in the LMM scheme, and the five-parameter ansatz (\ref{BtD5P}) in the Lambert $c_2=-4.9$ (Lamb.) scheme and $\MSbar$ scheme, giving in all cases  the same value $d_4(\MSbar,N_f=3)= 338.2$. For $\tal$ the corresponding central values in Eqs.~(\ref{tals}) were taken.}
\label{tab4P5P}
\begin{ruledtabular}
\begin{tabular}{r|rrrrr}
  scheme & $\tK$ & $\td_{2,1}^{\rm IR}$ & $\td_{3,2}^{\rm IR}$ &  $\td_{3,1}^{\rm IR}$ & $\td_{1,2}^{\rm UV}$ 
\\
\hline
LMM  & -0.770405 & -1.83066 & 11.0498 & -    & 0.00588513 
\\
Lamb. &  0.2228 & 4.74582 & -1.04837 & -5.89714 & 0.0276003 
\\
$\MSbar$ & 0.5190 & 1.10826 & -0.481538 & -0.511642 & -0.0117704

\end{tabular}
\end{ruledtabular}
\end{table}
This model for the Adler function will be applied for the $3\delta$ $\A$QCD \cite{4l3dAQCD} because this QCD variant was constructed in the LMM renormalization scheme.

An additional goal here is to make comparisons of the results obtained in this way with the analogous results obtained in $2\delta$ $\A$QCD \cite{2dAQCD}, which is a QCD variant constructed in the Lambert $c_2=-4.9$ renormalization scheme. Therefore, a model for the Adler function will now be constructed in this Lambert renormalization scheme, by requiring additionally that the same value $d_4(\MSbar,N_f=3)= 338.2$ be generated in this model. This now implies one more condition (in total five conditions) in this scheme. Hence, for the $c_2=-4.9$ Lambert renormalization scheme, the following ansatz with one more parameter is written (the IR $p=3$ renormalon will have the double and the single pole):
\bea
{\rm B}[\tD](u)^{\rm (5 P)} & = & \exp \left( \tK u \right) \pi {\Big \{}
\td_{2,1}^{\rm IR} \left[ \frac{1}{(2-u)} + \tal (-1) \ln \left( 1 - \frac{u}{2} \right) \right] + \frac{ \td_{3,2}^{\rm IR} }{(3 - u)^2} + \frac{ \td_{3,1}^{\rm IR} }{(3 - u)} + \frac{ \td_{1,2}^{\rm UV} }{(1 + u)^2} {\Big \}}.
\label{BtD5P}
\eea
Applying the five conditions, the five parameters of this ansatz ($\tK$ and the four renormalon residues) are determined uniquely, and this solution is given in the second line of Table \ref{tab4P5P}. The same procedure was repeated for the $\MSbar$ scheme (no comparable $\A$QCD version is available in that scheme, though), and the results are included in Table \ref{tab4P5P}.\footnote{The polynomial five-loop $\beta$ function \cite{5lMSbarbeta} was used in the $N_f=3$ regime, using as the $N_f=3$ reference value ${\bar a}_0 \equiv a(Q_0^2; \MSbar)_{Nf=3}= 0.0846346$ at $Q_0^2=(2 {\overline m}_c)^2$ (where ${\overline m}_c=1.27$ GeV). See Appendix \ref{app:ndAQCD} for more details, in particular footnote \ref{ftrefNf3} there for comparison of numerical values.}

Now we have the two models of Adler function, in 4LMM and Lambert schemes, the two models being presumably comparable because they give the same value of the $\sim a^5$ perturbation coefficient $d_4$, namely $d_4(\MSbar,N_f=3)= 338.2$.

\subsection{Construction of the characteristic distribution function $G_D(t)$}
\label{subs:ft}

Now the characteristic function $F_D(t)$ will be constructed, for the spacelike observables ${\cal D}$ with a rather generic form of the Borel transform ${\rm B}[\tD](u)$ of the correponding auxiliary quantity $\tD$. In this approach, mainly Neubert's construction \cite{Neubert} will be followed, who constructed such characteristic functions in the framework of the large-$\beta_0$ approximation in pQCD.

The characteristic function $F_D(t)$ is a function of a dimensionless parameter $t$ ($t> 0$) such that the integral
\be
{\cal D}_{\rm res.}(Q^2) = \int_0^{+\infty} \frac{dt}{t} F_D(t) a(t Q^2)
\label{Dres1}
\ee
represents the leading-twist part of the observable ${\cal D}(Q^2)$, in the sense that it generates the correct logarithmic perturbation expansion (\ref{Dkaplpt}) of the observable when the coupling $a(t Q^2)$ is Taylor-expanded around $a(Q^2)$
\be
a(t Q^2) = a( \kappa Q^2) + (-\beta_0) \ln (t/\kappa) \ta_2(\kappa Q^2) + \ldots + (-\beta_0)^n \ln^n (t/\kappa) \ta_{n+1}(\kappa Q^2) + \ldots
\label{alpt}
\ee
where the notation Eq.~(\ref{tan}) is used. This means that $F_D(t)$ must satisfy the following string of relations:
\be
(- \beta_0)^n \int_0^{+\infty} \frac{dt}{t} F_D(t) \ln^n \left( \frac{t}{\kappa} \right) = \td_n(\kappa) \quad (n=0,1,2,\ldots).
\label{FDSR}
\ee
Using these relations (with $\kappa=1$) and the expansion (\ref{BtD1}) for the Borel transform ${\rm B}[\tD](u)$, one obtains
\be
{\rm B}[\tD](u) = \int_0^{+\infty} \frac{dt}{t} F_D(t) t^{-u}.
\label{BtDMell}
\ee
This means that ${\rm B}[\tD](u)$ is Mellin transform of the characteristic function $ F_D(t)$. The inverse Mellin transform than gives the characteristic function in terms of ${\rm B}[\tD](u)$
\be
F_D(t) = \frac{1}{2 \pi i} \int_{u_0- i \infty}^{u_0 + i \infty} du {\rm B}[\tD](u) t^u,
\label{FDinvMell}
\ee
where the integration is in the complex $u$-plane paralell to the imaginary axis, and $u_0$ is any real value where the integral (\ref{BtDMell}) exists, i.e., in the case of the Adler function one can take $-1 < u_0 < +2$. One can choose $u_0=+1$, and write the above integral along the real axis in terms of the variable $z$ such that $u=1 - i z$
\be
F_D(t) = \frac{t}{2 \pi} \int_{-\infty}^{+\infty} dz   {\rm B}[\tD](u=1 - i z) \exp(-i z \ln t) .
\label{FDint}
\ee
For the Borel transforms  ${\rm B}[\tD]$ of the one-loop-type form (\ref{BtDp}) and (\ref{BtDp1UV}), the integrals (\ref{FDint}) for $F_D(t)$ can be evaluated in a straightforward way, by performing integration along judicially chosen contours and using Cauchy theorem. For the rather generic case of the Borel transform
\be
{\rm B}[\tD](u) =  \exp \left( \tK u \right) \pi {\Big \{}
\td_{2,1}^{\rm IR} \left[ \frac{1}{(2-u)} + \tal (-1) \ln \left( 1 - \frac{u}{2} \right) \right] + \frac{ \td_{N,2}^{\rm IR} }{(N - u)^2} + \frac{ \td_{N,1}^{\rm IR} }{(N - u)} + \frac{ \td_{M,2}^{\rm UV} }{(M + u)^2} + \frac{ \td_{M,1}^{\rm UV} }{(M + u)} {\Big \}},
\label{BtDgen}
\ee
the following result for the integrated observable ${\cal D}(Q^2)$ is obtained:
\be
{\cal D}(Q^2)_{\rm res} = \int_0^\infty \frac{dt}{t} G_D(t) a(t e^{-\tK} Q^2) + \int_0^\infty \frac{dt}{t} G^{\rm (SL)}_D(t) \left[ a(t e^{-\tK} Q^2) - a(e^{-\tK} Q^2) \right],
\label{Dres2a}
\ee
where the two characteristic functions are
\bes
\label{GDs}
\bea
 G_D(t) & = & G_D^{(-)}(t) \Theta(1 - t) +  G_D^{(+)}(t) \Theta(t-1),
\label{GD}
\\
G_D^{(-)}(t) & = &  \pi t^2 \left[\td_{2,1}^{\rm IR} - \td_{N,2}^{\rm IR} t^{N-2} \ln t +\td_{N,1}^{\rm IR} t^{N-2} \right],
\label{GDmi}
\\
G_D^{(+)}(t) & = &  \frac{\pi}{t^{M}} \left[\td_{M,2}^{\rm UV} \ln t + \td_{M,1}^{\rm UV} \right], 
\label{GDpl}
\\
G_D^{\rm (SL)}(t) & = & - \tal \td_{2,1}^{\rm IR} \frac{\pi t^2}{\ln t} \Theta(1-t) ,
\label{GDSL}
\eea
\ees
where $\Theta$ is the Heaviside step function, i.e., the result (\ref{Dres2a}) can be written as 
\be
{\cal D}(Q^2)_{\rm res} = \int_0^1 \frac{dt}{t} G^{(-)}_D(t) a(t e^{-\tK} Q^2) +
\int_1^{\infty} \frac{dt}{t} G^{(+)}_D(t) a(t e^{-\tK} Q^2) + \int_0^1 \frac{dt}{t} G^{\rm (SL)}_D(t) \left[ a(t e^{-\tK} Q^2) - a(e^{-\tK} Q^2) \right].
\label{Dres2b}
\ee
The characteristic function $G_D(t)$ ($ = F_D(e^{-\tK} t)$) is obtained by closing the integration path of $z$ with the large semicircle in the upper (lower) half plane for $t<1$ ($t>1$). For the subleading (SL) contribution, the function $G_D^{\rm (SL)}(t)$ is obtained by closing the $z$-path in the upper half plane and integrating there along both sides of the cut $(+i,+i \infty)$ in order not to enclose it. It can be further noted that the SL contribution contains in the integrand the subtraction $[a(t e^{-\tK} Q^2) - a(e^{-\tK} Q^2)]$ instead of simply $a(t e^{-\tK} Q^2)$, because the SL contribution starts at $~\sim a^2$ because there $\td_0=0$ (the corresponding Borel transform expansion starts at $\sim u^1$). The relations (\ref{FDSR}) can now be rewritten in terms of  $G_D(t)$ ($ = F_D(e^{-\tK} t)$) and include the SL contribution
\bea
\td_n(\kappa) & = & (- \beta_0)^n {\Bigg \{} \int_0^{+\infty} \frac{dt}{t} G_D(t) \ln^n \left( \frac{t}{\kappa} e^{-\tK} \right)
+ \int_0^{\infty}  \frac{dt}{t} G_D^{\rm (SL)}(t) \left[ \ln^n \left( \frac{t}{\kappa} e^{-\tK} \right) -  \ln^n \left( \frac{1}{\kappa} e^{-\tK} \right) \right] {\Bigg \}},
\label{GDLSLSR}
\eea
for $n=0,1,2,\ldots$. The form (\ref{BtDgen}) of ${\rm B}[\tD](u)$ includes in its form the cases of the four-parameter ansatz (\ref{BtD4P}) ($N=3$, $M=1$, $\td_{N,1}^{\rm IR}=0=\td_{M,1}^{\rm UV}$) which was applied in Sec.~\ref{subs:BtD} to the Adler function in the LMM renormalization scheme, and the five-parameter ansatz (\ref{BtD5P}) ($N=3$, $M=1$, $\td_{M,1}^{\rm UV}=0$) which was applied there to the Adler function in the Lambert $c_2=-4.9$ renormalization scheme. It can be explicitly checked, e.g. with {\it Mathematica} software \cite{Math}, that the integrals (\ref{GDLSLSR}) generate the very same perturbation (lpt) coefficients $\td_n(\kappa)$ as the Borel transform ${\rm B}[\tD](u)$ Eq.~(\ref{BtDgen}).

It should be pointed out that the coupling $a(t e^{-\tK} Q^2)$ in the integrand (\ref{Dres2a}) is, in principle, running to any chosen loop order, thus representing the (leading-twist of the) full observable ${\cal D}(Q^2)$. If the coupling $a(t e^{-\tK} Q^2)$ is one-loop running
\be
a^{\rm (1-l.)}(t e^{-\tK} Q^2) = \frac{a(Q^2)}{1 + a(Q^2) \beta_0 \ln (t e^{-\tK})},
\label{a1l}
\ee
then the integral (\ref{Dres2a})[$\Leftrightarrow$ (\ref{Dres2b})] reproduces the perturbation expansion (\ref{tD1}) of the auxiliary quantity $\tD(Q^2)$ when the coupling (\ref{a1l}) in the integral is expanded in powers of $a(Q^2)$.

The integral (\ref{Dres2a}) is in pQCD in general ambiguous for $Q^2>0$ because of the Landau singularities of the pQCD coupling $a(t e^{-\tK} Q^2)$ at low $t$ values. To avoid this ambiguity, the integral (for $Q^2>0$) is evaluated with the path scale slightly above the real positive axis, $a(t e^{-\tK} Q^2 + i \epsilon)$ and taking the real part of it (i.e., the generalized Principal Value)
\be
{\cal D}(Q^2)_{\rm pQCD res} = {\rm Re} {\Bigg \{} \int_0^\infty \frac{dt}{t} G_D(t) a(t e^{-\tK} Q^2+i \epsilon) + \int_0^\infty \frac{dt}{t} G^{\rm (SL)}_D(t) \left[ a(t e^{-\tK} Q^2 + i \epsilon) - a(e^{-\tK} Q^2 + i \epsilon) \right] {\Big \}}.
\label{Dres2pQCD}
\ee
The ambiguity is proportional to the imaginary part
\be
\delta {\cal D}(Q^2)_{\rm pQCD res} = \pm \frac{1}{\pi} {\rm Im} {\Bigg \{} \int_0^\infty \frac{dt}{t} G_D(t) a(t e^{-\tK} Q^2+i \epsilon) + \int_0^\infty \frac{dt}{t} G^{\rm (SL)}_D(t) \left[ a(t e^{-\tK} Q^2 + i \epsilon) - a(e^{-\tK} Q^2 + i \epsilon) \right] {\Big \}}.
\label{Dres2pQCDIm}
\ee
On the other hand, the situation is essentially different in the QCD variants with IR-safe coupling $\A(Q^{' 2})$ [the analog of $a(Q^{' 2})$]. In those variants, the coupling $\A(Q^{' 2})$ is in general holomorphic (analytic) function in the $Q^{'2}$-complex plane with the exclusion of (part of) the negative semiaxis: $Q^{'2} \in \mathbb{C} \backslash (-\infty, -M_{\rm thr}^2]$, where $M_{\rm thr} \sim 0.1$ GeV is a threshold scale generally of the order of the light meson mass. Two representative cases are the 2$\delta \A$QCD and 3$\delta \A$QCD, cf.~Appendix \ref{app:ndAQCD} for more details. In such QCD variants, the evaluation of the integrals (\ref{Dres2a}) [$\Leftrightarrow$ (\ref{Dres2b})] is unambiguous. In such $\A$QCD frameworks, these integrals represent an unambigous resummation of the leading-twist part of the spacelike observable ${\cal D}(Q^2)$, if one takes the position that the Borel transforms ${\rm B}[\tD](u)$ of the type (\ref{BtDgen}) represent the correct generators of all the $\td_n$ perturbation (lpt) coefficients of ${\cal D}(Q^2)$ Eq.~(\ref{Dkaplpt}). As argued in Appendix \ref{app:ndAQCD}, most of the relations in Sec.~\ref{subs:lpt} survive in these QCD variants, with the substitutions: $a \mapsto \A$, $\ta_n \mapsto \tA_n$, and $a^n \mapsto \A_n$. Since the coupling $\A(Q^2)$ in these $\A$QCD frameworks differs from the (underlying) pQCD coupling $a(Q^2)$ in the same renormalization scheme by nonperturbative contributions, we have one (crucial) difference, namely that the power analogs $\A_n$ are not simple powers of $\A$ ($\A_n \not= \A^n$), cf.~Appendix \ref{app:ndAQCD}. For example, the series (\ref{Dkapptlpt}) for the Adler function (and any other spacelike observable) in such $\A$QCD frameworks gets the form
\bes
\label{DAkapptlpt}
\bea
{\cal D}_{\A{\rm QCD}}(Q^2) &=& \td_0 \A(\mu^2) + \td_1(\kappa)  \tA_2(\mu^2) + \td_2(\kappa) \tA_3(\mu^2) + \ldots + \td_n(\kappa) \tA_{n+1}(\mu^2) + \ldots
\label{DAkaplpt}
\\
                         &=& d_0 \A(\mu^2) + d_1(\kappa)  \A_2(\mu^2) + d_2(\kappa) \A_3(\mu^2)^ + \ldots + d_n(\kappa) \A_{n+1}(\mu^2) + \ldots,
\label{DAkappt}
\eea
\ees
where $\td_0=d_0=1$, the definition of the couplings $\tA_k$ and $\A_k$ is given in Appendix \ref{app:ndAQCD} [Eqs.~(\ref{tAn}) and (\ref{AntAn})] and, as always, $\kappa \equiv \mu^2/Q^2$ is the (arbitrary) dimensionless renormalization scale parameter ($0<\kappa \sim 1$). The terms in the series (\ref{DAkapptlpt}) do not suffer from Landau singularities at low positive $Q^2$ ($0 \leq Q^2 \lesssim 1 \ {\rm GeV}^2$), in contrast to the terms in the pQCD series (\ref{Dkapptlpt}). However, the series (\ref{DAkapptlpt}) are asymptotically divergent, as are also the pQCD series (\ref{Dkapptlpt}). The corresponding resummation of the series ${\cal D}_{\A{\rm QCD}}(Q^2)$, with the characteristic functions $G_D(t)$ and $G_D^{\rm (SL)}(t)$, turns out to be completely similar to the resummation (\ref{Dres2b}) in pQCD, with the simple substitution $a \mapsto \A$
\be
{\cal D}(Q^2)_{\A{\rm res}} = \int_0^1 \frac{dt}{t} G^{(-)}_D(t) \A(t e^{-\tK} Q^2) +
\int_1^{\infty} \frac{dt}{t} G^{(+)}_D(t) \A(t e^{-\tK} Q^2) + \int_0^1 \frac{dt}{t} G^{\rm (SL)}_D(t) \left[ \A(t e^{-\tK} Q^2) - \A(e^{-\tK} Q^2) \right].
\label{DAres2b}
\ee
In contrast to pQCD, these integrals are unambiguous because the coupling $\A(Q^{'2})$ has no Landau singularities, and there is no need to employ the (generalized) Principal Value approach Eqs.~(\ref{Dres2pQCD})-(\ref{Dres2pQCDIm}).

\begin{figure}[htb] 
\begin{minipage}[b]{.49\linewidth}
  \centering\includegraphics[width=85mm]{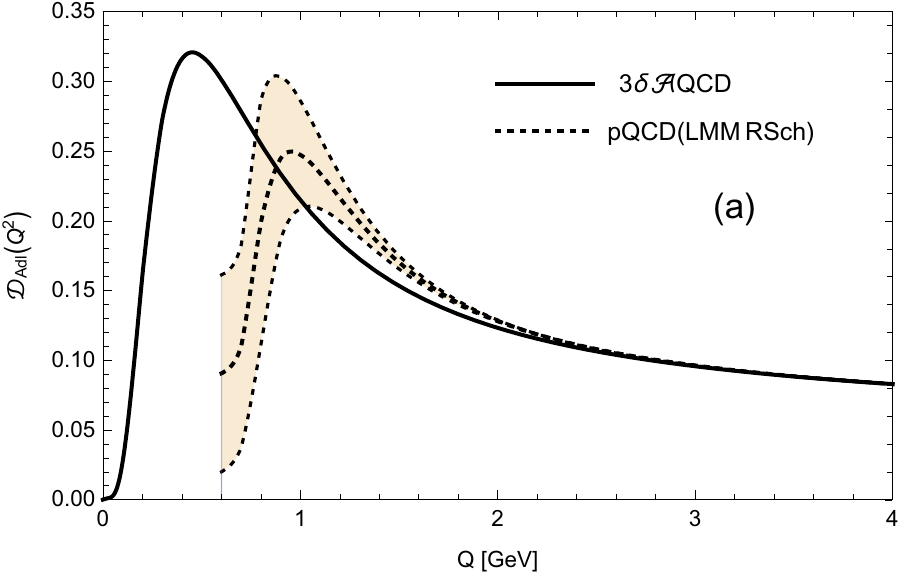}
  \end{minipage}
\begin{minipage}[b]{.49\linewidth}
  \centering\includegraphics[width=85mm]{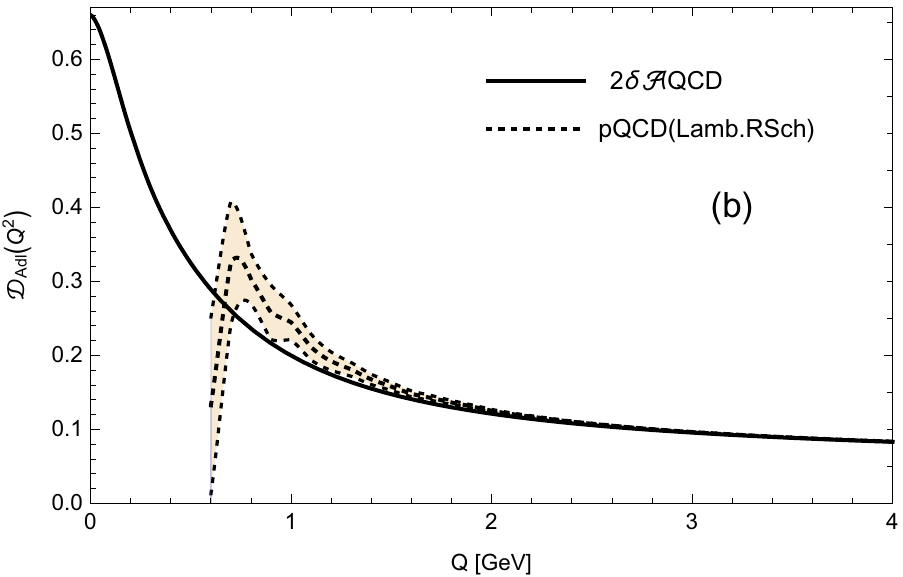}
\end{minipage}
\caption{\footnotesize  The radiative Adler function Eq.~(\ref{DAkapptlpt}) resummed with the characteristic function according to Eq.~(\ref{DAres2b}), as a function of $Q \equiv \sqrt{Q^2}$, for positive $Q^2$, in (a) $3\delta$ $\A$QCD (in the LMM renormalization scheme), and (b) $2\delta$ $\A$QCD (in the Lambert $c_2=-4.9$ renormalization scheme). Included are also the results Eqs.~(\ref{Dres2pQCD})-(\ref{Dres2pQCDIm}) in the underlying pQCD (in the same renormalization schemes). We refer to the text for more details.} 
\label{FigPlotDnd}
\end{figure}

In Figs.~\ref{FigPlotDnd}(a),(b) the resulting Adler function ${\cal D}(Q^2)_{\A{\rm res}}$ is presented for positive $Q^2$, as a function of $Q \equiv \sqrt{Q^2}$, in $3\delta$ $\A$QCD (in the LMM renormalization scheme) and in $2\delta$ $\A$QCD (in the Lambert $c_2=-4.9$ renormalization scheme), respectively. In both cases, the results in the corresponding underlying pQCD (i.e.,  pQCD in the LMM and the Lambert renormalization scheme) are included, using the resummation form Eq.~(\ref{Dres2pQCD}) with the uncertainty estimate Eq.~(\ref{Dres2pQCDIm}). The pQCD curves are not included for very low $Q < 0.6$ GeV because there they have more erratical behavior.

\begin{figure}[htb]
\centering\includegraphics[width=110mm]{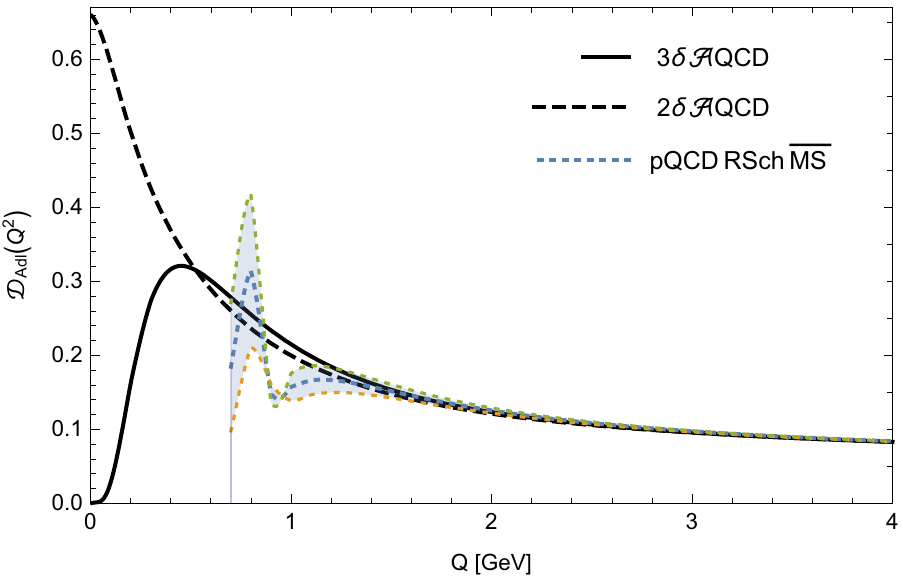}
\caption{\footnotesize The $\A$QCD curves for the Adler function ${\cal D}(Q^2)_{\A{\rm res}}$ of Figs.~\ref{FigPlotDnd}(a),(b). Included for comparison is the resummed pQCD Adler function ${\cal D}(Q^2)_{\rm pQCD res}$ in the (five-loop) $\MSbar$ scheme, using Eqs.~(\ref{Dres2pQCD})-(\ref{Dres2pQCDIm}). All the three frameworks correspond to $\alpha_s(M_Z^2;\MSbar)=0.1185$.}
\label{FigDComb}
\end{figure}

In Fig.~\ref{FigDComb}, the two mentioned $\A$QCD curves for the Adler function ${\cal D}(Q^2)_{\A{\rm res}}$ are presented in one figure, and, for comparison, the pQCD curves in the (five-loop) $\MSbar$ renormalization scheme are included according to Eqs.~(\ref{Dres2pQCD})-(\ref{Dres2pQCDIm}). All these curves were evaluated in ($\A$)QCD variants which correspond at the very high scale $Q^2=M_Z^2$ to the $\MSbar$ scheme value $\pi a(M_Z^2;\MSbar) = 0.1185$, and to the value of the $\tau$ lepton semihadronic decay ratio $r_{\tau}^{(D=0)}=0.201$ in $2\delta$ $\A$QCD \cite{2dAQCD} and in $3\delta$ $\A$QCD \cite{4l3dAQCD} in specific approaches.\footnote{Cf.~the next Sec.~\ref{subs:rtau} for more explanation on  $r_{\tau}^{(D=0)}$.}
The two programs used for the numerical evaluation are freely available at the www site \cite{MathPrgs}, and are written in {\it Mathematica\/} language. Further, as mentioned in Sec.~\ref{subs:BtD}, the higher order $\td_n$ coefficients of the Adler function in the three cases (LMM, Lambert and $\MSbar$) do not mutually agree when transformed to a common renormalization scheme (e.g., to $\MSbar$); however, the construction was such that the first five such coefficients mutually agree ($\td_0^{\MSbar},\ldots, \td_5^{\MSbar}$), the first four being the known coefficients ($\td_j^{\MSbar}$ $\Rightarrow$ $d_j^{\MSbar}$, where $j=0,\ldots,3$; $d_0=1$), and the fifth coefficient being mutually the same ($d_4^{\MSbar} = 338.19$).
One aspect in Fig.~\ref{FigDComb} that catches the attention is the significantly different behavior of the radiative Adler function at $Q<0.5$ GeV in the two $\A$QCD variants; this is a consequence of the fact that the $2\delta$ $\A$QCD coupling $\A(Q^2)$ freezes in the deep IR regime at the nonzero value $\A(0) \approx 0.66$, while the $3\delta$ $\A$QCD coupling goes to zero as $\A(Q^2) \sim Q^2$ when $Q^2 \to 0$, where the latter property is suggested by the results of the large volume lattice calculations of the dressing functions of the Landau gauge gluon and ghost propagators \cite{LattcouplNf0} (cf.~also \cite{LattcouplNf0b,LattcouplNf24,LattcouplDiscr}). The reader can refer to Appendix \ref{app:ndAQCD} for more details on the couplings $\A$ and $\tA_n$ in $2\delta$ and $3\delta$ $\A$QCD.

\subsection{Evaluation of the semihadronic $\tau$ decay ratio $r_{\tau}$}
\label{subs:rtau}

In the previous Section it was described how to evaluate the leading-twist part of the spacelike observables ${\cal D}(Q^2)$ from the Borel transforms ${\rm B}[\tD](u)$ of the auxiliary quantity $\tD(Q^2)$. This evaluation is unambiguous in $\A$QCD where no Landau singularities are present in the coupling $\A(Q^{'2})$ [and thus in $\tA_{n}(Q^{'2})$]. If ${\cal D}(Q^2)$ is the (massless) Adler function, this evaluation then allows one to evaluate, again unambiguously, the $\tau$ lepton semihadronic decay ratio $r_{\tau}^{(D=0)}$ [the strangeless and massless leading-twist ($D=0$) part]  
\bea
r_{\tau}^{(D=0)} &  = & \frac{1}{2 \pi} \int_{-\pi}^{\pi} d \theta (1 + e^{i \theta})^3 (1 - e^{i \theta}) {\cal D}( m_{\tau}^2 e^{i \theta} ),
\label{rtau1}
\eea
where the Adler function ${\cal D}(Q^2)$ has expansions of the form Eqs.~(\ref{Dkapptlpt}) with $d_0=\td_0=1$.\footnote{
This decay ratio is in the canonical form, i.e., its perturbation expansion starts with coefficient one: $(r_{\tau})_{\rm pt} = a + {\cal O}(a^2)$.} 
Using the expression (\ref{Dres2b}) for ${\cal D}$ leads to
\bea
r_{\tau}^{(D=0)} &  = & \frac{1}{2 \pi} \int_{-\pi}^{\pi} d \theta (1 + e^{i \theta})^3 (1 - e^{i \theta}) {\Bigg \{}
\int_0^{\infty} \frac{dt}{t} G_D(t) \A \left( t e^{-\tK} m_{\tau}^2 e^{i \theta} \right)
\nonumber\\ &&
+
\int_0^{\infty} \frac{dt}{t} G_D^{\rm (SL)}(t) \left[ \A \left( t e^{-\tK} m_{\tau}^2 e^{i \theta} \right) - \A \left( e^{-\tK} m_{\tau}^2 e^{i \theta} \right) \right] {\Bigg \}}.
\label{rtau2}
\eea
The (IR-safe and holomorphic) coupling $\A$ can be written in terms of the dispersion integral along its cut
\be
\A(Q^2)  =  \frac{1}{\pi} \int_0^{\infty} \frac{d \sigma \rho_{\A}(\sigma) }{( \sigma + Q^2 )},
\label{Adisp}
\ee
where $\rho_{\A}$ is the discontinuity (spectral) function of $\A$ along its cut, $\rho_{\A}(\sigma) = {\rm Im} \A(-\sigma - i \epsilon)$, and the integration starts generically at $\sigma=0$ [in 2$\delta \A$QCD and 3$\delta \A$QCD $\rho_{\A}$ is zero up to $\sigma=M^2_{\rm thr}=M_1^2$, cf.~Eq.~(\ref{rhoAnd})]. Substituting the dispersion integral (\ref{Adisp}) into Eq.~(\ref{rtau2}) and exchanging the order of integration, gives
\bea
r_{\tau}^{(D=0)} &  = & \frac{1}{\pi}
{\Bigg \{} \int_0^{\infty}  \frac{dt}{t} G_D(t) \int_0^{\infty} d \sigma \rho_{\A}(\sigma) {\cal G}(\sigma; t e^{-\tK} m_{\tau}^2)
\nonumber\\ &&
+ \int_0^{\infty}  \frac{dt}{t} G_D^{\rm (SL)}(t) \int_0^{\infty} d \sigma \rho_{\A}(\sigma) \left[  {\cal G}(\sigma; t e^{-\tK} m_{\tau}^2) -  {\cal G}(\sigma; e^{-\tK} m_{\tau}^2) \right] {\Bigg \}},
\label{rtau3}
\eea
where
\bea
{\cal G}(\sigma; t m^2) & \equiv & \frac{1}{2 \pi} \int_{-\pi}^{\pi} \frac{ d \theta (1 + e^{i \theta})^3 (1 - e^{i \theta})}{(\sigma + t m^2 e^{i \theta})}
\nonumber\\
& = & 
\Theta(\sigma-t m^2) \frac{1}{\sigma} +     
\Theta(t m^2 - \sigma) \frac{1}{t m^2} \left[2 - 2 \left( \frac{\sigma}{t m^2} \right)^2 + \left( \frac{\sigma}{t m^2} \right)^3 \right],
\label{calG}
\eea
where $m^2 \equiv e^{-\tK} m_{\tau}^2$. This then leads to the following expression for $r_{\tau}^{(D=0)}$ only in terms of the spectral function $\rho_{\A}(\sigma)$ and the characteristic functions $G_D(t)$ and $G_D^{\rm (SL)}(t)$ of the Adler function:
\bes
\label{rtau4}
\bea
r_{\tau}^{(D=0)} &  = & r_{\tau}^{\rm (L)} + r_{\tau}^{\rm (SL)},
\label{rtau4D0}
\\
r_{\tau}^{\rm (L)} &  = & \frac{1}{\pi}
{\Bigg \{} \frac{2}{m^2} \int_0^{\infty} \frac{dt}{t^2} G_D(t) \int_0^{t m^2} d \sigma \rho_{\A}(\sigma) - \frac{2}{m^6} \int_0^{\infty} \frac{dt}{t^4} G_D(t) \int_0^{t m^2} d \sigma \sigma^2 \rho_{\A}(\sigma)
\nonumber\\ &&
+ \frac{1}{m^8} \int_0^{\infty} \frac{dt}{t^5} G_D(t) \int_0^{t m^2} d \sigma \sigma^3 \rho_{\A}(\sigma)+\int_0^{\infty}  \frac{dt}{t} G_D(t) \int_{t m^2}^{\infty}  \frac{d \sigma}{\sigma} \rho_{\A}(\sigma) {\Bigg \}}, 
\label{rtau4L}
\\
r_{\tau}^{\rm (SL)} &=&  \frac{1}{\pi}
{\Bigg \{} \frac{2}{m^2} \int_0^{\infty}  \frac{dt}{t} G_D^{\rm (SL)}(t) \left[ \frac{1}{t} \int_0^{t m^2} d \sigma \rho_{\A}(\sigma) - \int_0^{m^2} d \sigma \rho_{\A}(\sigma) \right]
\nonumber\\ &&
  - \frac{2}{m^6} \int_0^{\infty}  \frac{dt}{t} G_D^{\rm (SL)}(t) \left[ \frac{1}{t^3} \int_0^{t m^2} d \sigma \sigma^2 \rho_{\A}(\sigma) - \int_0^{m^2} d \sigma \sigma^2 \rho_{\A}(\sigma) \right]
\nonumber\\ &&
+ \frac{1}{m^8} \int_0^{\infty}  \frac{dt}{t} G_D^{\rm (SL)}(t) \left[ \frac{1}{t^4} \int_0^{t m^2} d \sigma \sigma^3 \rho_{\A}(\sigma) -  \int_0^{m^2} d \sigma \sigma^3 \rho_{\A}(\sigma) \right]
+ \int_0^{\infty}  \frac{dt}{t} G_D^{\rm (SL)}(t) \int_{t m^2}^{m^2}  \frac{d \sigma}{\sigma} \rho_{\A}(\sigma) {\Bigg \}}.
\label{rtau4SL}
\eea
\ees
These double integrals can be reduced further to single integrals by integration by parts in $t$
\bes
\label{rtau5}
\bea
r_{\tau}^{\rm (L)} &  = & {\Bigg \{} -2 \int_0^{\infty} dt \rho_{\A}(t m^2) {\cal F}_1(t) + 2 \int_0^{\infty} dt \; t^2 \rho_{\A}(t m^2) {\cal F}_3(t)
\nonumber\\ &&
-  \int_0^{\infty} dt \; t^3 \rho_{\A}(t m^2) {\cal F}_4(t) + \left[\A(0) + \int_0^{\infty} \frac{dt}{t} \rho_{\A}(t m^2) {\cal F}_0(t) \right] {\Bigg \}}
\label{rtau5L}
\\
r_{\tau}^{\rm (SL)} &  = & {\Bigg \{} -2 \int_0^{1} dt \rho_{\A}(t m^2) \left[ {\cal F}_1^{\rm (SL)}(t) - C_1^{\rm (SL)} \right] + 2 \int_0^{1} dt \; t^2 \rho_{\A}(t m^2) \left[ {\cal F}_3^{\rm (SL)}(t) - C_3^{\rm (SL)} \right]
\nonumber\\ &&
-  \int_0^{1} dt \; t^3 \rho_{\A}(t m^2) \left[ {\cal F}_4^{\rm (SL)}(t) -  C_4^{\rm (SL)} \right] + \int_0^{1} \frac{dt}{t} \rho_{\A}(t m^2) \left[ {\cal F}_0^{\rm (SL)}(t)  - C_0^{\rm (SL)} \right] {\Bigg \}}
\label{rtau5SL}
\eea
\ees
where we recall that $m^2 = e^{-\tK} m_{\tau}^2$; the functions ${\cal F}_j$, ${\cal F}_j^{\rm (SL)}$ and the constants $C_j^{\rm (SL)}$ are
\bes
\label{cFjcFjSL}
\bea
 {\cal F}_j(t) & = & \frac{1}{\pi} \int_{+\infty}^t \frac{d u}{u^{j+1}} G_D(u) =
 \Theta(t-1) {\cal F}^{(+)}_j(t)+ \Theta(1-t) {\cal F}^{(-)}_j(t)
 \quad (j=0,1,3,4),
 \label{calFj}
 \\
 {\cal F}_j^{\rm (SL)}(t) & = &   \frac{1}{\pi}\int_{1/2}^t \frac{d u}{u^{j+1}} G_D^{\rm (SL)}(u)
 \quad (j=0,1,3,4),
\label{calFjSL}
\\
C_j^{\rm (SL)} & = & j \int_0^1 dt t^{j-1} {\cal F}_j^{\rm (SL)}(t) \quad (j=1,3,4),
\quad C_0^{\rm (SL)} = {\cal F}_0^{\rm (SL)}(0).
\label{CjSL}
\eea
\ees
Using the expressions (\ref{GDs}) for the characteristic functions $G_D$ and $G_D^{\rm (SL)}$ in the above integrations, the explicit expressions for the integrand functions ${\cal F}_j$ and ${\cal F}_j^{\rm (SL)}$ can be obtained and are given in Appendix \ref{app:FjFjSL}.

The quantity $r_{\tau}^{(D=0)}$ has been evaluated in the $2\delta$ $\A$QCD \cite{2dAQCD} and $3\delta$ $\A$QCD \cite{4l3dAQCD} frameworks in specific approximations. In $2\delta$ $\A$QCD, $r_{\tau}^{(D=0)}$ was obtained in \cite{2dAQCD} by evaluation of the leading-$\beta_0$ (LB) resummation analogous to the one described here [because all the coefficients $\td_j^{\rm (LB)}$ in the Adler function are known, cf.~Eqs.~(\ref{BtDLB})], and adding the beyond-leading-$\beta_0$ (bLB) contribution obtained from the known coefficients $\td_j - \td_j^{\rm (LB)}$ ($j \leq 3$). In $3\delta$ $\A$QCD, $r_{\tau}^{(D=0)}$ was obtained in \cite{4l3dAQCD} by evaluation of the known truncated series (\ref{DAkaplpt}) for the Adler function (i.e., truncated at $\td_3 \tA_4$), which happened to give almost the same value as the extended diagonal Pad\'e (dPA) approximation for the Adler function, ${\cal G}^{[M/M]}_{{\cal D}}(Q^2)_{\A{\rm QCD}}$ with $M=2$, cf.~the next Sec.~\ref{sec:dBG}. Here, the programs \cite{MathPrgs} were used for the $2\delta$ $\A$QCD (in the mentioned Lambert scheme) and $3\delta$ $\A$QCD (in the LMM scheme) where the parameters of the coupling $\A(Q^2)$ are adjusted so that at very high momenta the coupling corresponds to the $\MSbar$ value $\alpha_s(M_Z^2;\MSbar)=0.1185$, and gives the value $r_{\tau}^{(D=0)}=0.201$ obtained in the two aforementioned respective approximate approaches. These two programs were also used in the previous Section \ref{subs:ft} in the evaluation of the Adler function for positive $Q^2$, and the reader is referred to Appendix \ref{app:ndAQCD} for more details.

The obtained results for $r_{\tau}^{(D=0)}$ using the resummed expression obtained in this Section, Eqs.~(\ref{rtau5}), are
\bes
\label{rtauD0}
\bea
r_{\tau}^{(D=0)} & = & 0.2056 \qquad (3\delta \; \A{\rm QCD,} \; {\rm LMM \; scheme}),
\label{rtauD03d}
\\
r_{\tau}^{(D=0)} & = & 0.1973 \qquad (2\delta \; \A{\rm QCD, \; Lamb. \; scheme}).
\label{rtauD02d}
\eea
\ees
This is to be compared with the value $r_{\tau}^{(D=0)}=0.201$ obtained in $3\delta$ and $2\delta$ $\A$QCD by the approximate methods mentioned in the previous paragraph (cf.~also footnote \ref{footrtau} in Appendix \ref{app:ndAQCD}). For the SL coefficient $\tal$ the central values (\ref{tals}) $\tal=-0.14, -0.10$ were used, respectively. However, the results for $r_{\tau}^{(D=0)}$, and for the (resummed) Adler function ${\cal D}(Q^2)_{\A{\rm res}}$, depend only weakly on the choice of $\tal$. E.g., if $\tal=0$ is taken in  $3\delta$ $\A$QCD (in the LMM scheme), the result changes to $r_{\tau}^{(D=0)}=0.2066$, very close to the value in Eq.~(\ref{rtauD03d}); this is so because, with the change of $\tal$, the other parameters in the Borel transform (\ref{BtD4P}) change accordingly so as to reproduce the four known coefficients $\td_j$ ($j=0,\ldots,3$).\footnote{If $\tal=0$ is used in ${\rm B}[\tD](u)$, Eq.~(\ref{BtD4P}), the resulting value of $d_4^{\MSbar}$ becomes $300.92$ (for $\tal=-0.14$ it is $338.19$).} Nonetheless, the SL term proportional to $\tal$ was included in the Borel transform ans\"atze (\ref{BtD4P}) and (\ref{BtD5P}) because of the knowledge of the exact value of the parameter ${\hat c}_1^{(4)}$, Eq.~(\ref{hatc1}), which makes it possible to extract the value of $\tal$ [cf.~Eqs.~(\ref{tal1})-(\ref{al})], although in general with significant uncertainties, Eqs.~(\ref{tals}).

\section{From asymptotically divergent series to a convergent sequence}
\label{sec:dBG}

Here, another method of evaluation of the leading-twist part of spacelike observables ${\cal D}(Q^2)$ will be used, a method which grows increasingly effective when the number of known coefficients $\td_n$ in the logarithmic perturbation expansion (\ref{Dkaplpt}) increases. This method was proposed in Ref.~\cite{BGApQCD1} for the case when the number of known coefficients is even ($\td_0, \ldots, \td_{2 M-1}$), and was modified in Ref.~\cite{BGApQCD2} to be applicable when the number of known coefficients is odd. Later this method was applied in $\A$QCD variants where the $\A(Q^2)$ coupling is IR-safe and holomorphic outside the negative semiaxis in the complex $Q^2$-plane \cite{BGA,anOPE,Techn}. It is an extension of the diagonal Pad\'e (dPA) approach, where the latter gives a result which is renormalization scale independent at the one-loop level \cite{GardiPA}. The result of this extended dPA approach is exactly renormalization scale independent. The $2 M$-degree ($[M/M]$) approximant fo ${\cal D}(Q^2)$ is constructed from the knowledge of the coefficients $\td_j(\kappa)$ ($j=0,1,\ldots, 2 M - 1$) of the expansion (\ref{Dkaplpt}) by considering first the corresponding power expansion (\ref{tDkap}) of the auxiliary quantity $\tD(Q^2; \kappa)$ (truncated at $\sim a^{2 M-1}$) and constructing from it the diagonal Pad\'e (dPA) $[M/M]$, i.e., ratio of two polynomials of degree $M$
\be
P^{[M/M]}_{\tD}(a_{\mu}; Q^2) = \td_0 \frac{a_{\mu} \left( 1 + \sum_{j=1}^{M-1} N_j a_{\mu}^j \right)}{\left( 1 + \sum_{k=1}^M D_k a_{\mu}^k \right)},
\label{dPMM1}
\ee
where $a_{\mu} \equiv a(\mu^2)$ and the $2 M -1 $ coefficients $N_j, D_k$ are determined by the condition that the expansion of the above expression in powers of $a_{\mu}$ reproduce the (known) coefficients $\td_1(\kappa),\ldots, \td_{2 M-1}(\kappa)$. The ratio (\ref{dPMM1}) can always be decomposed into a sum of simple fractions
\bea
P^{[M/M]}_{\tD}(a_{\mu};Q^2) &=& \td_0 \sum_{k=1}^{M} \tal_j \frac{a_{\mu}}{(1+ \tu_j a_{\mu})}
\nonumber\\
&=&  \td_0 \sum_{k=1}^{M} \tal_j a^{\rm (1-l.)}_{(\mu^2)}(\kappa_j Q^2),
\label{dPMM2}
\eea
where $a^{\rm (1-l.)}_{(\mu^2)}(\kappa_j Q^2)$ is the value of the coupling $a$ RGE-evolved from the scale $\mu^2$ ($ \equiv \kappa Q^2$) to $\kappa_j Q^2$ by one-loop RGE
\be
a^{\rm (1-l.)}_{(\mu^2)}(\kappa_j Q^2) = \frac{a_{\mu}}{(1 + \beta_0 \ln(\kappa_j/\kappa) a_{\mu})},
\label{a1lkapj}
\ee
and the complex constants $\tal_j$ and $\kappa_j = \kappa \exp(\tu_j/\beta_0)$ in Eq.~(\ref{dPMM2}) are independent of the renormalization scale $\mu^2 \equiv \kappa Q^2$ and of the physical scale $Q^2$, \cite{BGApQCD1}. Further, we have
\be
\sum_{j=1}^M \tal_j = 1.
\label{sumtal}
\ee
The diagonal Pad\'e (\ref{dPMM2}) is independent of the initial renormalization scale $\mu^2$ in the approximation of one-loop, but it is not exactly $\mu^2$-independent. To extend this resummation in such a way as to get the exact $\mu^2$-independence, the crucial point was to realize that $\tal_j$ and $\kappa_j$ are $\mu^2$-independent constants. The one-loop RGE-evolved quantities $a^{\rm (1-l.)}(\kappa_j Q^2)$ are replaced by the exactly (n-loop) RGE-evolved couplings $a(\kappa_j Q^2)$, leading to
\bea
{\cal G}^{[M/M]}_{{\cal D}}(Q^2)_{\rm pQCD} & = &  \td_0 \sum_{k=1}^{M} \tal_j a(\kappa_j Q^2).
\label{GMMpQCD}
\eea
This is now exactly $\mu^2$-independent,\footnote{Note that the original power series (\ref{tDkap}) for $\tD(Q^2;\kappa)$, from which this approximant for ${\cal D}(Q^2)$ was constructed, is at an arbitrary normalization scale $\mu^2=\kappa Q^2$.}
 and can be shown  \cite{BGApQCD1} to agree with the perturbation expansion of the full observable ${\cal D}(Q^2)$ to the order $\sim a^{2 M}$ 
\be
{\cal G}^{[M/M]}_{{\cal D}}(Q^2)_{\rm pQCD} - {\cal D}(Q^2)_{\rm (l)pt} = {\cal O}(a^{2 M +1}).
\label{GMMpQCDapp}
\ee
This resummation was applied in pQCD \cite{BGApQCDex} with reduced success because some of the complex (or real) parameters $\kappa_j$ usually turn out to be small ($|\kappa_j|<1$), so that the scale $(\kappa_j Q^2)$ is in the vicinity or on the Landau singularities, making evaluation $a(\kappa_j Q^2)$ unrealistic or impossible. Later, in Refs.~\cite{BGA,anOPE,Techn}, it was realized that this resummation can be applied in the $\A$QCD frameworks, i.e., with IR-safe and holomorphic $\A(Q^2)$ coupling, with the same values of $\tal_j$ and $\kappa_j$
\bes
\label{GMMAQCD}
\bea
{\cal G}^{[M/M]}_{{\cal D}}(Q^2)_{\A{\rm QCD}} & = &  \td_0 \sum_{k=1}^{M} \tal_j \A(\kappa_j Q^2),
\label{GMMAQCDa}
\\
{\cal G}^{[M/M]}_{{\cal D}}(Q^2)_{\A{\rm QCD}} - {\cal D}(Q^2)_{\A{\rm QCD}} &=& {\cal O}(\tA_{2 M +1}) = {\cal O}(a^{2 M+1}),
\label{GMMAQCDb}
\eea
\ees
where ${\cal D}(Q^2)_{\A{\rm QCD}}$ is the expansion in logarithmic derivatives $\tA_n(\mu^2)$ in $\A$QCD, Eq.~(\ref{DAkaplpt}). In contrast to pQCD, in the $\A$QCD variants the small $|\kappa_j|$ values are not a problem, because there are no Landau singularities of $\A(Q^{'2})$ [and thus of $\tA_n(Q^{'2})$] in the $Q^{'2}$-complex plane. This approach was applied to the known four-term truncated series ${\cal D}_{\rm Adl}^{[4]}(Q^2)$ for the Adler function ($\td_j$ known up to $j_{\rm max}=2 M - 1 = 3$, i.e., $M=2$) for various $\A$QCD variants in Refs.~\cite{BGA,anOPE,Techn,4l3dAQCD}; and in Refs.~\cite{BGA,Techn} to the large-$\beta_0$ (LB) part of the Adler function, ${\cal D}_{\rm Adl}^{{\rm (LB)}}(Q^2)$, which is known to all orders [cf.~Eqs.~(\ref{BtDLB})]. However, the $\A$QCD series (\ref{DAkapptlpt}) is in general asymptotically divergent.\footnote{This is true even for the LB-part ${\cal D}_{\rm Adl}^{{\rm (LB)}}(Q^2)$, cf.~\cite{BGA,Techn}.}
Specifically, the sequence $\{ {\cal D}^{[2 M]}_{\A{\rm QCD}}(Q^2;\kappa); M=1,2,\ldots \}$ of the truncated $\A$QCD series (\ref{DNAkaplpt}) (for $N=2 M$)
\bea
{\cal D}^{[2 M]}_{\A{\rm QCD}}(Q^2; \kappa) &=& \td_0 \A(\kappa Q^2) + \td_1(\kappa)  \tA_2(\kappa Q^2) + \ldots + \td_{2 M-1}(\kappa) \tA_{2 M}(\kappa Q^2),
\label{DNAkapT2M}
\eea
is asymptotically divergent at any $Q^2$, because of the renormalon-type behavior of the coefficients $\td_n \sim n!$ at large $n$. Further, the extended dPA ${\cal G}^{[M/M]}_{{\cal D}}(Q^2)_{\A{\rm QCD}}$, Eq.~(\ref{GMMAQCDa}), is based on the truncated series (\ref{DNAkapT2M}), i.e., it is based on the knowledge of the first $2 M$ coefficients $\td_n$ ($n=0,1,\ldots, 2 M -1$). Therefore, applying the extended dPA method (\ref{GMMAQCD}) to ${\cal D}_{\rm Adl}(Q^2)$, and taking into account the relations Eq.~(\ref{GMMAQCDb}), one may conservatively expect that the sequence of the extended dPA's, $\{ {\cal G}^{[M/M]}_{{\cal D}}(Q^2)_{\A{\rm QCD}}; M=1,2,\ldots \}$ will be asymptotically divergent, too, for all or at least some values of $Q^2$. In the case of  LB-part ${\cal D}_{\rm Adl}^{{\rm (LB)}}(Q^2)$, applied in various $\A$QCD variants, the results of Refs.~\cite{BGA,Techn} strongly indicate that this is not so, and that the mentioned sequence of extended dPA's is a convergent sequence, for all complex $Q^2$ for which $\A(Q^2)$ is holomorphic, i.e., in the entire spacelike regime $Q^2 \in \mathbb{C} \backslash (-\infty, -M_{\rm thr}^2]$.

We recall that in Sec.~\ref{subs:BtD} the (leading-twist) Adler function ${\cal D}_{\rm Adl}(Q^2)$ was constructed in two models, i.e., the Adler function (\ref{DAkaplpt}) whose coefficients $\td_n$ are generated: (a) by the one-loop type Borel ${\rm B}[\tD](u)$ of Eq.~(\ref{BtD4P}), in the LMM renormalization scheme, where $3 \delta \A$QCD \cite{4l3dAQCD} is applied; (b) by ${\rm B}[\tD](u)$ of Eq.~(\ref{BtD5P}), in the $c_2=-4.9$ Lambert renormalization scheme, where $2 \delta \A$QCD \cite{2dAQCD} is applied. In Table \ref{tab4P5P} the parameters of these two Borel transforms were given. In both Adler function models, the first four known coefficients $\td_j$ ($\Leftrightarrow d_j$) are reproduced ($j=0,1,2,3$), and, in addition, both models give the same value of the next coefficient, which in the $\MSbar$ scheme is $d_4(\MSbar,N_f=3)= 338.2$. Further, we recall that in both models the full leading-twist Adler function can be evaluated (resummed) exactly, by integrations involving the corresponding characteristic functions $G_D(t)$ and $G_D^{\rm (SL)}$ and the coupling $\A(t e^{-\tK} Q^2)$, cf.~Sec.~\ref{subs:ft}. In view of the discussion in the previous paragraph, the natural question appearing at this point is whether the sequence of extended dPA's, $\{ {\cal G}^{[M/M]}_{{\cal D}}(Q^2)_{\A{\rm QCD}}; M=1,2,\ldots \}$, applied in these two models, is a convergent sequence; and, if it is, whether it converges to the exact values as determined by the mentioned integration with the characteristic functions. If the reply to both these questions is positive, the next natural question would be whether this sequence for the Adler function, when applied in the contour integral (\ref{rtau1}) for the $\tau$ semihadronic decay ratio $r_{\tau}^{(D=0)}$, leads to a sequence $r_{\tau}^{(D=0)}([M/M])$ converging to the exact value of $r_{\tau}^{(D=0)}$ as obtained with the method of characteristic functions Eqs.~(\ref{rtau5}).

\begin{figure}[htb] 
\begin{minipage}[b]{.49\linewidth}
  \centering\includegraphics[width=85mm]{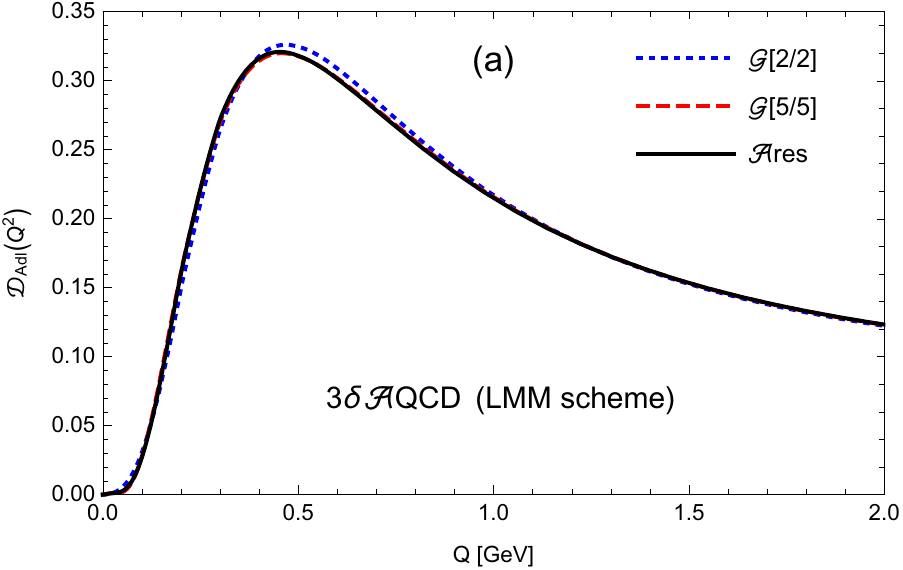}
  \end{minipage}
\begin{minipage}[b]{.49\linewidth}
  \centering\includegraphics[width=85mm]{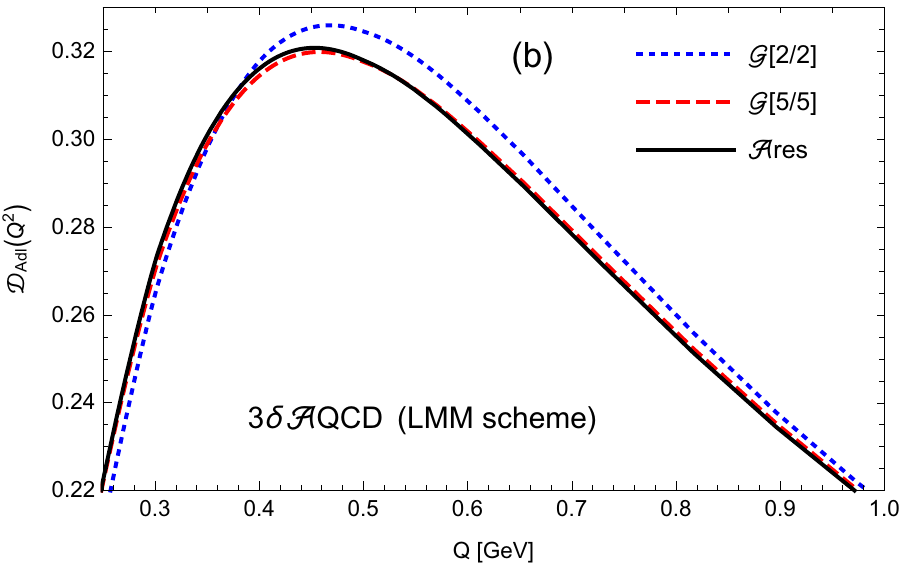}
\end{minipage}
\caption{\footnotesize  The Adler function Eq.~(\ref{DAkapptlpt}) as a function of $Q=\sqrt{Q^2}$ for positive $Q^2$, evaluated with the extended dPA method Eq.~(\ref{GMMAQCD}) for $M=2$ and $M=5$, in $3\delta$ $\A$QCD: (a) for $0 < Q < 2$ GeV; (b) zoomed version, for $0.25 \ {\rm GeV} < Q < 1$ GeV. The exact result resummed with the characteristic function according to Eq.~(\ref{DAres2b}) is included as a solid line, for comparison.}
\label{FigPlotD3ddBG}
\end{figure}
In Figs.~\ref{FigPlotD3ddBG}(a),(b) the Adler function for $3\delta$ $\A$QCD framework is presented, evaluated with the extended dPA method (\ref{GMMAQCD}) with $M=2$ and $M=5$. For comparison, we include also the exact evaluation, Eq.~(\ref{DAres2b}) using the characteristic function. We can see that it is almost impossible to distinguish by eye the extended dPA $M=5$ curve from the exact curve, even in the zoomed version Fig.~\ref{FigPlotD3ddBG}(b).

\begin{figure}[htb] 
\begin{minipage}[b]{.49\linewidth}
  \centering\includegraphics[width=85mm]{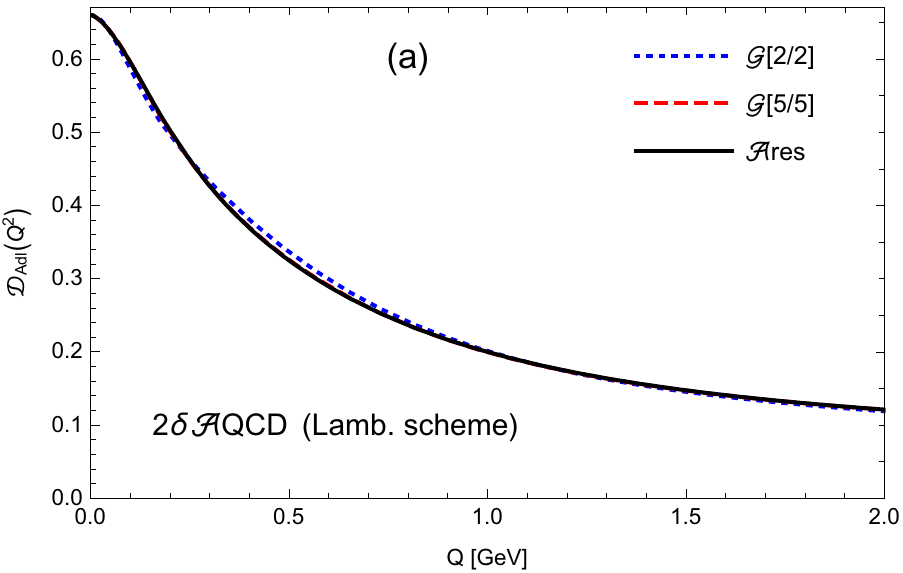}
  \end{minipage}
\begin{minipage}[b]{.49\linewidth}
  \centering\includegraphics[width=85mm]{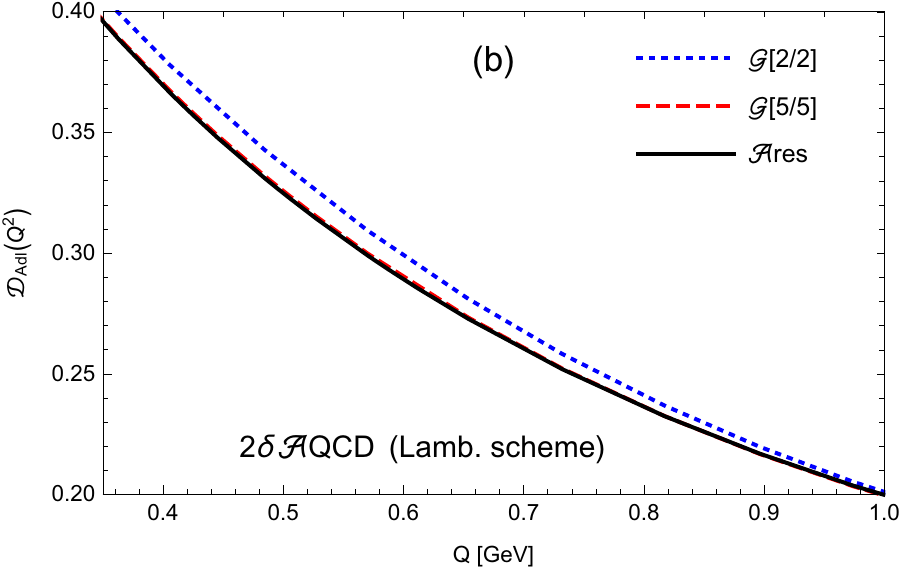}
\end{minipage}
\caption{\footnotesize  The analogous results as in Figs.~\ref{FigPlotD3ddBG}, but now for  $2\delta$ $\A$QCD: (a) for $0 < Q < 2$ GeV; (b) zoomed version, for $0.35 \ {\rm GeV} < Q < 1$ GeV.}
\label{FigPlotD2ddBG}
\end{figure}
In Figs.~\ref{FigPlotD2ddBG}(a),(b) the analogous results for the $2\delta$ $\A$QCD framework are presented, with similar conclusions.

Both Figures indicate that we have a converging sequence $\{ {\cal G}^{[M/M]}_{{\cal D}}(Q^2)_{\A{\rm QCD}}; M=1,2,\ldots \}$ when the index $M$ increases, for all positive $Q^2$, and that the sequence converges to the exact value as obtained with the characteristic functions of each Adler function model.

\begin{figure}[htb] 
\begin{minipage}[b]{.49\linewidth}
  \centering\includegraphics[width=85mm]{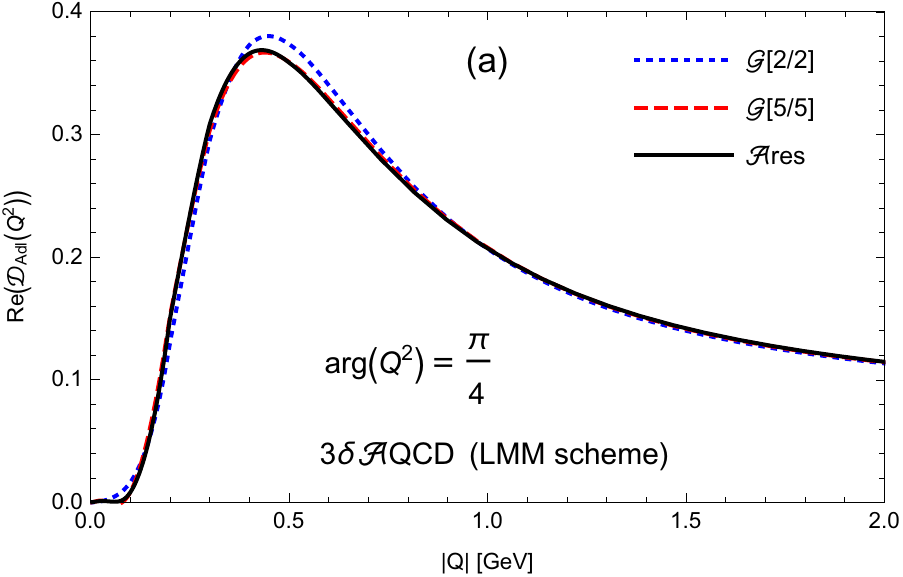}
  \end{minipage}
\begin{minipage}[b]{.49\linewidth}
  \centering\includegraphics[width=85mm]{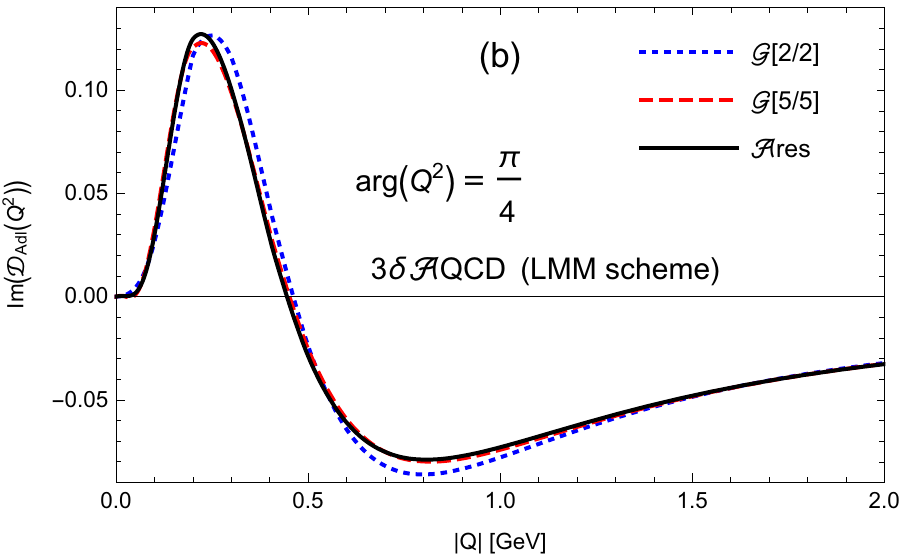}
\end{minipage}
\begin{minipage}[b]{.49\linewidth}
  \centering\includegraphics[width=85mm]{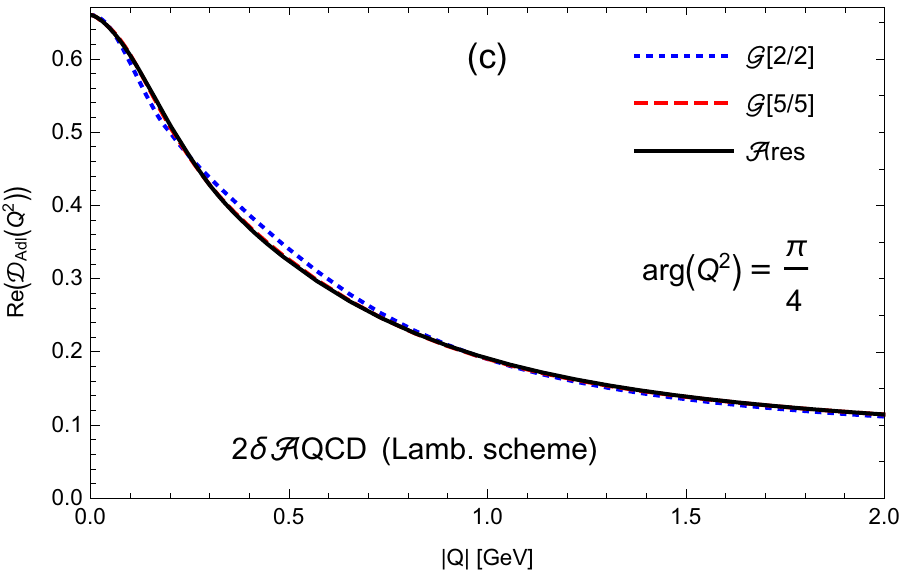}
  \end{minipage}
\begin{minipage}[b]{.49\linewidth}
  \centering\includegraphics[width=85mm]{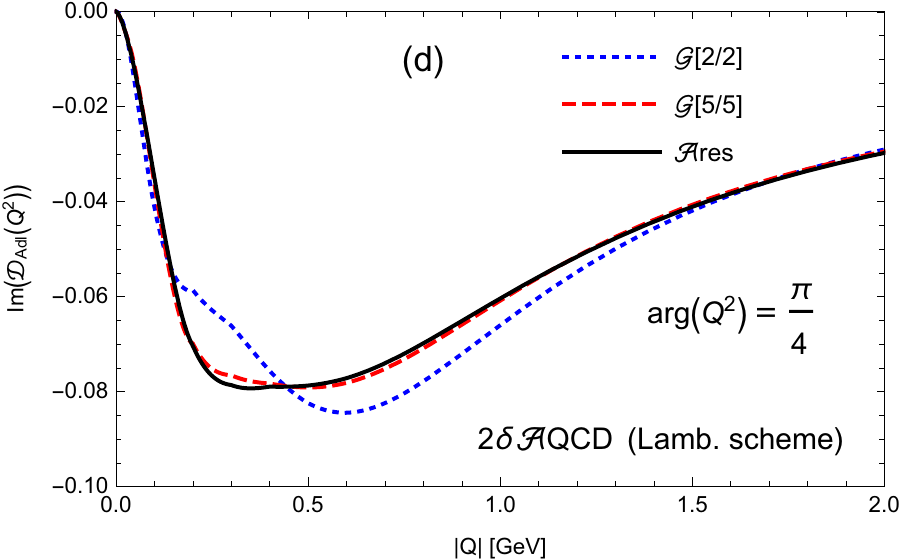}
\end{minipage}
\caption{\footnotesize   The Adler function Eq.~(\ref{DAkapptlpt}) as a function of $|Q| \equiv \sqrt{|Q^2|}$ for complex $Q^2=|Q^2| \exp( i \pi/4)$, evaluated with the extended dPA method Eq.~(\ref{GMMAQCD}) for $M=2$ and $M=5$: (a) real part, in $3\delta$ $\A$QCD; (b) imaginary part, in  $3\delta$ $\A$QCD; (c) real part, in $2\delta$ $\A$QCD; (d) imaginary part, in  $2\delta$ $\A$QCD. Included as the solid line is the exact result  resummed with the characteristic function according to Eq.~(\ref{DAres2b}).}
\label{FigPlotDdBGfiPi4}
\end{figure}

When $Q^2$ is complex, the convergence to the exact value persists, but gets slower for the values of $Q^2$ which are closer to the cut on the negative semiaxis. In Figs.~\ref{FigPlotDdBGfiPi4} analogous results as in the previous Figs.~\ref{FigPlotD3ddBG} and \ref{FigPlotD2ddBG} are presented, but now for complex $Q^2=|Q^2| \exp( i \pi/4)$, where the real and imaginary parts of the resulting Adler function ${\cal D}_{\rm Adl}(Q^2)$ are presented separately.

In Table \ref{tabdPA} the numerical results for ${\cal G}_{{\cal D}}^{[M/M]}(Q^2)_{{\A}{\rm QCD}}$ for $3\delta$ and $2\delta$ $\A$QCD are given, at $Q^2=0.5$, $1.0$ and $2.0 \ {\rm GeV}^2$, when the index $M$ increases. These results clearly show that we have indeed convergent sequences.
\begin{table}
  \caption{Numerical results for extended dPA evaluation of Adler function, Eq.~(\ref{GMMAQCD}), in $3\delta$ and $2\delta$ $\A$QCD, for increasing index $M$ at three different values of squared momenta $Q^2=0.5$, $1.0$ and $2.0 \ {\rm GeV}^2$. At each value, the value of the corresponding truncated series (\ref{DNAkapT2M}) is given in parentheses (with $\kappa=1$), for comparison.  Included below is the exact value obtained by the use of the characteristic function, Eq.~(\ref{DAres2b}), for comparison. At then end is given also the analogous value obtained in the underlying pQCD (i.e., in the LMM and Lambert scheme, respectively) by using the generalized Principal Value Eqs.~(\ref{Dres2pQCD})-(\ref{Dres2pQCDIm}).}
\label{tabdPA}
\begin{ruledtabular}
\begin{tabular}{r|lll|lll}
  $M$ & $3\delta : \; \; Q^2=0.5 \ {\rm GeV}^2$ & $1 \ {\rm GeV}^2$ & $2 \ {\rm GeV}^2$ &  $2\delta : \; \; Q^2=0.5 \ {\rm GeV}^2$ & $1 \ {\rm GeV}^2$ & $2 \ {\rm GeV}^2$
\\
\hline
2 & 0.28300 (0.28320) & 0.21682 (0.21700) &                 0.16031 (0.16035)    & 0.26518 (0.27341) & 0.20091 (0.20019)& 0.15193 (0.14884)
\\
5 & 0.27761 (0.92350) & 0.21542 (0.68605) &                 0.16085 (-0.02698)  & 0.25862 (0.03642) & 0.19940 (0.05317) & 0.15344 (0.12759)
 \\
10 & 0.27662 ($\sim 10^{6}$) & 0.21491 ($\sim 10^{6}$) & 0.16089 ($\sim 10^{5}$)  & 0.25809 ($\sim 10^{5}$) & 0.19976 ($\sim 10^{5}$) & 0.15385 ($\sim 10^{5}$)
\\
20 & 0.27665 ($\sim 10^{26}$) & 0.21484 ($\sim 10^{25}$) & 0.16087 ($\sim 10^{23}$) & 0.25828 ($\sim 10^{23}$) & 0.19969 ($\sim 10^{25}$) & 0.15377 ($\sim 10^{25}$)
\\
\hline
exact &  0.27666 & 0.21483 & 0.16087 & 0.25827 & 0.19968 & 0.15378 
\\
pQCD & $0.111 \pm 0.073$ & $0.247 \pm 0.039$ & $0.181 \pm 0.007$ &
$0.251 \pm 0.082$ & $0.245 \pm 0.024$ & $0.168 \pm 0.007$
\end{tabular}
\end{ruledtabular}
\end{table}
In Table \ref{tabdPA} the corresponding values of the truncated $\A$QCD series (\ref{DNAkapT2M}) were included in parentheses (with $\kappa=1$), which clearly show, as expected, that such series are asymptotically divergent, with divergence setting in at terms $\sim \tA_n(Q^2)$ with $2 \times 2 < n < 2 \times 5$.

\begin{figure}[htb] 
\begin{minipage}[b]{.49\linewidth}
  \centering\includegraphics[width=85mm]{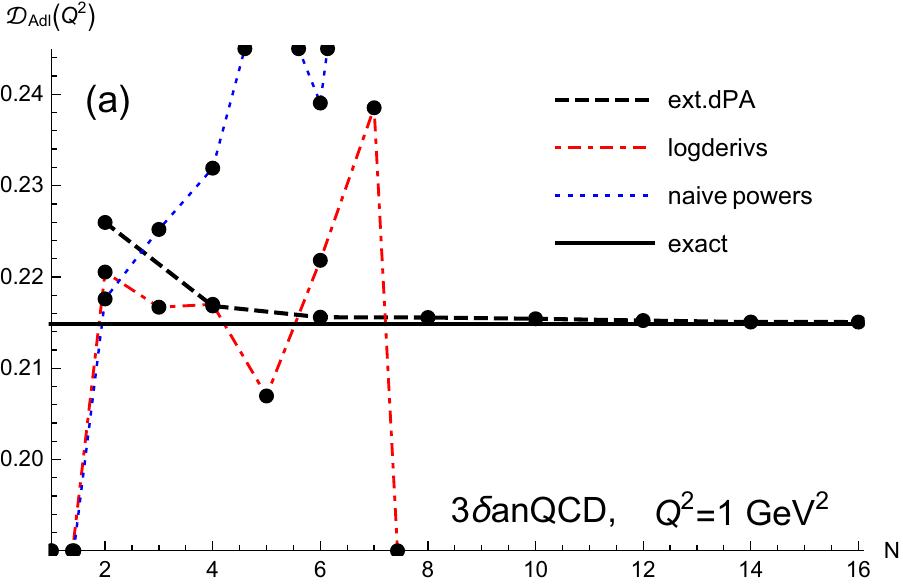}
  \end{minipage}
\begin{minipage}[b]{.49\linewidth}
  \centering\includegraphics[width=85mm]{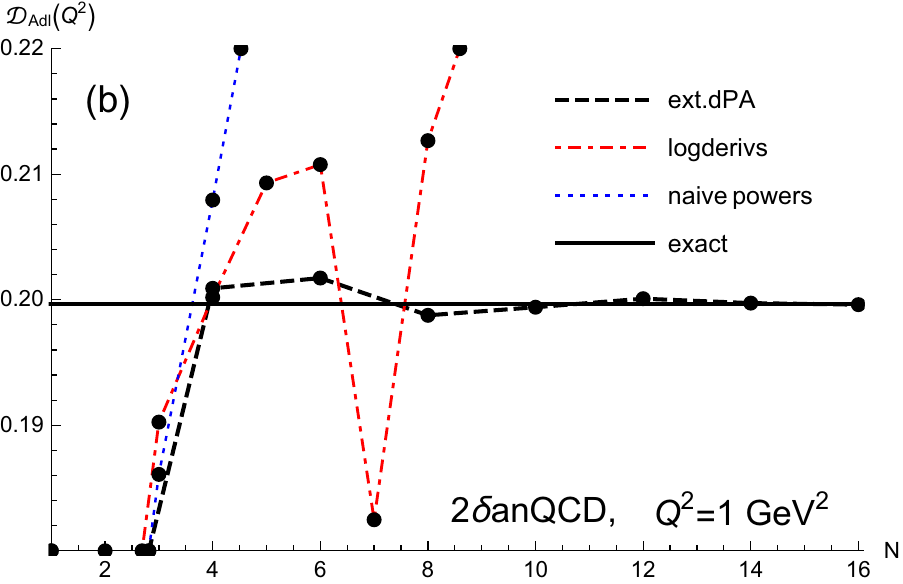}
\end{minipage}
\caption{\footnotesize  The convergence of the sequence of extended dPAs  $\{ {\cal G}^{[M/M]}_{{\cal D}}(Q^2)_{\A{\rm QCD}}; M=1,2,\ldots \}$, at $Q^2=1 \ {\rm GeV}^2$, as a function of index $N=2 M = 2, 4, \ldots, 16$: (a) in the considered $3\delta$ $\A$QCD; (b) in the considered $2\delta$ $\A$QCD. Included is, for comparison, the corresponding sequence $\{ {\cal D}^{[N]}_{\A{\rm QCD}}(Q^2; \kappa=1); N=1,2,\ldots \}$ of truncated $\A$QCD Eq.~(\ref{DNAkaplpt}), and the sequence of the truncated series with naive power terms $d_{n-1} \A(Q^2)^n$. The exact value, obtained in Sec.~(\ref{subs:ft}), is presented as the black horizontal line.}
\label{FigdBGvsN}
\end{figure}

In order to better visualise the convergence and/or asymptotic divergence with increasing index number, Figs.~\ref{FigdBGvsN}(a),(b) show the behavior of the extended dPA sequence $\{ {\cal G}^{[M/M]}_{{\cal D}}(Q^2)_{\A{\rm QCD}}; M=1,2,\ldots \}$ as a function of index $N=2 M$ ($ = 2, 4, \ldots, 16$), for $Q^2=1 \ {\rm GeV}^2$, in the considered $3\delta$ and $2\delta$ $\A$QCD, respectively.
For comparison, the exact value of the considered cases is included (as obtained with characteristic functions in Sec.~\ref{subs:ft}), as well as the corresponding values of the truncated $\A$QCD series ${\cal D}^{[N]}_{\A{\rm QCD}}(Q^2; \kappa=1)$ Eq.~(\ref{DNAkaplpt}), for $N=1,2,\ldots$ ('logderivs'); this series can be written also in the form of Eq.~(\ref{DNAkappt}) (with $\kappa=1$), where the power analogs $\A_n(Q^2)$ are not equal to the naive powers $\A(Q^2)^n$. For additional comparison, in Figs.~\ref{FigdBGvsN} were included the results of the truncated $\A$QCD series where the terms $d_{n-1} \A_n(Q^2)$ are replaced by the naive (and thus incorrect) power terms $d_{n-1} \A(Q^2)^n$ ('naive powers').\footnote{
In \cite{Techn} it was argued that the naive power terms $\A(Q^2)^n$ in $\A$QCD variants in general bring spurious uncontrollable nonperturbative contributions, in contrast to the logarithmic derivatives $\tA_n(Q^2)$ and the related power analogs $\A_n(Q^2)$; cf.~also Eqs.~(\ref{DNAkappt})-(\ref{AntAn}) in Appendix \ref{app:ndAQCD}.}$^,$\footnote{On the other hand, in Ref.~\cite{anpQCD} a perturbative (pQCD) coupling $a(Q^2)=\A(Q^2)$ was constructed which has no Landau singularities and reproduces the correct values of $r_{\tau}$. In this case, the construction (\ref{AntAn}) gives $\A_n \approx a^n$ (the equality becoming increasingly better when the truncation index $N$ increases), and the extended Pad\'e (\ref{GMMpQCD}) or equivalently (\ref{GMMAQCD}) can be applied, due to the absence of the Landau singularities, and gives a convergent sequence, presumably converging to the exact value. However, it is unclear how to construct a renormalon-motivated model (\ref{BtD5P}) for the higher order 'exact' coefficients $d_n$ of the Adler function in this case, because the pQCD scheme of this holomorphic pQCD coupling has the scheme coefficients $c_n =\beta_n/\beta_0$ growing so fast with increasing $n$ that the renormalon structure is severely affected by the transformation into this scheme (with the exception of the $p=1$ UV renormalon).}  
It can be seen in Figs.~\ref{FigdBGvsN} that the extended dPA sequence is consistently convergent, it converges to the exact value, and that the sequence of truncated $\A$QCD is asymptotically divergent, i.e., it approximately stabilizes at $3 \leq N \leq 6$, and for $6 < N$ is starts diverging. The sequence of the truncated series in naive powers, on the other hand, does not show any clear stabilization, it appears to be more divergent.    

The $\tau$ decay ratio parameter $r_{\tau}^{(D=0)}$ can also be evaluated with this method, by evaluating the Adler function ${\cal D}(m_{\tau}^2 \exp(i \theta))$ in the integrand of the contour integral Eq.~(\ref{rtau1}) with the described extended dPA method. If the latter method gives convergent sequence in principle for all complex $Q^2$ [including those not far from the negative semiaxis where $\A(Q^2)$ has a cut], then it is expected that the obtained sequence $r_{\tau}^{(D=0),[M/M]}$
\bea
r_{\tau}^{(D=0),[M/M]} &=& \frac{1}{2 \pi} \int_{-\pi}^{\pi} d \theta (1 + e^{i \theta})^3 (1 - e^{i \theta}) {\cal G}^{[M/M]}_{{\cal D}}(m_{\tau}^2 e^{i \theta})_{\A{\rm QCD}}
\label{rtaudPA}
\eea
converges to the exact value of $r_{\tau}^{(D=0)}$, Eqs.~(\ref{rtauD0}), although more slowly. This is really the case, as the obtained results presented in Table \ref{tabdPArtau} show.
\begin{table}
  \caption{Numerical results for the $\tau$ decay parameter $r_{\tau}^{(D=0)}$ as calculated with the extended dPA method, Eq.~(\ref{rtaudPA}), in $3\delta$ and $2\delta$ $\A$QCD, for increasing index $M$. Included below is the exact value obtained by the use of the characteristic function, Eqs.~(\ref{rtau5})-(\ref{cFjcFjSL}), for comparison.}
\label{tabdPArtau}
\begin{ruledtabular}
\begin{tabular}{r|r|r}
  $M$ & $r_{\tau}^{(D=0)}(3\delta)$ & $r_{\tau}^{(D=0)}(2\delta)$
\\
\hline
2 & 0.20102 & 0.19014
\\
5 & 0.20460 & 0.19436
 \\
10 & 0.20500 & 0.19770
\\
15 & 0.20511  & 0.19708
\\
20 & 0.20535   & 0.19748
\\
\hline
exact &   0.20559 & 0.19732
\end{tabular}
\end{ruledtabular}
\end{table}

\section{Conclusions}
\label{sec:concl}

In QCD we often face the problem of evaluation of the spacelike renormalization scale and scheme independent quantities ${\cal D}(Q^2)$, such as current correlators, nucleon structure functions and their sum rules, etc. The evaluation of the leading-twist part (which is usually dominant and strongly influences the values of the extracted parameters of the OPE higher-twist terms), has at least three aspects making the evaluation difficult and imprecise: (a) at low values $|Q^2| \lesssim 1 \ {\rm GeV}^2$, the evaluation of the pQCD coupling $a(\kappa Q^2) \equiv \alpha_s(\kappa Q^2)/\pi$ (where $0 < \kappa \sim 1$ is the renormalization scale parameter) cannot be performed reliably because the scale $\mu^2=\kappa Q^2$ is either close to, or in the regime of, the Landau singularitiers of the coupling; (b) the coefficients $d_n(\kappa)$ of the perturbation series of ${\cal D}(Q^2)$ are not known, with the exception of the first few; (c) even if we knew these coefficients, or had a reasonable physically motivated estimate for them, the resulting perturbation series $\sum d_n(\kappa) a(\kappa Q^2)^{n+1}$ would be asymptotically divergent, even in the high-momentum regime $|Q^2| > 1 \ {\rm GeV}^2$, due to the renormalon growth of the coefficients $d_n(\kappa) \sim n!$. This work addresses all three issues, and brings new insights in particular to the aspects (b) and (c).

The solution of the problem (a) has been known for some time, and it consists of two parts: (a1) replacing the pQCD coupling $a(\kappa Q^2)$ by a coupling $\A(\kappa Q^2)$ which has no Landau singularities, i.e., it is holomorphic (analytic) in the complex $Q^2$-plane with the exception of (a part of) the negative semiaxis, thus reflecting qualitatively the correct holomorphic properties of the spacelike observables ${\cal D}(Q^2)$ in the $Q^2$-complex plane; (a2) reorganizing the perturbation series ${\cal D}(Q^2)=\sum d_n(\kappa) a(\kappa Q^2)^{n+1}$ into a series ${\cal D}(Q^2)=\sum \td_n(\kappa) \ta_{n+1}(\kappa Q^2)$ with logarithmic derivatives $\ta_{n+1}$ [Eqs.~(\ref{tan}), (\ref{Dkapptlpt})], and replacing these pQCD logarithmic derivatives by the corresponding logarithmic derivatives $\tA_{n+1}$ of $\A$ [cf.~Eqs.~(\ref{DAkapptlpt}) and (\ref{tAn})]. Such QCD variants can be called (holomorphic) $\A$QCD frameworks.

In this work a solution to the problems (b) and (c) was proposed. In Sec.~\ref{sec:BTs}, for any spacelike observable ${\cal D}(Q^2)$, an auxiliary quantity $\tD(Q^2;\kappa)$ was introduced. Its perturbation (power) series $\tD(Q^2;\kappa)=\sum \td_n(\kappa) a(\kappa Q^2)^{n+1}$, Eq.~(\ref{tDkap}), agrees with the (logarithmic derivatives) perturbation series of the observable ${\cal D}(Q^2)=\sum \td_n(\kappa) \ta_{n+1}(\kappa Q^2)$ only in the one-loop approximation (in which $\ta_{n+1} = a^{n+1}$). The Borel transform ${\rm B}[\tD](u;\kappa)$ of this quantity was shown to have the one-loop-type renormalization scale dependence, and consequently physically motivated specific ans\"atze for  ${\rm B}[\tD](u;\kappa)$ were proposed which capture the known (one-loop-type, or large-$\beta_0$-type) renormalon structure of the observable ${\cal D}(Q^2)$. The parameters in this Borel transform were adjusted in such a way that the first few known coefficients $\td_n$ of the observable ${\cal D}(Q^2)$ were reproduced. Reexpansion of the obtained  ${\rm B}[\tD](u)$ in powers of $u$ then generated a (physically motivated) set of coefficients $\td_n$ at all $n$. Reorganizing backward the obtained logarithmic derivative expansion of ${\cal D}(Q^2) = \sum \td_n(\kappa) \ta_{n+1}(\kappa Q^2)$ into power expansion ${\cal D}(Q^2) = \sum d_n(\kappa) a(\kappa Q^2)^{n+1}$ then generated the perturbation (power) expansion coefficients $d_n$ at all $n$. The generation of $d_n$ coefficients from $\td_n$ coefficients, via the replacement $a^{n+1} \mapsto \ta_{n+1}$ in $\tD$,  can be viewed as a dressing procedure which incorporates an important part of the radiative beyond-one-loop corrections.\footnote{We notice that the $\td_n$ coefficients, although generated by the one-loop-type Borel transform ${\rm B}[\tD](u)$, contain also the radiative correction effects which are in general beyond-one-loop. Only the leading-$\beta_0$ (LB) part $\td_n^{\rm (LB)}$ ($\sim \beta_0^n$) of the coefficients $\td_n$, cf.~Eqs.~(\ref{BtDLB}), represents the contributions of the one-loop-chain Feynman diagrams. Nonetheless, since the Borel transforms ${\rm B}[\tD](u)$, which generate the (full) $\td_n$ coefficients, turn out to have the one-loop-type renormalization scale dependence [cf.~Eq.~(\ref{BtDkap})], the ans\"atze for ${\rm B}[\tD](u)$ were of the one-loop (large-$\beta_0$) type.}
  This was further motivated in Sec.~\ref{subs:fulloneloop} where the effect of this dressing was shown to give the full renormalon structure for the resulting Borel transforms ${\rm B}[{\cal D}](u)$. Specifically for the (massless) Adler function ${\cal D}(Q^2)$, the perturbation expansion coefficients $\td_n$ (and $d_n$) were generated in Sec.~\ref{subs:BtD} in three different renormalization schemes. This addresses the mentioned problem (b).

However, according to the mentioned aspect (c), the resulting perturbation series for ${\cal D}(Q^2)$, Eqs.~(\ref{Dkapptlpt}) in pQCD and Eq.~(\ref{DAkapptlpt}) in holomorphic $\A$QCD frameworks, are asymptotically divergent, primarily because of the fast growth of the coefficients $\td_n, d_n \sim n!$.  Therefore, the Neubert-type of characteristic distribution function $G_D(t)$ of the observable ${\cal D}(Q^2)$ was constructed in Sec.~\ref{subs:ft}, as the inverse Mellin transform of ${\rm B}[\tD](u)$, leading to the evaluation (resummation) of (the leading-twist part of) the observable ${\cal D}(Q^2)$ as an integral over $t$ of the product of $G_D(t)/t$ and the coupling $a(t k' Q^2)$ or $\A(t k' Q^2)$, Eqs.~(\ref{Dres2b}) and (\ref{DAres2b}) [where $k'=\exp(-\tK) > 0$]. In pQCD, due to the Landau singularities of the coupling $a(t k' Q^2)$, this evaluation becomes ambiguous at $Q^2>0$, cf.~Eqs.~(\ref{Dres2pQCD})-(\ref{Dres2pQCDIm}), while in the holomorphic $\A$QCD frameworks there is no ambiguity, cf.~Eq.~(\ref{DAres2b}). In Sec.~\ref{subs:ft} the (massless) Adler function as a function of $Q^2$ was explicitly evaluated in this way, in two $\A$QCD frameworks, namely $3\delta$ \cite{4l3dAQCD} and $2\delta$ $\A$QCD \cite{2dAQCD}, as well as in the corresponding underlying pQCD frameworks and in $\MSbar$ pQCD. This addresses the aforementioned problem (c).

In addition, the described formalism was extended to the evaluation of the timelike observables $R(s)$, when the latter can be expressed as integral transformations of the corresponding spacelike observables ${\cal D}(Q^2)$. Thus, in Sec.~\ref{subs:rtau} the semihadronic $\tau$ lepton decay ratio $r_{\tau}^{(D=0)}$ was evaluated in this way, which is a weighted contour integral of the Adler function ${\cal D}(Q^2)$ along the circle $Q^2=m_{\tau}^2 \exp(i \theta)$ in the $Q^2$-complex plane.

Finally, in Sec.~\ref{sec:dBG} a different evaluation of the (leading-twist, massless) Adler function ${\cal D}(Q^2)$ was applied, namely an earlier developed extended diagonal Pad\'e (dPA) evaluation method, which uses the truncated series $\sum_{n \leq 2 M -1} \td_n(\kappa) \tA_{n+1}(\kappa Q^2)$ of ${\cal D}(Q^2)$ in $\A$QCD to resum it into an extended dPA ${\cal G}^{[M/M]}_{{\cal D}}(Q^2)_{\A{\rm QCD}}$. It turned out that the corresponding sequence $\{ {\cal G}^{[M/M]}_{{\cal D}}(Q^2)_{\A{\rm QCD}}; M=1,2,\ldots \}$ is convergent, for any $Q^2$ outside the negative semiaxis (where we have the cut of the coupling $\A(k' Q^2)$, where $k'>0$); further, the sequence converges to the exact value of the (leading-twist) ${\cal D}(Q^2)$ as obtained by the aforementioned resummation via the characteristic function $G_D(t)$ in Sec.~\ref{subs:ft}. The same procedure was then repeated for the decay ratio $r_{\tau}^{(D=0)}$, where the extended dPA was applied to the Adler function ${\cal D}(m_{\tau}^2 \exp(i \theta))$ on the integration contour, and a sequence $\{ r_{\tau}^{(D=0),[M/M]}, M=1,2,\ldots \}$ was obtained which converged again, and it converged to the exact value $r_{\tau}^{(D=0)}$ obtained by the aforementioned resummation with the characteristic function $G_D(t)$ in Sec.~\ref{subs:rtau}.  

It should be pointed out that evaluations with the described methods can be applied also to other spacelike and timelike QCD observables, in the QCD variants with IR-safe (and holomorphic) coupling $\A(Q^2)$, and will presumably lead to comparably favorable results as they do for the Adler function and $\tau$ decay ratio.

\begin{acknowledgments}
The author acknowledges the support by FONDECYT (Chile) Grant No.~1180344.
\end{acknowledgments}

\appendix

\section{Recursion relations for $k_m(n)$ and $\tk_m(n)$}
\label{app:ktkrr}

Here, the recursion relations are presented for the coefficients $k_m(n)$ and $\tk_m(n)$, introduced in Eqs.~(\ref{tanan1}) and (\ref{antan1}) in Sec.~\ref{subs:lpt}. The procedure for obtaining these recursion relations is based on the reasoning explained in Sec.~\ref{subs:lpt}, where some explicit expressions for these coefficients are also given for low $m$ and $n$, cf.~Eqs.~(\ref{tanan2}) and (\ref{antan}) [cf.~also Eqs.~(\ref{tdns})].

The convention $c_0 =1$ is taken, and $c_j = \beta_j/\beta_0$ ($j \geq 1$), cf.~Eqs.~(\ref{RGE}). We recall that the coefficients $c_j$ ($j \geq 2$) characterize here the renormalization scheme (the momentum scaling convention is fixed throughout this work by using the usual $\MSbar$ scaling, i.e., ${\overline \Lambda}^2$).  The recursion relations for $k_m(n+1)$ in terms of $k_s(n)$ are
\bes
\label{krr}
\bea
k_m(2) & = & c_m \qquad (m=0,\ldots,N-2),
\label{km2}
\\
k_m(n+1) & = & \frac{1}{n} \sum_{s=0}^m (n+s) \; c_{m-s} \; k_s(n) \qquad
(m=0,\ldots,N-n-1; \; n=2,\ldots,N-1),
\label{kmn1}
\\
\eea
\ees
where the index $N$ means that the recursion relations (\ref{tanan1}) and (\ref{antan1}) are considered truncated at $\sim a^N$ and at $\sim \ta_N$, respectively. In practice, in this work $N$ acquired values up to $N=71$. This means that the simple one-loop-type generated coefficients $\td_n$ [Eqs.~(\ref{BtDtdns}) and (\ref{tdnUVp1})] were used to generate the corresponding coefficients $d_n$ via the relations (\ref{dntdn}) up to $n=N-1$ (i.e., up to $n=70$).

Once knowing the coefficients $k_i(j)$, the recursion relations for $\tk_{m+1}(n)$ in terms of $\tk_p(r)$ ($p \leq m$) are
\bes
\label{tkrr}
\bea
\tk_0(n) & = & 1 \qquad (n=2,\ldots,N),
\label{tkm0}
\\
\tk_{m+1}(n) & = &  -\sum_{s=1}^{m+1} k_s(n) \; \tk_{m+1-s}(n+s) \qquad
(m=0,\ldots,N-n-1; \; n=2,\ldots,N-1).
\label{tkm1n}
\\
\eea
\ees

\section{IR-safe and holomorphic $\A$QCD variants}
\label{app:ndAQCD}  

The pQCD running coupling $a(Q^2) \equiv \alpha_s(Q^2)/\pi$ is a solution of the perturbative RGE (pRGE) Eq.~(\ref{RGE}), where the first two $\beta$-coefficients,  $\beta_0 = (1/4)(11- 2 N_f/3)$ and $\beta_1=(1/16)(102 - 38 N_f/3)$, are universal, i.e., scheme independent, in mass independent schemes. The other coefficients $c_j =\beta_j/\beta_0$ ($j \geq 2$) characterize in pQCD the renormalization scheme \cite{Stevenson}. Stated differently, the form of the function $\beta(a; c_2, c_3,\ldots)$ can be regarded as the definition of the renormalization scheme. The momentum scale parameter $\Lambda_{\rm QCD}$ is not considered here as a scheme parameter, but rather as the definition of the momentum (re)scaling, and a scaling change can be equivalently described as a change of the renormalization scale. In this work, the $\MSbar$ scaling definition ($\Lambda^2_{\rm QCD}={\overline \Lambda}^2$) is used throughout.

When integrating the pRGE in a given or chosen renormalization scheme, the resulting pQCD running coupling $a(Q^2)$ usually acquires singularities on the positive axis in the $Q^2$-complex plane, $0 \leq Q^2 \lesssim \Lambda_{\rm QCD}^2$ ($\sim 0.01$-$1 \ {\rm GeV}^2$), in addition to the otherwise expected singularities on the negative axis. On the other hand, the general principles of Quantum Field Theories imply that the spacelike QCD observables ${\cal D}(Q^2)$, such as current correlators and nucleon structure functions and their sum rules, are holomorphic (analytic) functions in the $Q^2$-complex plane with the exception of a part of the negative semiaxis, $Q^2 \in \mathbb{C} \backslash (-\infty, -M_{\rm thr}^2]$, where $M_{\rm thr} \sim 0.1$ GeV \cite{BS,Oehme}. The pQCD running coupling $a(Q^2)$ therefore usually does not reflect qualitatively these properties, because of the mentioned singularities (cut and branching points) on the positive axis, $0 \leq Q^2 \leq \Lambda_{\rm Lan.}^2$. This aspect of $a(Q^2)$  is considered unfortunate, especially if the coupling $a(Q^2)$ [or $a(\mu^2)$ with $\mu^2=\kappa Q^2 \sim Q^2$] is to be used to evaluate ${\cal D}(Q^2)$ at low values $|Q^2| \lesssim 1 \  {\rm GeV}^2$. These singularities are called Landau singularities or Landau ghosts, and the point $Q^2 =\Lambda_{\rm Lan.}^2$ is usually called the Landau branching point. The application of the Cauchy theory to the integrand $a(Q^{'2})/(Q^{'2}-Q^2)$ in the $Q^{'2}$-complex plane leads then to the following dispersion integral representation of the pQCD coupling $a(Q^2)$:
\be
a(Q^2) = \frac{1}{\pi} \int_{-\Lambda_{\rm Lan.}^2- \eta}^{+\infty} d \sigma \frac{\rho_a(\sigma)}{(\sigma + Q^2)}, \qquad (\eta \to +0),
\label{dispa}
\ee
where $\rho_a(\sigma) = {\rm Im} a(Q^{' 2} = -\sigma - i \epsilon)$ is called the discontinuity or spectral function of $a$ along its cut. The simplest elimination of the disturbing Landau singularities is to eliminate in the above dispersion integral the integration part along the positive $\sigma$ (negative $Q^{' 2}$), this leading to the minimal analytic (or Analytic Perturbation Theory - APT) coupling \cite{ShS,MS,Sh1Sh2}
\be
\A^{\rm (APT)}(Q^2) = \frac{1}{\pi} \int_{0}^{+\infty} d \sigma \frac{\rho_a(\sigma)}{(\sigma + Q^2)}.
\label{dispAAPT}
\ee
This coupling has the cut threshold $\sigma_{\rm min} (\equiv M_{\rm thr}^2)=0$, and the difference between this coupling and the underlying pQCD coupling $a(Q^2)$ (i.e., the pQCD coupling in the same renormalization scheme) at large $|Q^2| > 1 \ {\rm GeV}^2$ is appreciable, namely $\A^{\rm (APT)}(Q^2) - a(Q^2) \sim (\Lambda^2_{\rm QCD}/Q^2)^N$ with the index $N=1$.

We do know that at low squared momenta ($|Q^2|, \sigma \lesssim 1 \ {\rm GeV}^2$) the pQCD coupling $a(Q^2)$ and its spectral function $\rho_a(\sigma)$ do not describe correctly the physics. Therefore, it appears natural to replace in the dispersion integral (\ref{dispAAPT}) the pQCD spectral function $\rho_a(\sigma)$ at $\sigma \lesssim 1 \ {\rm GeV}^2$ by another, unknown spectral function $\rho_{\A}(\sigma) \not= \rho_a(\sigma)$. An efficient way of parametrizing this spectral function $\rho_{\A}(\sigma)$ in the low-momentum regime $\sigma \lesssim 1 \ {\rm GeV}^2$ is to represent it as a linear combination of Dirac delta functions. This suggests then the following form for the spectral function of the coupling $\A(Q^2)$:
\be
\rho_{\A}^{(n \delta)}(\sigma) =  \pi \sum_{j=1}^{n} {\cal R}_j \; \delta(\sigma - M_j^2)  + \Theta(\sigma - M_0^2) \rho_a(\sigma) \ ,
\label{rhoAnd}
\ee
where we expect $0 < M_1^2 < \ldots < M_n^2 < M_0^2$, and $M_0^2 \sim 1 {\rm GeV}^2$ can be called the pQCD-onset scale. The corresponding coupling is now
\bea
\A^{(n \delta)}(Q^2) \left( \equiv \frac{1}{\pi} \int_0^{\infty} d \sigma \frac{\rho_{\A}(\sigma)}{(\sigma + Q^2)} \right) & = &  \sum_{j=1}^n \frac{{\cal R}_j}{(Q^2 + M_j^2)} + \frac{1}{\pi} \int_{M_0^2}^{\infty} d \sigma \frac{ \rho_a(\sigma) }{(Q^2 + \sigma)} \ .
\label{AQ2}
\eea
The $n$ Dirac delta functions in the spectral function thus give $\Delta \A_{\rm IR}(Q^2)$ which is a linear combination of $n$ simple fractions $\sim 1/(Q^2+M_j^2)$, and this can be represented as a near diagonal Pad\'e approximant $\Delta \A_{\rm IR}(Q^2) = [n/n-1](Q^2)$. Such Pad\'e approximants are known to approximate usually the holomorphic functions in the $Q^2$-complex plane increasingly well when the index $n$ increases.

In Refs.~\cite{2dAQCD,2dCPC} and \cite{3l3dAQCD,4l3dAQCD}, such couplings were constructed, with two ($n=2$) and three ($n=3$) Dirac delta functions, respectively, in specific renormalization schemes of the underlying pQCD coupling $a$. The $(2 n +1)$ parameters ($M^2_j$, ${\cal R}_j$, $j=1,\ldots,n$; and $M_0^2$) were then fixed by several physically motivated conditions. Four of these conditions were obtained by requiring that the $\A(Q^2)$ coupling at high $|Q^2| > 1 \ {\rm GeV}^2$ practically coincides with the underlying pQCD
\be
\A^{(n \delta)}(Q^2) - a(Q^2) \sim \left( \frac{\Lambda^2}{Q^2} \right)^N \quad
{\rm where:} \; N=5.
\label{diffAaN5}
\ee
In addition, at moderate momenta $|Q^2| \sim m_{\tau}^2$ ($\sim 1 \ {\rm GeV}^2$) the requirement was imposed that the well measured physics of the semihadronic $\tau$ lepton decays be reproduced correctly, i.e., that the (massless and strangeless) $\tau$ decay ratio $r_{\tau}^{(D=0)}$, defined theoretically in Eq.~(\ref{rtau1}), gives the values\footnote{It is known that the higher dimensional (higher-twist) terms $D>0$ contribute only little to $r_{\tau}$, cf.~\cite{4l3dAQCD} and references therein.}
\be
r_{\tau}^{(D=0)} = 0.201 \pm 0.002.
\label{rtauD0exp}
\ee
These values were imposed in a specific chosen evaluation procedure that was considered relatively reliable.\footnote{\label{footrtau} In $3 \delta$ $\A$QCD, which was constructed in the LMM renormalization scheme, $r_{\tau}^{(D=0)}$ was evaluated by the contour integration (\ref{rtau1}) with the Adler function evaluated with the truncated series in logarithmic derivatives, Eq.~(\ref{DAkaplpt}), truncated at the last fully known coefficient [$\td_3 \tA_4(Q^2)$]; this was considered reasonably reliable, because the application of the corresponding extended dPA (\ref{GMMAQCD}) with $M=2$ to the Adler function gave practically the same result for $r_{\tau}^{(D=0)}$ \cite{4l3dAQCD}. On the other hand, in $2 \delta$ $\A$QCD, which was constructed in the ($c_2=-4.9$) Lambert renormalization scheme, the two mentioned methods give substantially different results (namely, $0.177$ and $0.190$, respectively, for the $2 \delta$ parameter values given  in Table \ref{tab2d3lres}. Since in this scheme the leading-$\beta_0$ (LB) parts $\td_j^{\rm (LB)}$ are relatively dominant in the known coefficients $\td_j$ ($j=1,2,3$) of the Adler function, and all $\td_n^{\rm (LB)}$ coefficients of the Adler function are known, the corresponding LB parts of the Adler function were resummed, in the procedure analogous to Eq.~(\ref{DAres2b}), and the remaining terms [beyond-LB (bLB): $\sum_{j=1,2,3} (\td_j - \td_j^{\rm (LB)})  \tA_{j+1}(Q^2)$] were added; the use of such (LB+bLB) Adler function then led \cite{2dAQCD,2dCPC}, via the contour integration (\ref{rtau1}), to the result $r_{\tau}^{(D=0)}( {\rm LB+bLB})=0.203-0.002=0.201$ where $0.203$ and $-0.002$ are the LB and bLB contributions, respectively.}
Finally, the two additional conditions needed in $3 \delta$ $\A$QCD were at very low momenta, namely that $\A^{(3 \delta)}(Q^2) \sim Q^2$ when $Q^2 \to 0$, and that $A^{(3 \delta)}(Q^2)$ acquires at positive $Q^2$ a local maximum at about $Q^2 \approx 0.135 \ {\rm GeV}^2$, in the Lambert MiniMOM (LMM) renormalization scheme. These two conditions are suggested by the large volume lattice calculations  \cite{LattcouplNf0} (cf.~also \cite{LattcouplNf0b,LattcouplNf24,LattcouplDiscr}) of the dressing functions $Z_{\rm gl}(Q^2)$ and $Z_{\rm gh}(Q^2)$ of the Landau gauge gluon and ghost propagators in the MiniMOM (MM) renormalization scheme \cite{MM1,MM2}, where the lattice coupling was defined naturally as the product of these dressing functions: $\A_{\rm latt.} \propto Z_{\rm gl}(Q^2) Z_{\rm gh}(Q^2)^2$.

The Lambert MM (LMM) scheme, in which the $3\delta$ $\A$QCD coupling was constructed, is just the MM scheme of the lattice calculations [with the  corresponding  RGE scheme coefficients $c_j({\rm MM})$, $j \geq 2$] but with the usual ($\MSbar$-type) scaling (i.e., $\Lambda_{\rm QCD} = {\overline \Lambda}$). In Ref.~\cite{3l3dAQCD} the $\A^{(3 \delta)}$ coupling was constructed in a scheme which agrees with the perturbative LMM scheme up to three-loop level (i.e., up to $c_2$); in Ref.~\cite{4l3dAQCD} the construction was refined to agree with the known LMM scheme up to the four-loop level (i.e., up to $c_3$), i.e., the level known at present \cite{MM1} for the MM scheme. On the other hand, the $2\delta$ $\A$QCD coupling $\A^{(2 \delta)}(Q^2)$ was constructed in the underlying Lambert scheme with $c_2=-4.9$ \cite{2dCPC}.

The $\beta$-functions $\beta(a)$ defining the two schemes (Lambert and four-loop LMM), and thus the underlying pQCD running coupling $a(Q^2)=\alpha_s(Q^2)/\pi$, have the Pad\'e form
\bes
\label{betas}
\bea
\beta(a) & \equiv & \frac{d}{d \ln Q^2} a(Q^2),
\label{defbeta}
\\
\beta_{\rm Lamb.}(a; c_2) & = & - \beta_0 a^2 \frac{[1 + (c_1 - c_2/c_1) a]}{[1 - (c_2/c_1)a]},
\label{betaLamb}
\\
\beta_{\rm LMM}(a; c_2, c_3) & = & - \beta_0 a^2 \frac{ \left[ 1 + a_0 c_1 a + a_1 c_1^2 a^2 \right]}{\left[ 1 - a_1 c_1^2 a^2 \right] \left[ 1 + (a_0-1) c_1 a + a_1 c_1^2 a^2 \right]} \ ,
\label{betaLMM}
\eea
\ees
where in Eq.~(\ref{betaLMM}) we denoted
\be
a_0  =  1 + \sqrt{c_3/c_1^3}, \quad 
a_1 = c_2/c_1^2 +   \sqrt{c_3/c_1^3} .
\label{a0a1}
\ee
In $2\delta$ $\A$QCD, the value of the scheme parameter $c_2 = \beta_2/\beta_0$ in the Lambert scheme $\beta$-function (\ref{betaLamb}) can be varied in a restricted interval, $-5.9 < c_2 < -2.0$ if we require that the pQCD onset scale $M_0$ is $M_0 \lesssim 1$ GeV and $\A_{\rm max}$ ($= \A(0)$) $\lesssim 1$ \cite{2dCPC}; the value $c_2=-4.9$ is chosen. In $3\delta$ $\A$QCD, the values of the scheme parameters $c_2$ and $c_3$ in the LMM $\beta$-function (\ref{betaLMM}) are adjusted to coincide with the known values of MM scheme \cite{MM1,MM2}, namely $c_2^{\rm (MM)}=9.2970$ and $c_3^{\rm (MM)} = 71.4538$ (at $N_f=3$ which is used throughout). RGEs with the $\beta$-functions (\ref{betas}) can be solved explicitly, giving the underlying pQCD running coupling $a(Q^2)$ in terms of the Lambert functions $W_{\pm 1}(z)$
\bes
\label{as}
\bea
a_{\rm Lamb.}(Q^2) & = & - \frac{1}{c_1} \frac{1}{[1 - c_2/c_1^2 + W_{\mp}(z)]},
\label{aLamb}
\\
a_{\rm LMM}(Q^2) & = & \frac{2}{c_1} \left[ - \sqrt{\omega_2} - 1 - W_{\mp 1}(z) + 
\sqrt{(\sqrt{\omega_2} + 1 + W_{\mp 1}(z))^2 - 4(\omega_1 + \sqrt{\omega_2})} \right]^{-1}.
\label{aLMM}
\eea
\ees
In Eq.~(\ref{aLMM}) the notations $\omega_1= c_2/c_1^2$, $\omega_2=c_3/c_1^3$ were used. The squared momentum $Q^2 \equiv -q^2$ is in general complex, $Q^2 = |Q^2| \exp(i \phi)$. The Lambert function $W_{-1}$ is used when $0 \leq \phi < \pi$, and $W_{+1}$ when $-\pi \leq \phi < 0$. The argument $z = z(Q^2)$ in the Lambert functions $W_{\pm 1}(z)$ is
\be
z \equiv z(Q^2) = -\frac{1}{c_1 e} \left( \frac{\Lambda_{\rm L}^2}{Q^2} \right)^{\beta_0/c_1} \ ,
\label{zexpr}
\ee
where $\Lambda_{\rm L}$ is called the Lambert scale ($\Lambda_{\rm L} \lesssim 1 \ {\rm GeV}^2$).\footnote{In Ref.~\cite{GCIK}, the expression (\ref{zexpr}) without the factor $1/( c_1 e)$ was used for $z$, which just redefines the Lambert scale $\Lambda_{\rm L}$.} The solution (\ref{aLamb}) was obtained first in \cite{GGK}, and the solution (\ref{aLMM}) in \cite{GCIK}. The solution (\ref{aLamb}) has one adjustable scheme parameter ($c_2$), and the solution (\ref{aLMM}) has two ($c_2, c_3$). These solutions, and their spectral functions $\rho_a(\sigma) = {\rm Im} a(Q^2 = -\sigma - i \epsilon)$, can be evaluated very efficiently with {\it Mathematica}\footnote{In {\it Mathematica}, $W_{\mp}(z)$ is called by the command ${\rm ProductLog}[\mp 1, z]$.}, even at high values of $|Q^2|$ and $\sigma$, allowing thus for an efficient evaluation of the dispersion integrals in Eq.~(\ref{AQ2}). In this work, as the reference point for the underlying coupling constant the value $\alpha_s(M_Z^2;\MSbar)=0.1185$ was taken, which then determines the value of the Lambert scale $\Lambda_{\rm L}$ in the $N_f=3$ regime of low $|Q^2| \lesssim 1 \ {\rm GeV}^2$.\footnote{\label{ftrefNf3} The underlying running pQCD coupling $a(Q^2)$ in the $N_f=3$ regime is obtained in the following way: in the four-loop $\MSbar$ scheme the coupling $a(M_Z^2;\MSbar,N_f=5))=0.1185/\pi$ is RGE-evolved downwards in $Q^2$, using the three-loop quark threshold relations \cite{CKS} at $Q^2 = (2 {\overline m}_q)^2$ ($q=b, c$; ${\overline m}_b=4.20$ GeV and ${\overline m}_c=1.27$ GeV), giving ${\bar a}_0 \equiv a(Q_0^2; \MSbar)_{Nf=3}= 0.0846346$ at $Q_0^2=(2 {\overline m}_c)^2$. Then at this scale, the change of the renormalization scheme is made [e.g., cf.~Eq.~(13) of Ref.~\cite{4l3dAQCD}], giving  $a(Q_0^2)= 0.0792708$ in the $c_2=-4.9$ Lambert scheme, and $a(Q_0^2)=0.0897919$ in the LMM scheme. The five-loop effects in the mentioned RGE evolution are small; namely, in order to reproduce the aforementioned value  ${\bar a}_0=0.0846346$ when using the polynomial five-loop $\MSbar$ $\beta$-function \cite{5lMSbarbeta} and the corresponding four-loop quark mass thresholds  \cite{4lquarkthresh}, the starting value of $\alpha_s(M_Z^2;\MSbar)=0.118577$ is needed which is close to $0.1185$.}

The obtained values of the parameters of the discussed $2\delta$ and $3\delta$ $\A$QCD are given in Table \ref{tab2d3lres}.\footnote{Note that the values of the $2\delta$ $\A$QCD parameters differ somewhat from those of Table 2 of Ref.~\cite{2dCPC} (the case $c_2=-4.9$ there), because there the high-momentum reference point was taken to be $a(M_Z^2;\MSbar,N_f=5)=0.1184/\pi$, while here it is  $a(M_Z^2;\MSbar,N_f=5)=0.1185/\pi$.}
\begin{table}
\caption{Values of the parameters of 2$\delta$ and 3$\delta$ $\A$QCD coupling used here, for $N_f=3$: the dimensionless parameters $s_j \equiv M_j^2/\Lambda_{{\rm L}}^2$ and $r_j \equiv {\cal R}_j/\Lambda_{{\rm L}}^2$; the Lambert scale $\Lambda_{{\rm L}}$ (in GeV); and the maximum value of the coupling for positive $Q^2$. The ``input'' parameter choice is $\alpha_s(M_Z^2;\MSbar)=0.1185$ and $r_{\tau}^{(D=0)}=0.201$ (see the text for details).}
\label{tab2d3lres}
\begin{ruledtabular}
\centering
\begin{tabular}{rrr|lllllllll}
 $\A$QCD & ${\overline \alpha}_s(M_Z^2)$ & $r_{\tau}^{(D=0)}$ & $s_1$ & $s_2$ & $s_3$ & $f_1$ & $f_2$ & $f_3$ & $s_0$ & $\Lambda_{{\rm L}}$ [GeV] & $\pi \A_{\rm max}$ 
\\
\hline
2$\delta$ & 0.1185 & 0.201 & 18.727 & 1.0351 & - & 0.2928 & 0.5747 & - & 25.600 & 0.2564 & 2.0727 \\
\hline
3$\delta$ & 0.1185 & 0.201 & 3.970 & 18.495 & 474.20 & -2.8603 & 11.801 & 5.2543 & 652. & 0.11564 & 0.9156 \\
\end{tabular}
\end{ruledtabular}
\end{table}

The scheme parameters $c_j$ ($j \geq 2$) of the LMM and Lambert schemes, as well as the (five-loop) $\MSbar$ scheme, all at $N_f=3$, are given in Table \ref{tabcj}. 
\begin{table}
\caption{The values of scheme parameters $c_j = \beta_j/\beta_0$ in the LMM and in the $c_2=-4.9$ Lambert (Lamb.) scheme (note that in the latter $c_j=c_2^{j-1}/c_1^{j-2}$); Nf=3, and $c_1=16/9$. For comparison, the known $\MSbar$ coefficients (with $N_f=3$) are given as well.}
\label{tabcj}
\begin{ruledtabular}
\centering
\begin{tabular}{r|rrr}
  $j$ & $c_j({\rm LMM})$ & $c_j({\rm Lamb.})$ & $c_j(\MSbar)$
 \\
\hline
2 & 9.297 & -4.9 & 4.47106 \\
3 & 71.4538 &  13.5056 & 20.9902 \\
4 & 201.843 & 37.2249 & 56.5876 \\
\hline
5 & 684.698 &  102.601 & -\\ 
10 & $1.57996 \times 10^{6}$ &  -16320.9 & -\\
15 & $4.53556 \times 10^{7}$ & $2.5962 \times 10^{6}$ & - \\
20 & $7.22843 \times 10^{12}$ & $-4.12982 \times 10^{8}$ & - \\
25 & $1.2884 \times 10^{16}$ & $6.56937 \times 10^{10}$ & - \\
30 & $5.3876 \times 10^{19}$ &   $-1.045 \times 10^{13}$ & - \\
35 & $7.02458 \times 10^{22}$ & $1.6623 \times 10^{15}$ & - \\
\hline
65 & $3.08253 \times 10^{42}$ &  $2.6932 \times 10^{28}$ & - \\
70 & $1.02063 \times 10^{46}$ & $-4.28412 \times 10^{30}$ & - 
\end{tabular}
\end{ruledtabular}
\end{table}

Once a specific coupling $\A(Q^2)$ is obtained ($a \mapsto \A$), the analogs $\A_n(Q^2)$ of the powers $a(Q^2)^n$ of the underlying pQCD coupling, in general holomorphic $\A$QCD, can be obtained by the construction presented in Ref.~\cite{CV12} for integer $n$, and in Ref.~\cite{GCAK} for general (noninteger) $n$. The construction of $\A_n(Q^2)$ from $\A(Q^2)$ for integer $n$ goes via the logarithmic derivatives of $\A(Q^2)$, in close analogy with the pQCD definitions and relations in Sec.~\ref{subs:lpt}.  

Here, the construction given in Ref.~\cite{CV12} for integer $n$ will be summarized. Since the coupling $\A(Q^2)$ is the holomorphic analog of the corresponding underlying pQCD coupling $a(Q^2)$ (in the same renormalization scheme), the linearity of the ``analytization'' $a(Q^2) \mapsto \A(Q^2)$ implies that the logarithmic derivatives $\ta_{n+1}(Q^2)$ of $a(Q^2)$, Eq.~(\ref{tan}), get replaced (i.e., ``analytized'') in $\A$QCD by the completely analogous logarithmic derivatives $\tA_{n+1}(Q^2)$ of $\A(Q^2)$
\be
\tA_{n+1}(Q^2) \equiv \frac{(-1)^n}{\beta_0^n n!} \left( \frac{d}{d \ln Q^2} \right)^n \A(Q^2) \qquad (n=0,1,2,\ldots).
\label{tAn}
\ee
This construction is already enough to evaluate the (truncated) $\A$QCD series Eq.~(\ref{DAkaplpt})
\bea
{\cal D}^{[N]}_{\A{\rm QCD}}(Q^2; \kappa) &=& \td_0 \A(\kappa Q^2) + \td_1(\kappa)  \tA_2(\kappa Q^2) + \ldots + \td_{N-1}(\kappa) \tA_{N}(\kappa Q^2),
\label{DNAkaplpt}
\eea
where a weak renormalization scale dependence ($\kappa$-dependence) now appears due to the truncation effect. The pQCD analog of this expression is the truncated version of the series Eq.~(\ref{Dkaplpt}), truncated at $\td_{N-1}(\kappa) \ta_{N}(\kappa Q^2)$. Formally, the truncated series (\ref{DNAkaplpt}) differs from the full result ${\cal D}(Q^2)$ by a term $\sim \tA_{N+1}$ ($ \sim \ta_{N+1} \sim a^{N+1}$). The above truncated series can be rewritten in terms of the coefficients $d_n(\kappa)$ of the original perturbation (power) series (\ref{Dkappt})
\bea
{\cal D}^{[N]}_{\A{\rm QCD}}(Q^2; \kappa) &=& d_0 \A(\kappa Q^2) + d_1(\kappa)  \A_2(\kappa Q^2) + \ldots + d_{N-1}(\kappa) \A_{N}(\kappa Q^2),
\label{DNAkappt}
\eea
where the power analogs $\A_n$ are linear combinations of logarithmic derivatives $\tA_{n+m}$ in complete analogy with the pQCD relations (\ref{antan1})
\be
\A_n  =  \tA_n + \sum_{m=1}^{N-n} \tk_n(m) \tA_{n+m} \qquad (n=2,\ldots,N),
\label{AntAn}
\ee
where the truncation is performed consistently at $\tA_N$; note that $\A_N = \tA_N$ in this truncation. We recall that the truncated series (\ref{DNAkappt}) has its pQCD analog in the original perturbation (power) series (\ref{Dkappt}) truncated at  $d_{N-1}(\kappa) a(\kappa Q^2)^N$. We point out that, as long as $\A(Q^2)$ has some nonperturbative contributions in comparison to its underlying pQCD coupling [such as terms $\sim 1/(Q^2+M^2)^k$], we have $\A_n(Q^2) \not= \A(Q^2)^n$ ($n \geq 2$). In such cases, even if the truncation index $N$ in the relations (\ref{AntAn}) is very high, we do not have $\A_n(Q^2) \approx \A(Q^2)^n$ at low values $|Q^2| \lesssim 1 \ {\rm GeV}^2$.\footnote{
At high $|Q^2| > 1 \ {\rm GeV}^2$ we have in general $\A_n(Q^2) \approx \tA_n(Q^2) \approx \A(Q^2)^n \approx a(Q^2)^n $, due to the relation (\ref{diffAaN5}), i.e., ($2\delta$ and $3\delta$) $\A$QCD in the high-momentum regime is practically indistinguishable from the underlying pQCD.} 
  In \cite{Techn} it was argued that if in  Eq.~(\ref{DNAkappt}) the naive powers $\A(Q^2)^n$ were used instead of $\A_n(Q^2)$, this would bring into the series spurious uncontrollable nonperturbative contributions at $|Q^2| \lesssim 1 \ {\rm GeV}^2$. It is therefore important to use the series in logarithmic derivatives instead, i.e., Eq.~(\ref{DNAkaplpt}) [$=$Eq.~(\ref{DNAkappt})], and the related extended dPA expressions as explained here in Sec.~\ref{sec:dBG}.

\begin{figure}[htb] 
\begin{minipage}[b]{.49\linewidth}
  \centering\includegraphics[width=85mm]{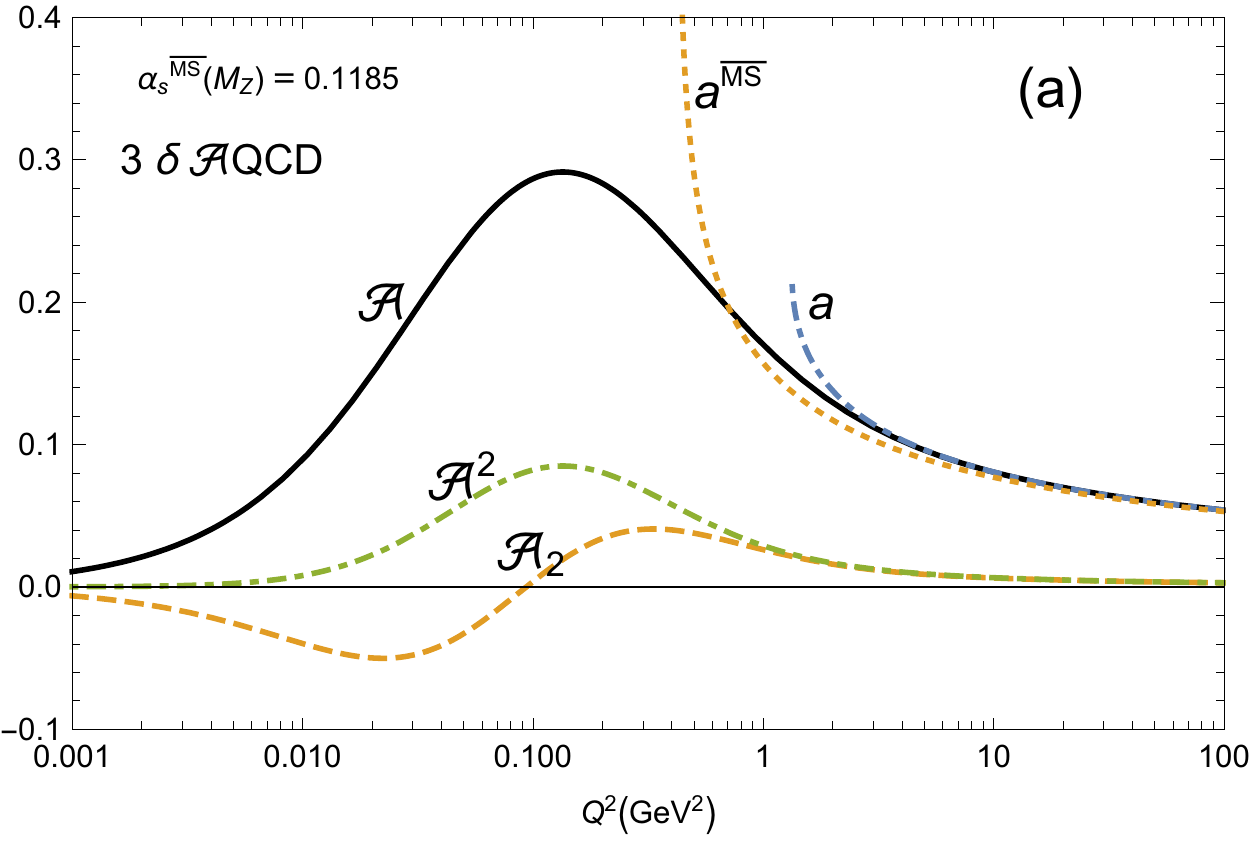}
  \end{minipage}
\begin{minipage}[b]{.49\linewidth}
  \centering\includegraphics[width=85mm]{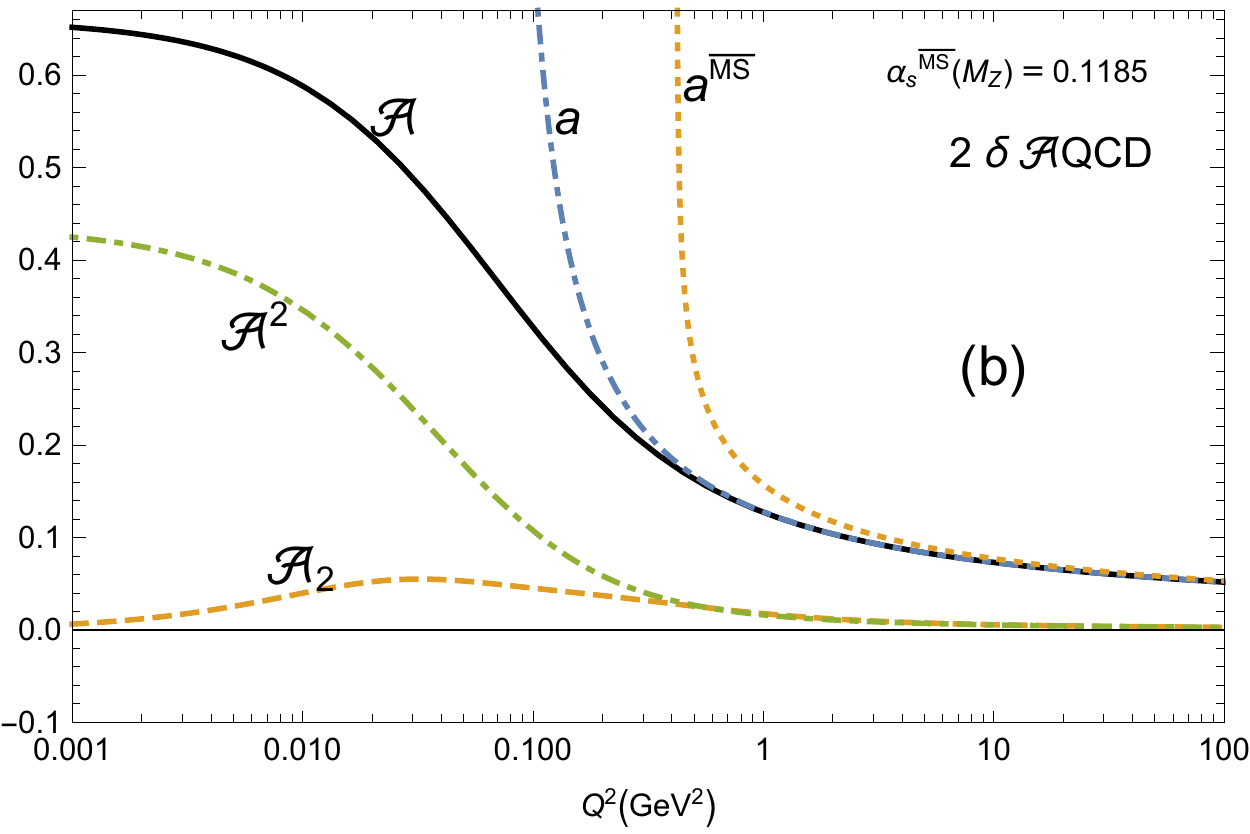}
\end{minipage}
\caption{\footnotesize  The running couplings $\A(Q^2)$, $\A_2(Q^2)$, and the underlying pQCD coupling $a(Q^2)$, as a function of positive $Q^2$: (a) in the considered $3 \delta$ $\A$QCD case (in the LMM scheme); (b) in the considered  $2 \delta$ $\A$QCD case (in the $c_2=-4.9$ Lambert scheme). For comparison, the naive power $\A(Q^2)^2$ [$\not= \A_2(Q^2)$] is included, as well as the five-loop $\MSbar$ coupling $a^{\MSbar}(Q^2)$ [normalized to $a^{\MSbar}(Q_0^2) \approx 0.08463$, cf.~footnote \ref{ftrefNf3}]. In all cases $N_f=3$. }
\label{FigAtA2}
\end{figure}

In Figs.~\ref{FigAtA2}(a),(b) the couplings $\A(Q^2)$, $\A_2(Q^2)$ are presented, as a function of $Q^2>0$, for the considered $2\delta$ and $3\delta$ $\A$QCD, respectively. The corresponding underlying pQCD coupling $a(Q^2)$ and the $\MSbar$ coupling ${\bar a}(Q^2)$ are included for comparison (all are for $N_f=3$).
The coupling $\A_2(Q^2)$ is obtained by Eq.~(\ref{AntAn}) with the truncation index $N=4$ (i.e., $n=2, N=4$). In these Figures the naive power $\A(Q^2)^2$ is included, and we can see clearly that $\A_2(Q^2) \not\approx \A(Q^2)^2$ at low $Q^2$. Further, it can be noted that pQCD coupling $a(Q^2)$ in the LMM scheme has the branching point at a rather large value $Q_{\rm br}^2= 1.33 \ {\rm GeV}^2$, and it is not a pole. In the Lambert scheme, $a(Q^2)$ has $Q^2_{\rm br} = 0.068 \ {\rm GeV}^2$ (it is a pole), and in the $\MSbar$ scheme $Q^2_{\rm br}=0.42 \ {\rm GeV}^2$ (it is a pole). These curves can be obtained by using the programs \cite{MathPrgs} written in {\it Mathematica}.

\section{Evaluation of ${\cal C}_{i,j}^{(D)}$ coefficients}
\label{app:calC}

Here it will be outlined how the values (and their ranges) of the coefficients ${\cal C}_{j,k}^{(2 p)}$, appearing in the expressions (\ref{BDdns}) and (\ref{BDdnsUV}) for the coefficients $d_n$, were estimated. In each of those considered cases, the coefficient $d_n$ has a specific $n$-dependence
\bea
d_n & = & S_{\gamma} \Gamma(\gamma+1+n) \left( \frac{\beta_0}{p} \right)^n \left[ 1 + \frac{a_1}{(\gamma+n)} + \frac{a_2}{(\gamma+n)(\gamma-1+n)} + {\cal O}\left( \frac{1}{n^3} \right) \right].
\label{dnform}
\eea
Here, $p=2, 3$ in the cases of IR $p=2,3$; and the UV $p=1$ case has formally $p=-1$ in the above expression.\footnote{The $N_f$-dependence of such expressions, for large $N_f$, is not well understood, as pointed out in Ref.~\cite{KatMol}, because the factor $S_{\gamma}$ has, in general, some $N_f$-dependence.}
The ratios can then be formed where the leading order (LO) $n$-dependence is factored out; and then the logarithm thereof is taken
\bes
\label{Sn}
\bea
{\cal S}(n) &\equiv& \ln \left( \frac{d_n}{  \Gamma(\gamma+1+n) \left( \frac{\beta_0}{p} \right)^n } \right)
\label{Sndef}
\\
& = & \ln S_{\gamma} + \frac{a_1}{n} + \left(a_2 - \frac{1}{2} a_1^2 - \gamma a_1 \right) \frac{1}{n^2} + {\cal O} \left( \frac{1}{n^3} \right).
\label{Snexp}
\eea
\ees
It is then straightforward to check that the coefficients $a_1$ and $a_2$ can be extracted in the following approximate forms:
\bes
\label{a1a2}
\bea
a_1 & = & {\cal H}_1(n) - \frac{1}{n} {\cal H}_2(n) + {\cal O}\left( \frac{1}{n^2} \right),
\label{a1est}
\\
a_2 & = & \frac{1}{2}  {\cal H}_2(n) + \frac{1}{2} \left( {\cal H}_1(n) \right)^2 + \gamma {\cal H}_1(n) + {\cal O}\left( \frac{1}{n} \right),
\label{a2est}
\eea
\ees
where the expressions ${\cal H}_j(n)$ are specific derivatives of ${\cal S}(n)$
\bes
\label{H1H2}
\bea
{\cal H}_1(n) & = & - n^2 \frac{d}{dn} {\cal S}(n),
\label{H1def}
\\
{\cal H}_2(n) & = & \left(n^2 \frac{d}{dn}\right)^2 {\cal S}(n).
\label{H2def}
\eea
\ees
These derivatives must be evaluated in discrete form, because the ratios ${\cal S}(n)$ are known only at discrete (positive) integer values of $n$
\bes
\label{H1H2app}
\bea
{\cal H}_1(n) & = & - \frac{1}{2} n^2 \left( {\cal S}(n+1) -  {\cal S}(n-1) \right),
\label{H1app}
\\
  {\cal H}_2(n) & = & \frac{1}{4} n^2 \left[ (n+1)^2 {\cal S}(n+2) - 2 (n^2+1) {\cal S}(n) + (n-1)^2 {\cal S}(n-2) \right].
\label{H2app}
\eea
\ees
It turns out that the relation (\ref{a1est}) is frequently useful to estimate $a_1$ and thus the NLO coefficients  ${\cal C}_{1,k}^{(2 p)}$. On the other hand, the rleation (\ref{a2est}) gives only sometimes a useful (but rough) estimate of $a_2$ and thus of the NNLO coefficients  ${\cal C}_{2,k}^{(2 p)}$.

Nonetheless, often it is necessary to complement these estimates, or even to replace them, with the estimates obtained with a combination of two additional indicators which were regarded as necessary conditions and will be explained in the following.

The following LO, NLO and NNLO ratios will be considered:
\bes
\label{Ps}
\bea
P^{\rm (LO)}(n) & \equiv & \frac{d_n}{  \Gamma(\gamma+1+n) \left( \frac{\beta_0}{p} \right)^n },
\label{PLO}
\\
P^{\rm (NLO)}(n;e_1) & \equiv & \frac{d_n}{ \Gamma(\gamma+1+n)  \left( \frac{\beta_0}{p} \right)^n \left[ 1 + \frac{e_1}{(\gamma+n)} \right]},
\label{PNLO}
\\
P^{\rm (NNLO)}(n;e_1,e_2) & \equiv & \frac{d_n}{ \Gamma(\gamma+1+n) \left( \frac{\beta_0}{p} \right)^n \left[ 1 + \frac{e_1}{(\gamma+n)} + \frac{e_2}{(\gamma+n)(\gamma-1+n)}  \right]}.
\label{PNNLO}
\eea
\ees
Then the differences are formed
\be
{\rm diff}^{\rm (X)}(n) =  P^{\rm (X)}(n+1) - P^{\rm (X)}(n)
\label{diffdef}
\ee
where X=LO, NLO, NNLO. The idea is to adjust the value of $e_1$ to the correct value $e_1=a_1$, and the value of $e_2$ to the correct value $e_2=a_2$. For X=NLO (as a representative  example), we have the following expansion:
\bea
{\rm diff}^{\rm (NLO)}(n; e_1) &=& -\frac{S_{\gamma}}{n^2} \left[(a_1-e_1) +
  \frac{1}{n} \left( 2 a_2 + {\cal O}(a_1-e_1) \right) + {\cal O}\left( \frac{1}{n^2} \right) \right].
\label{diffexp1}
\eea
When we are approaching with the $e_1$ estimate to the true value $a_1$, the first term in brackets, $(a_1-e_1)$, becomes smaller, and becomes comparable with the second term at large $n$ (which is $\sim 1/n$). This means that we can have, at a large $n$, cancellation of these two terms. This means that we have a sign change in these differences ${\rm diff}^{\rm (NLO)}(n)$ at large $n$ when $n$ increases. Therefore, such a sign change at increasing $n$ indicates that we have not sufficiently approximated $e_1$ to the true value $a_1$. On the other hand, when $e_1$ is adjusted to $a_1$ with a sufficiently high precision, namely when $|a_1-e_1| \ll 2 |a_2|/n$, the first term $(a_1-e_1)$ in brackets of Eq.~(\ref{diffexp1}) becomes negligible even at large $n$ (we go up to $n=70$), and the difference becomes
\bea
{\rm diff}^{\rm (NLO)}(n; e_1 \approx a_1) & \approx & -\frac{S_{\gamma} 2 a_2}{n^3} \left[1 + {\cal O}\left( \frac{1}{n} \right) \right].
\label{diffexp2}
\eea
This means that we must adjust the value $e_1$ in such a way that the sequence $\{ {\rm diff}^{\rm (NLO)}(n; e_1); n=30, \ldots \}$ has no sign change (in practice we go up to $n=70$). Analogously, the value $e_2$ must be adjusted so closely to $a_2$ that the sequence $\{ {\rm diff}^{\rm (NNLO)}(n; a_1, e_2); n=30, \ldots \}$ has no sign change. These requirements of no sign change represent one of the two mentioned necessary conditions for the determination of $a_1$ and $a_2$, and thus of ${\cal C}_{1,k}^{(2 p)}$ and ${\cal C}_{2,k}^{(2 p)}$.

The other necessary condition is a somewhat arbitrary (although a rather conservative) requirement that at each next order we have at increasing $n$ ($40 < n \leq 70$) the velocity of convergence better by at least a factor of two. Specifically, this means that at thus large $n$, one requires
\be
|{\rm diff}^{\rm (NNLO)}(n; e_1,e_2)| < \frac{1}{2}|{\rm diff}^{\rm (NLO)}(n; e_1)| < \frac{1}{4} |{\rm diff}^{\rm (LO)}(n)|.
\label{2ndcond}
\ee

With a combination of all these conditions, the values, or rather the ranges of values, of the coefficients ${\cal C}_{1,k}^{(2 p)}$ and ${\cal C}_{2,k}^{(2 p)}$ appearing in Tables \ref{tabcalCrat} and \ref{tabcalCratT} in Sec.~\ref{subs:fulloneloop} were estimated.

\section{Explicit expressions for ${\cal F}_j$ and ${\cal F}_j^{\rm (SL)}$}
\label{app:FjFjSL}

The integrations (\ref{cFjcFjSL}) with the characteristic functions $G_D$ and $G_D^{\rm (SL)}$ of Eqs.~(\ref{GD})-(\ref{GDSL}) [corresponding to the generic Borel transform ${\rm B}[\tD](u)$, Eq.~(\ref{BtDgen}), which is the generator of coefficients $\td_n$] can be performed explicitly. This gives
\bes
\label{FplFmi}
\bea
{\cal F}_j^{(+)}(t) & = & \left[ - \td_{M,2}^{\rm UV} \frac{1}{(M+j)^2 t^{M+j}}
\left( 1 + (M+j) \ln t \right) - \td_{M,1}^{\rm UV} \frac{1}{(M+j) t^{M+j}} \right],
\label{Fpl}
\\
{\cal F}_j^{(-)}(t) & = & {\Bigg [} - \td_{M,2}^{\rm UV} \frac{1}{(M+j)^2}- \td_{M,1}^{\rm UV} \frac{1}{(M+j)} - \td_{2,1}^{\rm IR} \frac{1}{(2-j)} (1 - t^{2-j})
\nonumber\\ &&
- \td_{N,2}^{\rm IR} \frac{1}{(N-j)^2} \left( 1 - t^{N-j} + (N-j) t^{N-j} \ln t \right) - \td_{N,1}^{\rm IR} \frac{1}{(N-j)} (1 - t^{N-j}) {\Bigg ]},   
\label{Fmi}
\eea
\ees
where $j=0,1,3,4$. The case when $N=j$ (e.g., when $N=3=j$), the limiting value of the corresponding expression is implied
\be
\lim_{j \to N}  \td_{N,2}^{\rm IR} \frac{1}{(N-j)^2} \left( 1 - t^{N-j} + (N-j) t^{N-j} \ln t \right) = \td_{N,2}^{\rm IR} \; \frac{1}{2} \ln^2 t.
\label{lim}
\ee
Similarly, the corresponding integrals in the SL case (\ref{calFjSL}) are (note that $0<t<1$)
\bes
\label{FCSL}
\bea
{\cal F}_j^{\rm (SL)}(t) & = & \tal \td_{2,1} (-1) \int_{1/2}^t \frac{d u \; u}{u^j \ln u}  \qquad (j=0,1,3,4),
\label{FjSL}
\\
{\cal F}_1^{\rm (SL)}(t) & = & \tal \td_{2,1} \left[ -{\rm li}(t) + {\rm li}(1/2) \right],
\label{F1SL}
\\
C_1^{\rm (SL)} &=& \int_0^1 dt {\cal F}_1^{\rm (SL)}(t) = \tal \td_{2,1} \left[ \ln 2 + {\rm li}(1/2) \right],
\label{C1SL}
\\
{\cal F}_3^{\rm (SL)}(t) & = & \tal \td_{2,1} \left[ -{\rm Ei}(- \ln t) + {\rm Ei}(\ln 2) \right],
\label{F3SL}
\\
C_3^{\rm (SL)} &=& 3 \int_0^1 dt \; t^2 {\cal F}_3^{\rm (SL)}(t) = \tal \td_{2,1} \left[ \ln 2 + {\rm Ei}( \ln 2 ) \right],
\label{C3SL}
\\
{\cal F}_4^{\rm (SL)}(t) & = & \tal \td_{2,1} \left[ -{\rm Ei}(- 2 \ln t) + {\rm Ei}(\ln 4) \right],
\label{F4SL}
\\
C_4^{\rm (SL)} &=& 4 \int_0^1 dt \; t^3 {\cal F}_4^{\rm (SL)}(t) = \tal \td_{2,1} {\rm Ei}( \ln 4 ),
\label{C4SL}
\\
{\cal F}_0^{\rm (SL)}(t) & = & \tal \td_{2,1} \left[ -{\rm Ei}(2 \ln t) + {\rm Ei}(- \ln 4) \right],
\label{F0SL}
\\
C_0^{\rm (SL)} &=& {\cal F}_0^{\rm (SL)}(0) = \tal \td_{2,1} {\rm Ei}( -\ln 4 ),
\label{C0SL}
\eea
\ees
Here, ${\rm li}(z)$ and ${\rm Ei}$ are known functions whose integral representations are
\bea
{\rm li}(z) & = & \int_0^z \frac{u}{\ln u},
\label{li}
\\
{\rm Ei}(z) & = & - \int_{-z}^{\infty} \frac{du e^{-u}}{u}.
\label{Ei}
\eea
These two functions are evaluated in {\it Mathematica} \cite{Math} very fast (there, they are evaluated with the commands: ${\rm LogIntegral}[z]$ and  ${\rm ExpIntegralEi}[z]$, respectively).

\end{document}